\documentclass[%
prd,
reprint,nodate,
superscriptaddress,
nofootinbib,
 amsmath,amssymb,
 aps,
]{revtex4-1}

\pdfoutput=1

\usepackage[utf8]{inputenc} 

\usepackage{xcolor}
 
\usepackage{graphicx}% Include figure files
\usepackage{dcolumn}% Align table columns on decimal point
\usepackage{bm}% bold math
\usepackage{microtype}% micro-improvements to the formatting
\usepackage{threeparttable} %table with notes
\usepackage{mathtools}
\usepackage[colorlinks=true]{hyperref}

\newcommand{\rf}[1]{Eq.\,(\ref{#1})}

\newcommand{\be}{\begin{eqnarray}}
\newcommand{\ee}{\end{eqnarray}}

\begin{document}

\author{Michal P.\ Heller} \email{michal.p.heller@ugent.be}
\affiliation{Department of Physics and Astronomy, Ghent University, 9000 Ghent, Belgium}
\affiliation{Max Planck Institute for Gravitational Physics (Albert Einstein Institute),
14476 Potsdam-Golm, Germany}

\author{Alexandre Serantes}
\email{alexandre.serantes@ub.edu}
\affiliation{Departament de Física Quàntica i Astrofísica, Institut de Ciències del Cosmos (ICCUB), Facultat de Física, Universitat de Barcelona, Martí i Franquès 1, E-08028 Barcelona, Spain}
\affiliation{National Centre for Nuclear Research, 02-093 Warsaw, Poland}

\author{Micha\l\ Spali\'nski}
\email{michal.spalinski@ncbj.gov.pl}
\affiliation{National Centre for Nuclear Research, 02-093 Warsaw, Poland}
\affiliation{Physics Department, University of Bia{\l}ystok, 15-245 Bia\l ystok, Poland}

\author{\\Viktor Svensson}
\email{viktor.svensson@ftf.lth.se}
\affiliation{Division of Solid State Physics and NanoLund, Lund University, S-221 00 Lund, Sweden}
\affiliation{National Centre for Nuclear Research, 02-093 Warsaw, Poland}
\affiliation{Max Planck Institute for Gravitational Physics (Albert Einstein Institute),
14476 Potsdam-Golm, Germany}

\author{Benjamin Withers}
\email{b.s.withers@soton.ac.uk}
\affiliation{Mathematical Sciences and STAG Research Centre, University of Southampton, Highfield, Southampton SO17 1BJ, UK}

\title{Relativistic Hydrodynamics: A Singulant Perspective}

\begin{abstract}
There is growing evidence that the hydrodynamic gradient expansion is factorially divergent. We advocate for using Dingle's singulants as a way to gain analytic control over its large-order behaviour for nonlinear flows. Within our approach, singulants can be viewed as new emergent degrees of freedom which reorganise the large-order gradient expansion. We work out the  physics of singulants for longitudinal flows, where they obey simple evolution equations which we compute in M\"uller-Israel-Stewart-like models, holography and kinetic theory. These equations determine the dynamics of the large-order behaviour of the hydrodynamic expansion,  which we confirm with explicit numerical calculations. One of our key findings is a duality between singulant dynamics and a certain linear response theory problem. Finally, we discuss the role of singulants in optimal truncation of the hydrodynamic gradient expansion. A by-product of our analysis is a new M\"uller-Israel-Stewart-like model, where the qualitative behaviour of singulants shares more similarities with holography than models considered hitherto.
\end{abstract}

\maketitle

\section{Introduction}\label{sec:introduction}

The past two decades have been a true golden age for our understanding of dissipative relativistic fluids. This surge of interest has been primarily driven by the interplay between growing sophistication in hydrodynamic modelling of the newly discovered quark-gluon plasma at RHIC (and later also at LHC)~\cite{Heinz:2013th,Busza:2018rrf} and unprecedented progress in studying nonequilibrium phenomena with hydrodynamic tails at strong and weak interaction strength using, respectively, holography and relativistic kinetic theory~\cite{Florkowski:2017olj,Romatschke:2017ejr}. Today the domain of relativistic hydrodynamics encompasses also time-dependent black hole phenomena via the holographic fluid-gravity duality~\cite{Bhattacharyya:2007vjd}, neutron star modelling in the context of gravitational wave physics~\cite{Shibata:2017jyf,Alford:2017rxf,Bemfica:2019cop}, as well as condensed matter phenomena~\cite{Hartnoll:2016apf}, including electron flow in graphene~\cite{Lucas:2017idv}. Furthermore, recent developments indicate that relativistic hydrodynamics can apply also to far-from-equilibrium situations, which is a subject of the hydrodynamics attractors research program~\cite{Heller:2015dha,Romatschke:2017vte,Soloviev:2021lhs}.

The key notion underlying relativistic hydrodynamics is that of constitutive relations. As an effective field theory of transport of conserved currents, the fundamental object of interest is the expectation value of the energy-momentum tensor~$T^{\mu \nu}$. Hydrodynamic constitutive relations provide an ansatz for this quantity as an infinite series in gradients
\begin{subequations}\label{intro:master}
\begin{eqnarray}\label{intro:T_Landau_frame}
T_{\mu\nu} &=& \mathcal{E}\, U_\mu U_\nu + \mathcal{P}(\mathcal{E})\,(g_{\mu\nu} + U_\mu U_\nu) + \Pi_{\mu\nu},\\
\label{intro:PiSeries}\Pi_{\mu\nu} &=& \sum_{n=1}^\infty \epsilon^n \Pi_{\mu\nu}^{(n)},
\end{eqnarray}
\end{subequations}
where $\cal E$ and $U_{\mu}$ ($U_{\mu} U^{\mu} = -1$) are the local energy density and the local velocity encapsulating slow degrees of freedom and $\Pi^{\mu \nu}$ is a gradient-expanded dissipative part of the energy-momentum tensor. Here and in the following, $\epsilon$ is a formal bookkeeping parameter counting the number of gradients, and equalities involving infinite power series in $\epsilon$ are to be understood in the sense of a formal ansatz. To remove ambiguities in the gradient expansion \eqref{intro:master} associated to field redefinitions, we work in the Landau frame and hence demand that $\Pi_{\mu\nu}$ is transverse to the fluid velocity, $\Pi_{\mu\nu}U^\nu = 0$. We refer the reader to Refs.~\cite{Florkowski:2017olj,Romatschke:2017ejr} for contemporary reviews on relativistic hydrodynamics.

The gradient expansion \eqref{intro:master} can be constructed as long as the energy-momentum tensor at the spacetime point being considered has a single real timelike eigenvector \cite{Arnold:2014jva}. Provided that this condition is met the gradient expansion \eqref{intro:master} can be written down, and thus it is natural to ask whether it can capture processes arbitrarily far away from global thermal equilibrium. The answer to this question depends crucially on whether the gradient expansion is a convergent series. If convergent, the gradient expansion \eqref{intro:master} can in principle provide a full description of the underlying microscopic theory as it pertains to the expectation value of the energy-momentum tensor. If this were the case, there would be no obstruction in obtaining an approximation to $T_{\mu\nu}$ with an error as small as desired: one just needs to truncate the gradient expansion at a sufficiently high order. On the other hand, if the gradient expansion is a divergent series, this is not possible, even in principle. In this situation, the first step becomes that of finding the optimal truncation order.

Results for highly symmetric flows obtained in recent years have provided indications that the hydrodynamic gradient expansion has a vanishing radius of convergence~\cite{Heller:2013fn,Heller:2015dha,Basar:2015ava,Aniceto:2015mto,Denicol:2016bjh,Florkowski:2016zsi,Heller:2016rtz,Buchel:2016cbj,Denicol:2017lxn,Behtash:2017wqg,Casalderrey-Solana:2017zyh,Denicol:2018pak,Baggioli:2018bfa,Buchel:2018ttd,Aniceto:2018uik,Heller:2018qvh,Blaizot:2019scw,Denicol:2019lio,Behtash:2019qtk,Du:2021fok}. This was further corroborated in situations with less~\cite{Heller:2021oxl} and even without any symmetry~\cite{Heller:2020uuy}. In these highly symmetric flows~\cite{Heller:2013fn,Heller:2015dha,Basar:2015ava,Aniceto:2015mto,Denicol:2016bjh,Florkowski:2016zsi,Heller:2016rtz,Buchel:2016cbj,Denicol:2017lxn,Behtash:2017wqg,Casalderrey-Solana:2017zyh,Denicol:2018pak,Baggioli:2018bfa,Buchel:2018ttd,Aniceto:2018uik,Heller:2018qvh,Blaizot:2019scw,Denicol:2019lio,Behtash:2019qtk,Du:2021fok}, as well as in other settings~\cite{Marino:2012zq,Aniceto:2013fka,Aniceto:2018bis}, the dominant large-order behaviour of the series coefficients in question, $J^{(n)}$, takes on a factorial-over-power form 
\be
\label{factopower}
J^{(n)} \sim A \frac{\Gamma(n+\alpha)}{\chi^{n+\alpha}},  \label{singulant_ansatz_basic}
\ee
where $\chi$ and $\alpha$ are constants. This form of the expansion coefficients $J^{(n)}$ at large order is motivated by the work of Dingle~\cite{dingle1973asymptotic} and subsequent studies of factorially divergent series. A key role in these considerations in played by the parameter $\chi$, for which Dingle introduced the term {\em singulant}.  Indeed, the remaining parameters in \rf{factopower} play no role at leading order for large $n$. Asymptotic expansions of the  form~\eqref{factopower} appear supplemented by exponential corrections. Such generalised series (transseries) involve intricate resurgence relations between their coefficients.

In this paper we propose to systematically apply the idea of the singulants to the gradient expansion~\eqref{intro:PiSeries} regardless of the presence of any flow symmetries. This leads to the large-order ansatz \eqref{factopower}
\be\label{singulant_ansatz_general}
\Pi^{(n)}_{\mu\nu}(t,\vec{x}) \sim A_{\mu\nu}(t,\vec{x}) \frac{\Gamma(n + \alpha(t,\vec{x}))}{\chi(t,\vec{x})^{n + \alpha(t,\vec{x})}}, 
\ee
with the singulant itself becoming a scalar field in spacetime (see also Refs.~\cite{chapman1998exponential,chapman2005exponential}). Again, $A_{\mu\nu}(t,\vec{x})$ and $\alpha(t,\vec{x})$ play no role at leading order at large $n$. Our analysis will focus on longitudinal flows \cite{Florkowski:2016kjj,Heller:2021oxl}, where we explicitly confirm the validity of the ansatz \eqref{singulant_ansatz_general} in a number of models. 
We will see that, in general, there will be multiple singulants which contribute additively to Eq.~\eqref{singulant_ansatz_general}. From the point of view of the hydrodynamic gradient expansion alone, singulants can be thought of as a novel emergent phenomenon in relativistic hydrodynamics, not necessarily related to a single contribution to the expansion at a given order, but rather to a reorganisation of the whole series. 
\begin{figure}
\label{fig:chiHA}
\centering
\includegraphics[width=0.6\columnwidth]{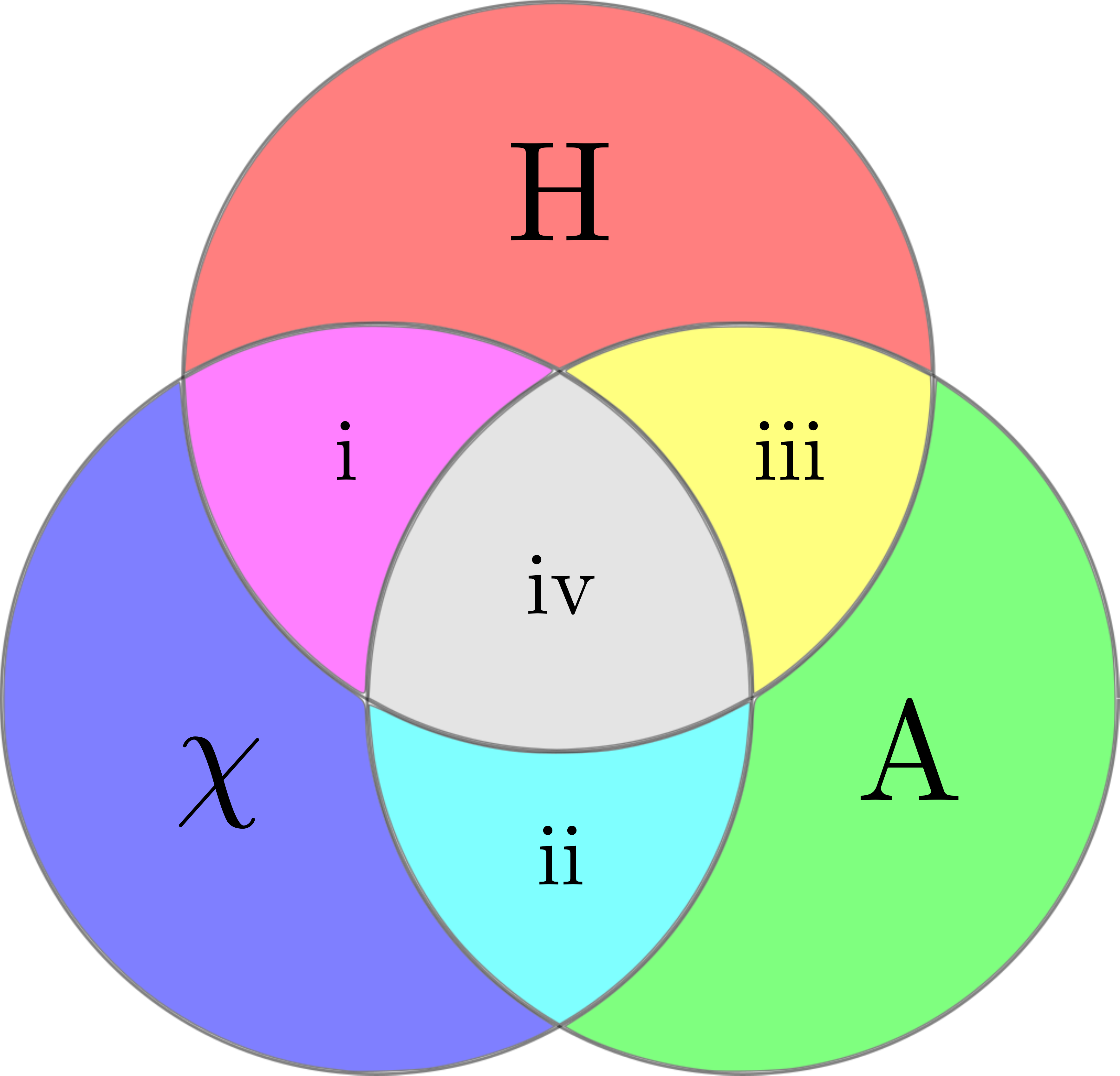}
\caption{Hydrodynamics ($\mathrm{H}$) and linear response theory ($\mathrm{A}$) are well-known techniques to study collective states of matter. Our paper introduces in this context a new perspective based on singulants ($\chi$). The main text discusses the meaning of different overlaps between these domains represented as disks.}
\end{figure}

For systems near equilibrium (or other special solutions such as attractors) quasinormal modes describe the decay of non-hydrodynamic degrees of freedom and the approach to equilibrium. Far from equilibrium however, quasinormal modes can no longer be used and it is not obvious what concept or object replaces them, if anything. We advocate that singulants can fill this role. The reason for this is that singulants are fundamentally non-hydrodynamic collective fields whose contribution to the energy-momentum tensor decays over time. To see that they are non-hydrodynamic we note that our large-order ansatz is intimately related to the appearance of contributions to the transseries generalisation of the hydrodynamic gradient expansion which are non-perturbative in $\epsilon$. In all examples considered in this paper, the value of $|\chi|$ grows after enough time has passed. This leads to a growth in the optimal truncation order and a decrease in the truncation error, thus gradually extending the applicability of the hydrodynamic expansion over time.

Our results indicate that singulants ($\chi$) provide a new perspective on collective states of matter that complements and connects in a novel way the existing paradigms of the hydrodynamic gradient expansion ($\mathrm{H}$) and the amplitude expansion exemplified by linear response theory~($\mathrm{A}$). This is represented by Fig.\,\ref{fig:chiHA}. In this work, we primarily consider the region $\mathrm{i}$, where the large-order behaviour of the gradient expansion \eqref{intro:master} for nonlinear fluids is governed by singulants. One of our most relevant findings is a duality between the singulant dynamics, as computed in region ($\mathrm{i}$), and a particular linear response theory problem defined in region ($\mathrm{ii}$). Region ($\mathrm{iii}$) corresponds to the realm of linearised hydrodynamics, where one considers the gradient expansion \eqref{intro:master} for infinitesimal perturbations away from global thermal equilibrium, but does not focus on its large-order behaviour. Finally, in region ($\mathrm{iv}$), when utilising the three approaches---singulants, the gradient expansion and linear response theory---we gain complete analytic control of the singulants. Appendix~\ref{app:D} in the Supplemental Material provides an example of such a study.

This paper is organised as follows. In Sec.\,\ref{sec:singulants}, we introduce longitudinal flows and discuss several results pertaining to the large-order behaviour of the gradient expansion \eqref{intro:master} and singulants in this context. Then, in Sec.\,\ref{sec:phenomenological_models}, we test the general results for longitudinal flows put forward in Sec.\,\ref{sec:singulants} in a series of phenomenological models of the M{\"u}ller-Israel-Stewart (MIS) class. These include the Baier-Romatschke-Son-Starinets-Stephanov (BRSSS) model \cite{Baier:2007ix}, the Heller-Janik-Spaliński-Witaszczyk (HJSW) model \cite{Heller:2014wfa}, and a new model we introduce for the first time in this work. These and related models originate from the need to have equations of motion for relativistic fluids that give rise to causal and stable evolutions~\cite{Romatschke:2009im,Bemfica:2020xym}. In particular, the original MIS model is the workhorse for the vast majority of hydrodynamic simulations of quark-gluon plasma in nuclear collisions. In the three MIS-like models we consider, we demonstrate that the singulant dynamics can be mapped to a linear response theory problem that consists in computing the poles of a momentum-dependent sound attenuation length $\gamma_s$. In Sec.\,\ref{sec:holography}, we explore the large-order behaviour of the gradient expansion for longitudinal flows in the context of holography and show that, if a factorial divergence is present, then the singulant equation of motion is also determined by the poles of $\gamma_s$. Sec.\,\ref{sec:kt} discusses singulants in kinetic theory, focusing on the gradient expansion of the distribution function. We test our analytic predictions by explicit numerical computations for Bjorken flow in the relaxation time approximation (RTA). Sec.\,\ref{sec:optimal} explores the interplay between singulants and optimal truncation in the context of BRSSS theory. The paper closes with a discussion of the physical interpretation of singulants and key open problems in Sec.\,\ref{sec:outlook}. 

Several computations supporting the results presented in the main body of the paper have been relegated to Supplemental Material as appendices. Appendix~\ref{app:A} provides the demonstration of the general results introduced in Sec.\,\ref{sec:singulants}, while Appendix~\ref{app:B} discusses the large-order behaviour of the gradient expansion \eqref{intro:master} beyond longitudinal flows in the context of MIS-like models. Appendix~\ref{app:C} contains the detailed computation of $\gamma_s$ in holography. Appendix~\ref{app:D}---which has been mentioned before---explores the large-order behaviour of the gradient expansion \eqref{intro:master} in linearised hydrodynamics. Finally, in Appendix~\ref{app:E} we discuss the causality and stability of the new MIS-like model we introduced in this work, at the level of linearised perturbations of global thermal equilibrium.

\section{Singulants in longitudinal flows}\label{sec:singulants}

In this paper we develop the paradigm put forward in the Introduction for a particular class of fluid flows which we refer to as longitudinal. Longitudinal flows simplify drastically the technical aspects of our analysis and, at the same time, keep intact the essential features of far-from-equilibrium fluid dynamics. In particular, as originally shown in Ref.\,\cite{Heller:2021oxl}, for longitudinal flows in MIS-like theories it is feasible to compute the gradient expansion \eqref{intro:PiSeries} numerically up to an order sufficiently large to assess its asymptotic behaviour. 

A longitudinal flow in $d$-dimensional Minkowski spacetime is defined by singling out one spatial direction, $x$, and demanding translational invariance and isotropy in the hyperplane spanned by the remaining spatial coordinates $x_\perp^{(1)},\ldots,x_\perp^{(d-2)}$. This requirement implies that that the nontrivial dynamics is confined to the plane spanned by $t$ and $x$, which we refer to as the longitudinal plane. With these symmetry restrictions, both the fluid velocity $U^\mu$ and any two-tensor $A^{\mu\nu}$ which is symmetric, transverse to $U^\mu$, and traceless can be parameterised in terms of a single degree of freedom. Specifically, 
\begin{subequations}
\begin{equation}
U^\mu\partial_\mu = \cosh{u} \, \partial_t + \sinh{u} \, \partial_x,
\end{equation}
\begin{equation}
A^{\mu\nu} = (2-d)
\left(
\eta^{\mu\nu} + U^{\mu}U^\nu - \frac{d-1}{d-2}P^{\mu\nu}_T
\right)
A_\star,  
\end{equation}
\end{subequations}
where 
$P^{\mu\nu}_T$ is the projector into the transverse hyperplane and $u$,~$A_\star$ depend solely on $t$ and $x$. In this work, we consider only longitudinal flows in conformal theories, in such a way that $T_{\mu\nu}$ is traceless. This entails that the equation of state of the fluid is given by $\mathcal{P}(\mathcal{E}) = \mathcal{E}/(d-1)$ and that $\Pi_{\mu\nu}$ is also traceless. Using the energy density we define $T(t,x) = (\mathcal{E}(t,x)/\mathcal{E}_0)^\frac{1}{d}$ and refer to this field as the effective temperature. In equilibrium $T(t,x)$ becomes the temperature of the system, as determined by the equation of state.

In a longitudinal flow, the gradient expansion \eqref{intro:PiSeries} becomes a gradient expansion for $\Pi_\star$, 
\begin{equation}\label{singulants:Pi_series}
\Pi_\star(t,x) = \sum_{n=1}^\infty \Pi_\star^{(n)}(t,x) \epsilon^n, \end{equation}
and the asymptotic ansatz \eqref{singulant_ansatz_general} reads
\begin{equation}\label{factorial_ansatz}
\Pi_\star^{(n)}(t,x) \sim A(t,x) \frac{\Gamma(n+\alpha(t,x))}{\chi(t,x)^{n+\alpha(t,x)}}. 
\end{equation}
The singulant field, $\chi(t,x)$, controls the subleading geometric correction to the leading-order factorial growth of the gradient expansion.

In this work, we explore the gradient expansion \eqref{singulants:Pi_series} in the context of MIS-like models (Sec.~\ref{sec:phenomenological_models}), holography (Sec.~\ref{sec:holography}) and RTA kinetic theory (Sec.~\ref{sec:kt}). We note the following. 
\begin{itemize}
\item In MIS-like theories, we work directly at the level of $\Pi_\star$ and the gradient expansion \eqref{singulants:Pi_series}.
\item In holography, we work with the gradient expansion of the bulk metric, 
\begin{equation}\label{singulants:metric}
g_{AB}(X) = \sum_{n=0}^\infty g_{AB}^{(n)}(X)\epsilon^n,  
\end{equation}
from which the gradient expansion \eqref{singulants:Pi_series} descends by holographic renormalisation. In Eq.~\eqref{singulants:metric}, $X=(r,x)$ and $r$ parameterises the radial direction of the higher-dimensional geometry. The singulant field governing the large-order behaviour of the gradient expansion \eqref{singulants:Pi_series} follows from the singulant field governing the large-order behaviour of the gradient expansion \eqref{singulants:metric}: 
\begin{equation}\label{factorial_ansatz_metric}
g_{AB}^{(n)} \sim A_{AB}(X) \frac{\Gamma(n+\alpha(X))}{\chi(x)^{n+\alpha(X)}}. 
\end{equation}
Note that we have assumed that the singulant field is $r$ independent. We will show that this assumption is self-consistent in Sec.~\ref{sec:holography}.  
\item In kinetic theory, we work with the gradient expansion of the distribution function, 
\begin{equation}\label{singulants:f}
f(x,p) = \sum_{n=0}^\infty f^{(n)}(x,p)\epsilon^n, 
\end{equation}
from which the gradient expansion \eqref{singulants:Pi_series} descends by computing the second-order moments. In Eq.~\eqref{singulants:f}, $p$ is the momentum. The singulant field governing the large-order behaviour of the gradient expansion \eqref{singulants:Pi_series} follows from the singulant field governing the large-order behavior of the gradient expansion \eqref{singulants:f}:
\begin{equation}\label{factorial_ansatz_f}
f^{(n)} \sim A(x,p) \frac{\Gamma(n+\alpha(x,p))}{\chi(x,p)^{n+\alpha(x,p)}}. 
\end{equation}
\end{itemize}

In every theory being considered in this work, the exact $\Pi_\star$ can be computed in terms of the energy density $\mathcal{E}$ and fluid velocity $U \equiv U^\mu\partial_\mu$ associated to a given out-of-equilibrium state by solving a nonlinear system of integro-PDEs (partial differential equations). This nonlinear system of integro-PDEs is specific to the theory in question. One has the following.
\begin{itemize}
\item In MIS-like models, it corresponds to the relaxation equation obeyed by $\Pi_{\mu\nu}$ [cf. Eqns.~\eqref{BRSSS:pistar}, \eqref{HJSW:dynamical_constitutive_relation}, and \eqref{new_model_def} in Sec.~\ref{sec:phenomenological_models}]. 
\item In holography, it corresponds to a subset of the Einstein equations governing the higher-dimensional bulk geometry [cf. Eq.~\eqref{eq.dyneinst} in Sec.~\ref{sec:holography}]. 
\item In kinetic theory, it corresponds to the Boltzmann equation [cf. Eq.~\eqref{kt:boltzmann} in Sec.~\ref{sec:kt}].
\end{itemize}
Given the hydrodynamic fields $\mathcal{E}$ and $U$, we determine the coefficients of the gradient expansions \eqref{singulants:Pi_series}, \eqref{singulants:metric}, and \eqref{singulants:f} through the following systematic procedure. 
\begin{itemize}
\item[(i)] We introduce the bookkeeping parameter $\epsilon$ into the nonlinear system of integro-PDEs determining $\Pi_\star$ mentioned above by means of a homogeneous rescaling of the longitudinal plane coordinates: 
\begin{equation}\label{singulants:rescaling}
t \to \frac{t}{\epsilon}, \quad x \to \frac{x}{\epsilon}.  
\end{equation}
\item[(ii)] We plug in the ansatz between Eqs.~\eqref{singulants:Pi_series}, \eqref{singulants:metric}, \eqref{singulants:f} appropriate for the theory in question. 
\item[(iii)] We solve in a small-$\epsilon$ expansion. 
\end{itemize}
The end result of this procedure is a set of recursion relations that fix the $n$th order coefficient of the relevant gradient expansion in terms of the previous lower-order ones. Explicit examples are provided in the rest of the paper whenever necessary. We emphasise that, once the recursion relations have been written down, any ambiguity in the gradient expansion associated with the usage of the conservation equations is removed.

The singulant equation of motion is determined by the recursion relations, and it follows straightforwardly from the requirement that the asymptotic ansatz \eqref{factorial_ansatz} [\eqref{factorial_ansatz_metric} and  \eqref{factorial_ansatz_f} resp.] for the coefficients of the gradient expansion \eqref{singulants:Pi_series} [ \eqref{singulants:metric} and \eqref{singulants:f} resp.] solves the associated recursion relations at leading order at large $n$. A fact that will be crucial for our analysis is that, for a factorially divergent gradient expansion, the recursion relations simplify drastically in this large $n$ regime. The two major simplifications taking place at large $n$ are the following.
\begin{itemize}
\item[(a)] The recursion relations become linear and, 
\item[(b)] terms in the recursion relations associated with gradients of the hydrodynamic fields, $\mathcal{E}$ and $U$, drop out.
\end{itemize}
Hence, rather than being sensitive to every possible term in the recursion relations, the singulant dynamics only depends on a subset of dominant terms, which are characterised by points (a) and (b) above.

To illustrate the general discussion of the previous paragraph, we consider the case of MIS-like models. The cases of holography and kinetic theory are similar and are discussed in Secs.~\ref{sec:holography} and \ref{sec:kt}, respectively. In MIS-like models, when one introduces the large-order ansatz \eqref{factorial_ansatz} into the recursion relations and takes the $n \to \infty$ limit, one finds that the recursion relations simplify to 
\begin{equation}\label{singulants:surviving_terms}
\sum_{p=0}^2 f_{(p)}^{\mu_1\ldots\mu_p}(T,U)\partial_{\mu_1}\ldots\partial_{\mu_p} \Pi_{\star}^{(n-p)} \sim 0, \quad n\to\infty.  
\end{equation}
In accordance with property (a), Eq.~\eqref{singulants:surviving_terms} is linear in $\Pi_\star$; in accordance with property (b), the tensors $f_{(p)}^{\mu_1\ldots\mu_p}$ only depend on the local values of the hydrodynamic fields at the spacetime point being considered. Terms involving gradients of these fields and/or nonlinear in $\Pi_\star$ result in subleading contributions at large $n$ and are therefore not included in Eq.~\eqref{singulants:surviving_terms}. The singulant equation of motion follows immediately from the fact that the factorial-over-power ansatz \eqref{factorial_ansatz} has to solve the simplified form of the recursion relations \eqref{singulants:surviving_terms} and the observation that, at leading order at large $n$, the following identity holds: 
\begin{equation}
\partial_{\mu_1}\ldots\partial_{\mu_p}\Pi_\star^{(n-p)} \sim (-1)^p \partial_{\mu_1}\chi\ldots\partial_{\mu_p}\chi \Pi_\star^{(n)}.     
\end{equation}
One important consequence of the linearisation of the recursion relations is that the most general large-order behaviour of a factorially divergent gradient expansion of the form \eqref{singulants:Pi_series} (resp. \eqref{singulants:metric} and \eqref{singulants:f}) is described by a linear combination of contributions of the form \eqref{factorial_ansatz} (resp. \eqref{factorial_ansatz_metric} and \eqref{factorial_ansatz_f}),
\begin{equation}\label{singulants:factorial-over-power_ansatz}
\Pi_\star^{(n)}(t,x) \sim \sum_q A_q(t,x) \frac{\Gamma(n+\alpha_q(t,x))}{\chi_q(t,x)^{n+\alpha_q(t,x)}}. 
\end{equation}
where each $\chi_q$ satisfies the same equation of motion but with different initial conditions. Note that, in every theory being considered, the reality of the gradient expansion coefficients appearing in the left hand side of Eq.~\eqref{singulants:factorial-over-power_ansatz} implies that every singulant contribution appearing in the right hand side is either real or accompanied by a complex-conjugated partner. In the specific models where we computed the gradient expansion numerically, we always find that these complex-conjugated singulant pairs are present. When multiple singulant contributions are present, we refer to the one with the smallest norm as the \emph{dominant singulant}.

A complementary viewpoint on the singulant dynamics is provided by the observation that, when upgrading the perturbative series \eqref{singulants:Pi_series} to a transseries, the singulants weight the nonperturbative transseries sectors, 
\begin{equation}\label{singulants:transseries}
\Pi_\star = \sum_{n=1}^\infty \Pi_\star^{(n)} \epsilon^n + \sum_q e^{-\frac{\chi_q}{\epsilon}}\sum_{n=1}^\infty \tilde{\Pi}_{\star,q}^{(n)} \epsilon^n + \ldots,    
\end{equation}
where the ellipsis represents possible nonperturbative transseries sectors associated to nonlinear interactions between different singulants $\chi_q$. In the holography and kinetic theory cases, there exist transseries analogous to Eq.\,\eqref{singulants:transseries} for $g_{AB}$ and $f$, from which Eq.\,\eqref{singulants:transseries} descends naturally. 

The transseries approach allows for an alternative derivation of the singulant equation of motion. In this alternative derivation, one starts with the nonlinear system of integro-PDEs appropriate for the theory in question after the bookkeeping parameter $\epsilon$ has been introduced but, rather than introducing the relevant gradient expansion among \eqref{singulants:Pi_series}, \eqref{singulants:metric}, and \eqref{singulants:f}, one plugs in its associated transseries ansatz and expands around $\epsilon = 0$. Demanding that the coefficient multiplying $e^{-\frac{\chi_q}{\epsilon}}$ vanishes at leading order in the $\epsilon \to 0$ limit gives directly the singulant equation of motion. From this perspective, the singulant equation of motion can be interpreted as the eikonal equation coming from a WKB analysis of the nonlinear integro-PDE system determining $\Pi_\star$.

As we discuss in detail in Appendix~\ref{app:A} in the Supplemental Material, the procedure described in the previous paragraph is equivalent to 
\begin{itemize}
\item[(i)] take the original nonlinear system of integro-PDEs and linearise it around the zeroth-order term of the corresponding gradient expansion, 
\item[(ii)] neglect terms associated to gradients of the hydrodynamic fields, and 
\item[(iii)] perform the replacement, 
\begin{equation}
\partial_{\mu_1}\ldots\partial_{\mu_n} \to (-1)^n\partial_{\mu_1}\chi\ldots \partial_{\mu_n}\chi.
\end{equation}
\end{itemize}
The end result of (i)--(iii) is again the singulant equation of motion. One of the main benefits of the transseries approach, as embodied in these three steps, is that, in the cases of MIS-like models and holography, it allows one to identify in a straightforward way a linear response theory computation dual to the singulant equation of motion. Indeed, steps (i)--(iii) are respectively equivalent to first taking the nonlinear system of integro-PDEs that determines $\Pi_\star$, then setting the hydrodynamic fields, $T$ and $U$, to spacetime-independent constants, $T_0$ and $U_0$, and finally finding the dispersion relation for linearised perturbations of $\Pi_\star$: 
\begin{equation}\label{singulants:plane-wave}
\delta\Pi_\star = \delta\hat{\Pi}_{\star} e^{i k_\mu x^\mu}.      
\end{equation}
There exists a map involving the identifications, 
\begin{equation}\label{singulants:map}
T_0 \to T(t,x), \quad U_0 \to U(t,x), \quad i k_\mu \to - \partial_\mu \chi,     
\end{equation}
that transforms the equations determining the dispersion relation of the linearised perturbation \eqref{singulants:plane-wave} into the equations of motion determining the singulant dynamics. This map is discussed in further detail in Appendix~\ref{app:A} in the Supplemental Material and we will see explicit incarnations of it in Secs.~\ref{sec:phenomenological_models} and \ref{sec:holography}.

A linear response theory problem that features prominently in studies of nonequilibrium physics in MIS-like theories, holography and kinetic theory is the computation of the modes of the system \cite{Baier:2007ix, Horowitz:1999jd, Kovtun:2005ev, Romatschke:2015gic}. A mode is a singularity of a retarded thermal two-point correlator in Fourier space, and can be thought of as being located at a frequency $\omega \in \mathbb{C}$ that depends nontrivially on the the spatial momentum $\textbf{k} \in \mathbb{C}^{d-1}$ as determined by the dispersion relation $\omega = \omega(\textbf{k})$. In linear response theory, hydrodynamic modes are associated with long-lived and slowly varying excitations such that $\omega(\textbf{k}) \to 0$ as $|\textbf{k}| \to 0$, while nonhydrodynamic modes are associated with excitations that do not obey the latter property.

A fact that the reader should keep in mind is that the linear response theory problem dual to the singulant equation of motion may not correspond to a mode computation. The reason is that, in a mode computation, the hydrodynamic fields $T$ and $U$ are treated dynamically. These two problems are therefore equivalent if and only if the dynamics of the hydrodynamic fields decouple from the dynamics of $\delta\Pi_\star$. For general longitudinal flows, this will be the case in the MIS-like models discussed in Secs.~\ref{subsec:BRSSS} and \ref{subsec:HJSW}, but not in the in the new model introduced in Sec.~\ref{subsec:new_model}, nor in holography.

To conclude our overview of singulants in longitudinal flows, we comment briefly on how singulants can be extracted from the numerical values of the gradient expansion coefficients. For illustrative purposes, we focus on a gradient expansion of the form \eqref{singulants:Pi_series} with the large-order behaviour \eqref{singulants:factorial-over-power_ansatz}, and note that identical strategies exist for the gradient expansions of the metric \eqref{singulants:metric} and the distribution function \eqref{singulants:f}.

First, in the restricted case in which one is solely interested in the absolute value of the dominant singulant, $|\chi_d|$, it follows from the relation \begin{equation}\label{singulants:root-test_master_formula}
|\Pi_\star^{(n)}|^\frac{1}{n} \sim \frac{n}{e |\chi_d|}, \quad n\to\infty, 
\end{equation}
that, asymptotically, a root-test plot of gradient expansion coefficients results in a straight line of slope $1/(e |\chi_d|)$, in such a way that $|\chi_d|$ is fixed once this slope is known.\footnote{If the dominant singulant belongs to a pair invariant under complex conjugation there are oscillations of frequency $\arg \chi$ superimposed to the linear growth; however, these oscillations are subleading in the $n \to \infty$ limit.} On the other hand, if one wishes to determine the values of the subdominant singulants as well, the best strategy is to consider the analytical continuation of the Borel transform of the gradient expansion.  In this case, each singulant contribution $\chi_q$ appears as a singularity in the Borel plane located at $\zeta = \chi_q$, where $\zeta$ is a complex-valued variable used to parameterise the Borel plane throughout this paper.

\section{MIS-like models}\label{sec:phenomenological_models}

\subsection{Introduction}

In this section, we compute the singulants contributing to the large-order behaviour of the gradient expansion for longitudinal flows in a class of phenomenological MIS-like models. As mentioned in the Introduction, these are the BRSSS model \cite{Baier:2007ix}, the HJSW model \cite{Heller:2014wfa}, and the new model introduced for the first time in this work. The rationale behind these models is embedding hydrodynamics in a framework compatible with relativistic causality. To achieve this, the dissipative tensor $\Pi_{\mu\nu}$ is promoted to a set of independent dynamical degrees of freedom obeying their own equation of motion. While such models are inspired by hydrodynamics, it is crucial to note that unlike hydrodynamics they are \emph{exact}, in the sense that they are not defined using a perturbative expansion in the number of gradients.

Different models in this class are distinguished by the different equation of motion obeyed by $\Pi_{\mu\nu}$ which we detail case by case in the sections that follow; however, each case obeys the same current conservation equations, $\nabla_\mu T^{\mu\nu} = 0$, which for longitudinal flows are
\begin{eqnarray}
D\mathcal{E} + \left[\frac{d \mathcal{E}}{d-1} - (2-d)\Pi_\star \right] (\nabla\cdot U) &=& 0,\label{currentcons:1}\\
\frac{\nabla^\mu \mathcal{E}}{d-1} + \left[ \frac{ d \mathcal{E}}{d-1} + (2-d)\Pi_\star \right] DU^\mu &=& 0, \label{currentcons:2}
\end{eqnarray}
where $D = U^\mu \nabla_\mu$ is a longitudinal derivative. 

As a final comment, we note that the MIS-like models we consider in this work are causal and stable at the level of linear response around global thermal equilibrium, however the reader should keep in mind that these are not sufficient conditions for their causality and well-posedness at the fully nonlinear level \cite{Bemfica:2020xym,Bemfica:2020zjp}.

\subsection{BRSSS model}\label{subsec:BRSSS}
\begin{figure}[h!]
\begin{center}
\includegraphics[width=0.92\columnwidth]{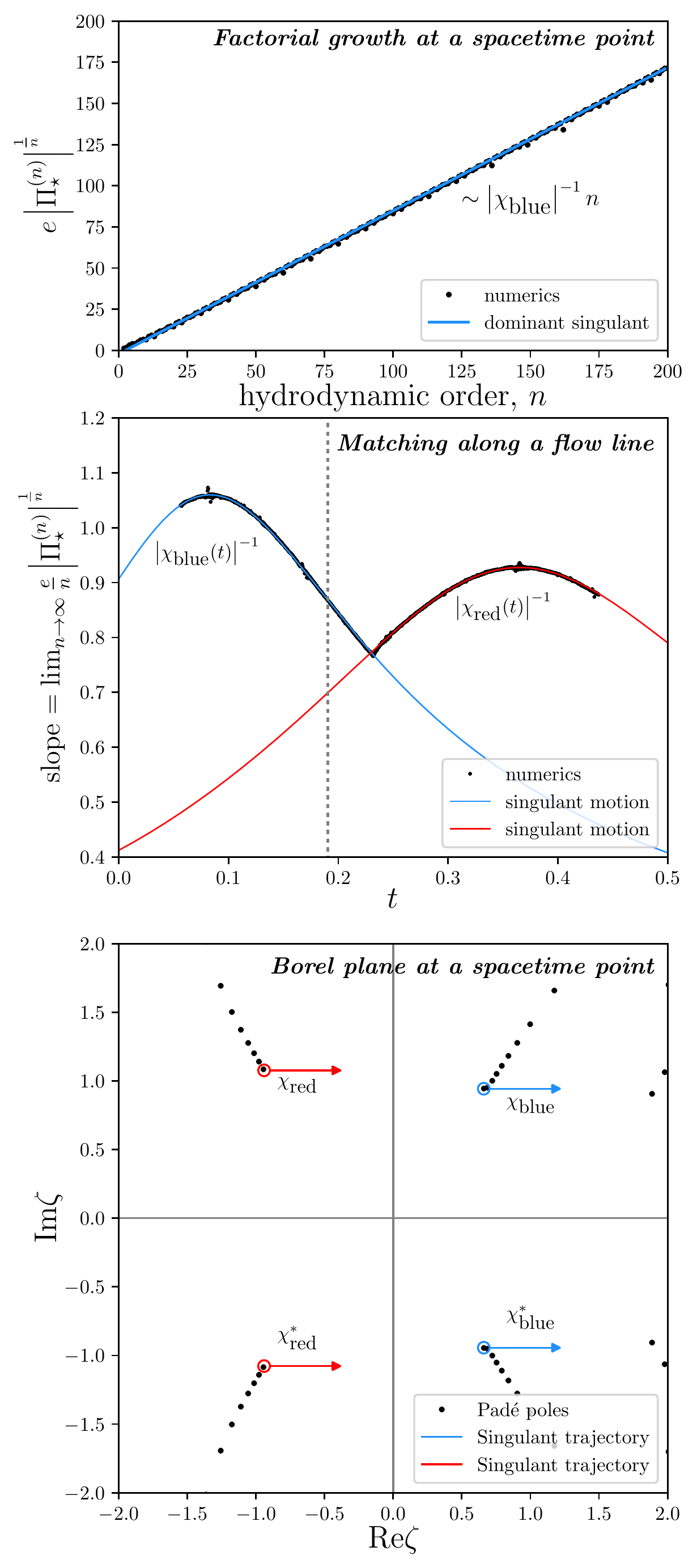}
\caption{Singulants in the BRSSS model for a nonlinear longitudinal flow. \textbf{Top panel:} root-test for a numerical evaluation of the series $\Pi_\star^{(n)}$ (black), and the slope predicted by the dominant singulant at this point $\chi_\text{blue}$ (blue). \textbf{Middle panel:} fitted slope of the root-test plot evolving along a fluid flow line (black), with singulant predictions (red and blue). The crossing corresponds to an exchange of singulant dominance. The dashed line marks $t$ used in the other panels. \textbf{Bottom panel:} singularities in the Borel plane $\zeta$, indicated by Pad\'e poles (black), illustrating agreement with singulant values (red and blue circles). }
\label{fig:brsss}
\end{center}
\end{figure}

In the BRSSS model the exact dynamical equation governing $\Pi^{\mu\nu}$, when specialised to longitudinal flows, is given by
\begin{equation}\label{BRSSS:pistar}
\Pi_\star = {-}\eta \sigma_\star{-} \tau_\Pi\left[ D\Pi_\star{+}\frac{d (\nabla \cdot U)}{d{-1}} \Pi_\star \right] {-} \frac{\lambda_1}{\eta^2}(d{-}3)\Pi_\star^2,
\end{equation}
from which the series \eqref{singulants:Pi_series} can be obtained by solving the following recursion relation: 
\begin{subequations}\label{BRSSS:recursion}
\begin{align}
&\Pi_\star^{(1)} = -\eta \sigma_\star, \\
&\Pi_\star^{(n+1)} = -\tau_\Pi (U \cdot \partial) \Pi_\star^{(n)} - \frac{d(\partial \cdot U)}{d{-}1}  \tau_\Pi \Pi_\star^{(n)} \nonumber \\ &-(d{-}3)\frac{\lambda_1}{\eta^2} \sum_{m=1}^{n} \Pi_\star^{(m)}\Pi_\star^{(n+1-m)}, \quad n > 1. 
\end{align}
\end{subequations}
This recursion relation is solved at large orders by the factorial-over-power ansatz \eqref{singulants:factorial-over-power_ansatz}, provided the singulant equation is obeyed,
\begin{equation}
D\chi_q(t,x) = \frac{1}{\tau_\Pi(T(t,x))}. \label{BRSSS:singulant_eom}
\end{equation}
In earlier work \cite{Heller:2021oxl} we established the factorial growth of $\Pi_\star^{(n)}$ for general nonlinear longitudinal flows in BRSSS, by numerically solving Eqs.~\eqref{currentcons:1}--\eqref{BRSSS:pistar} and evaluating Eq.~\eqref{BRSSS:recursion} on them. An example of this is given in the top panel of Fig.\,\ref{fig:brsss} where the $n!$ behaviour is illustrated.\footnote{The initial data corresponds to a periodic overdensity in $T$ that locally resembles a Gaussian, see Ref.\,\cite{Heller:2021oxl} for more details on this function and the numerical method. We have set $\mathcal{E}/T^4 =1$, $\eta/s = 1/(4\pi)$ and $\tau_{\Pi}T = 1/4$.} The presence [Eq.~\eqref{singulants:factorial-over-power_ansatz}] and motion [Eq.~\eqref{BRSSS:singulant_eom}] of the singulants governing the large-order behaviour of such flows can be readily confirmed for such solutions, and we now present two such ways of doing so.

First, according to \rf{singulants:factorial-over-power_ansatz} the singulant field $\chi_q$ with the smallest $|\chi_q|$ at any given spacetime point dominates the large-order behaviour yielding the prediction $|\Pi_\star^{(n)}|^\frac{1}{n} \sim \frac{n}{e |\chi_q|}$ where $\chi_q$ obeys \rf{BRSSS:singulant_eom}. Solving \rf{BRSSS:singulant_eom} on a given background is unique up to a choice of complex integration constant per flow line. Two such solutions are presented as the blue and red curves in the middle panel of Fig.\,\ref{fig:brsss} alongside the black dots which correspond to fitting a straight line to the numerical series data, $|\Pi_\star^{(n)}|^\frac{1}{n}$, showing excellent agreement. Here we have chosen to determine the singulant integration constants by adjusting for the best fit on the entire flow line.

Second, the presence of $\chi_q$ can be seen as singularities appearing in the Borel transform of \rf{singulants:Pi_series}. This is demonstrated in the bottom panel of Fig.\,\ref{fig:brsss}. This is a snapshot at a time labelled by the dashed vertical line in the middle panel, where the blue singulant dominates. Given the integration constants as determined in the previous paragraph, there is a precise match between $\chi_q$ and the location of a singularity in the Borel plane inferred by a Pad\'e approximant of the $\Pi^{(n)}_\star$ series. Along a flow line these singulants move from left to right in the Borel plane according to \rf{BRSSS:singulant_eom} as indicated by the arrows in the lower panel. There is a time for which they are at their point of closest approach to the origin, corresponding to the maxima in the middle panel of Fig.\,\ref{fig:brsss}. Similarly there is a time for which $|\chi_{\text{red}}| = |\chi_{\text{blue}}|$ corresponding to an exchange of dominance as seen by the crossing of the red and blue singulant trajectories in the middle panel.

\subsection{HJSW model}\label{subsec:HJSW}

The HJSW model \cite{Heller:2014wfa} is a generalisation of conformal BRSSS theory such that the equation of motion for $\Pi^{\mu\nu}$ includes second-order derivatives along the fluid velocity $U$, 
\begin{align}\label{HJSW:dynamical_constitutive_relation}
&\left(\left(\frac{\mathcal{D}}{T}\right)^2 {+}2 \Omega_I\left(\frac{\mathcal{D}}{T}\right){+}|\Omega|^2 \right)\Pi^{\mu\nu}{=}\nonumber\\ 
&{-}\eta |\Omega|^2 \sigma^{\mu\nu}{-} \frac{c_\sigma}{T} \mathcal{D}(\eta \sigma^{\mu\nu}),
\end{align}
where $\Omega = \Omega_R+i\Omega_I \in \mathbb C$, $c_\sigma \in \mathbb R$, and we are neglecting possible nonlinear terms in $\Pi^{\mu\nu}$. As the temperature $T$ is the only dimensionful scale of the theory, the shear viscosity takes the same functional form as it did in BRSSS theory. 

The physical motivation behind the construction of the HJSW theory was upgrading the original conformal BRSSS theory to a model with a nonhydrodynamic sector closer to the AdS/CFT one, in the sense of having two nonhydrodynamic sound modes, $\omega_\textrm{NH}^{(\pm)}(k)$, with opposite real parts at zero momentum. Indeed, in the HJSW model, $\omega_\textrm{NH}^{(\pm)}(k=0)$ are controlled by $\Omega$ as 
\begin{equation}\label{HJSW:NH_modes_k=0}
\omega_\textrm{NH}^{(\pm)}(k) = T_0 (\pm \Omega_R - i \Omega_I) + O(k^2). 
\end{equation}
For linearised perturbations around thermal equilibrium (and in the local rest frame of the fluid) the HJSW model is known to be causal and stable provided that the parameters $\Omega$, $\eta/s$ and $c_\sigma$ are chosen appropriately. Linear stability, in particular, always requires a finite $c_\sigma$~\cite{Heller:2014wfa}. Demonstrating the causality of this model at the fully nonlinear level is a challenging open problem. It should be kept in mind that even in the case of MIS-type theories such a proof has only become available recently~\cite{Bemfica:2020xym}. That being said, exploratory studies of numerical evolution in the HJSW model, with appropriate parameters, have not exposed any pathologies.

\begin{figure}[h]
\begin{center}
\includegraphics[width=\linewidth]{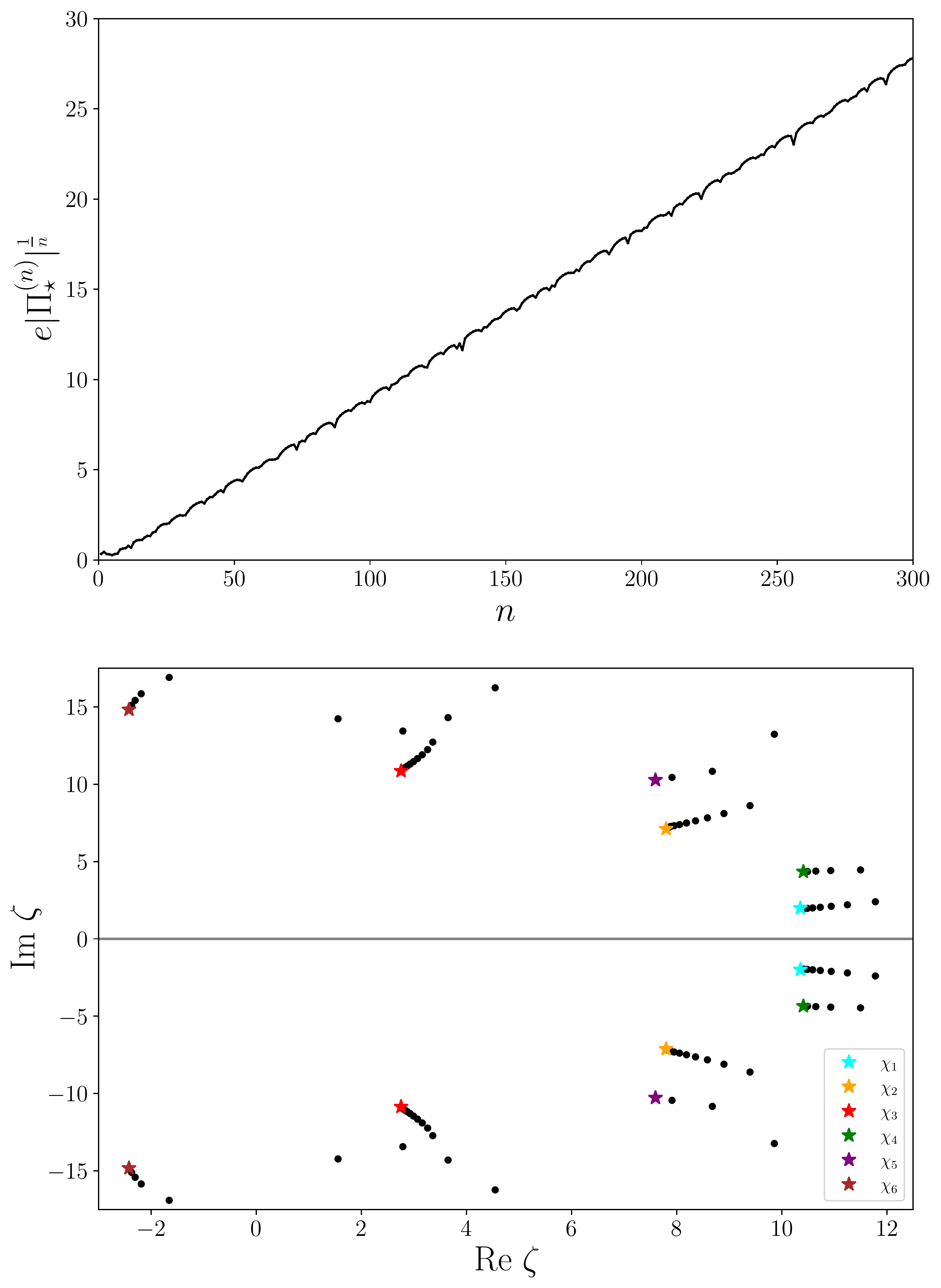}
\caption{\textbf{Upper panel}: large-order behaviour of the gradient expansion at $x=0$, $t=0.5$ for initial data of the form \eqref{HJSW:initial_data} as quantified by a root-test plot. The factorial growth is manifest. \textbf{Bottom panel}: singularities of the Padé approximant of the Borel transform of the gradient expansion at $x=0$, $t=0.5$. The points at which a line of pole accumulation starts are singulants and have been highlighted as stars. Cyan, orange, red, green, purple and brown stars correspond to candidate singulant pairs $\chi_1, \ldots, \chi_6$ of progressively increasing norm.}
\label{fig:HJSW_divergence}
\end{center}
\end{figure}

It is immediate to compute the recursion relation obeyed by $\Pi_\star^{(n)}$ once  \rf{HJSW:dynamical_constitutive_relation} is known. The only structural difference with respect to the BRSSS case is that, since \rf{HJSW:dynamical_constitutive_relation} is second order, $\Pi_\star^{(n)}$ now depends both on $\Pi_\star^{(n-1)}$ and $\Pi_\star^{(n-2)}$ for $n > 2$. In the following, we work in $d=4$. We have that 
\begin{subequations}
\begin{equation}\label{HJSW:recursion_relations_n=1}
\Pi_\star^{(1)} = \frac{2}{3}\eta\theta,
\end{equation}
\begin{equation}\label{HJSW:recursion_relations_n=2}
\begin{split}
&|\Omega|^2 T \Pi_\star^{(2)} = - 2\Omega_I D\Pi_\star^{(1)}-\frac{8}{3}\Omega_I \theta \Pi_\star^{(1)} \\
&+\frac{2}{3}c_\sigma D(\eta \theta) + \frac{8}{9}c_\sigma\eta\theta^2,     
\end{split}
\end{equation}
\begin{equation}\label{HJSW:recursion_relations_n=3}
\begin{split}
&|\Omega|^2T^2\Pi_\star^{(n)} = -D^2\Pi_\star^{(n-2)}- 2\Omega_I T D\Pi_\star^{(n-1)} \\
&+\left(D\log T - \frac{8}{3}\theta \right)D\Pi_\star^{(n-2)}
-\frac{8}{3}\Omega_I T \theta \Pi_\star^{(n-1)} \\
&+\frac{4}{9}\left(3 \theta D\log T - 4 \theta^2- 3 D\theta\right)\Pi_\star^{(n-2)}, 
\end{split}
\end{equation}
\end{subequations}
where $\theta$ is the expansion of the flow, $\theta = \nabla_\alpha U^\alpha$. 
Empirically, we always find that the gradient expansion defined by the recursion relations \eqref{HJSW:recursion_relations_n=1}--\eqref{HJSW:recursion_relations_n=3} is factorially divergent when evaluated on any longitudinal flow. In Fig.~\ref{fig:HJSW_divergence}, we provide an example for initial data of the form 
\begin{equation}\label{HJSW:initial_data}
\begin{split}
T(0,&x) = 1+e^{-\frac{x^2}{2\sigma^2}}, \quad u(0,x) = 0,\\ 
&\Pi_\star(0,x) = \partial_t \Pi_\star(0,x) = 0,
\end{split}
\end{equation}
with $\sigma=1$. We have chosen $\mathcal{E}/T^4 =1$, $\eta/s = 1/(4\pi)$, $\Omega_R = 2$, $\Omega_I = 4$, and $c_\sigma = \pi$. The root test applied to the gradient expansion associated to these initial data at $x=0$ and $t=0.5$ is represented in the upper panel. The asymptotic behaviour at large $n$ is clearly linear, implying that the gradient expansion is factorially divergent. In the lower panel, we represent the singularities of the Padé approximant to the Borel transform of the gradient expansion as black dots. There are several well-defined lines of pole condensation, grouped in complex-conjugated pairs. The poles from which these lines emanate are candidate singulants, and have been highlighted as stars. The singulants with the smallest norm correspond to the cyan stars. 

According to the general results presented in Sec.~\ref{sec:singulants}, the singulant equation of motion is determined by the first three terms in \rf{HJSW:recursion_relations_n=3}. One finds that 
\begin{equation}\label{HJSW:singulant_eom}
D \chi = (\Omega_I \pm i \Omega_R)T.    \end{equation}
Hence, just as it happened in the conformal BRSSS model, the singulant trajectory is determined by $\omega_\textrm{NH}^{(\pm)}(k=0)$ evaluated at the effective temperature. Equation \eqref{HJSW:singulant_eom} implies that, when one moves along a particular flow line, the trajectories described by the singulants are inclined lines in the Borel plane, with slopes of magnitude $|\Omega_R/\Omega_I|$. This is in stark contrast with the BRSSS case, where the singulant trajectories were given by horizontal lines. 
\begin{figure}[h]
\begin{center}
\includegraphics[width=\linewidth]{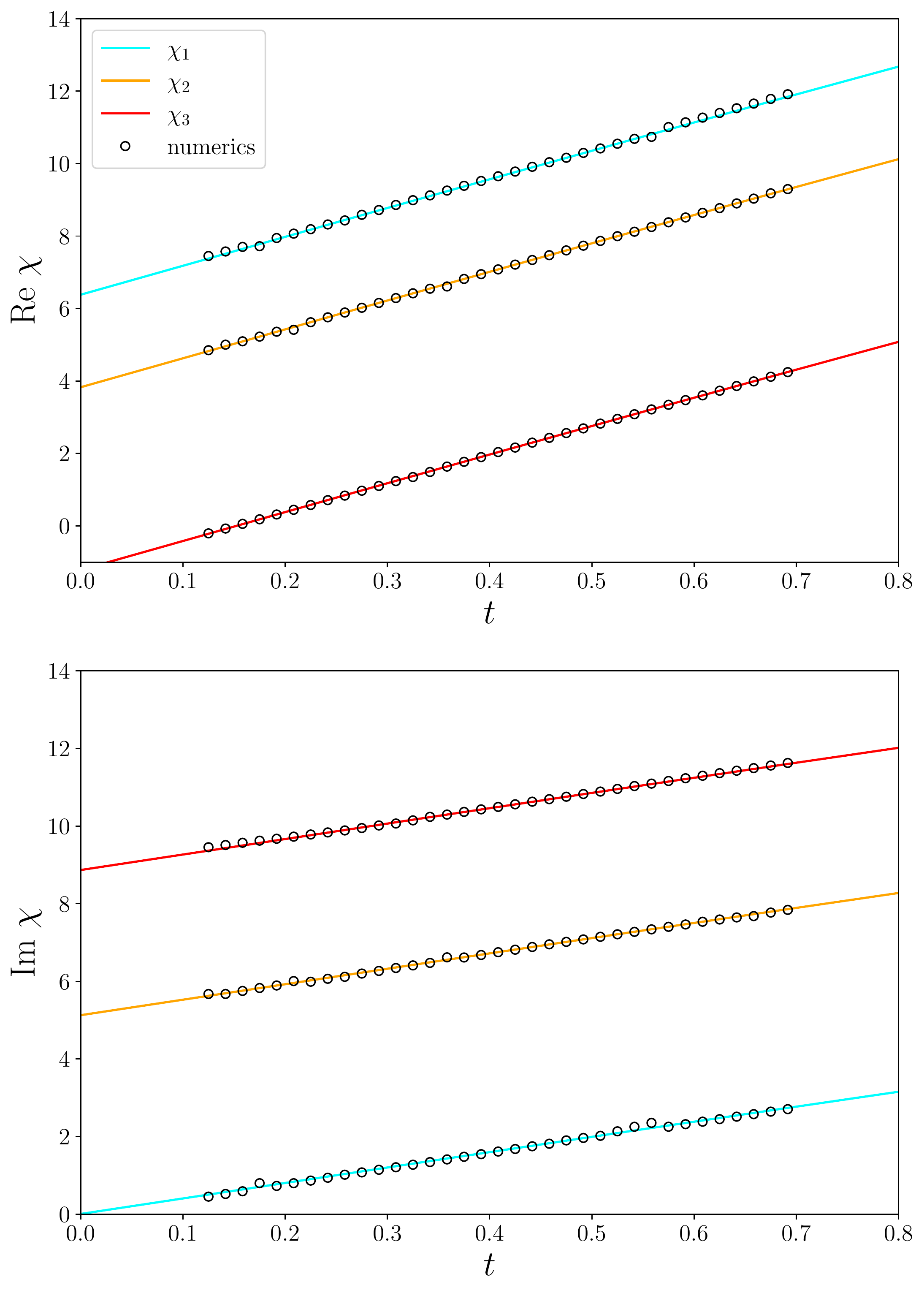}
\caption{Comparison between the time evolution of the three singulants of smallest norm as determined numerically (open circles) and the prediction \eqref{HJSW:singulant_trajectory} (solid lines). The matching has been performed at $t=0.5$. The colour coding of the lines coincides with the colour coding of the stars in the bottom panel of Fig.\,\ref{fig:HJSW_divergence}. 
\textbf{Upper panel}: real parts. \textbf{Lower panel}: imaginary parts.}
\label{fig:HJSW_singulants}
\end{center}
\end{figure}

Let us select the flow line passing through $x = 0$ at $t=0$, which corresponds to the $t$ axis in the longitudinal plane. In Fig.~\ref{fig:HJSW_singulants}, we plot the real and imaginary parts of the three singulants of smallest norm as a function of time along this particular flow line. We have selected the singulants with positive imaginary part. These singulants, which have been computed by means of a Padé approximant to the Borel transform of the gradient expansion, have to be compared with the prediction of \rf{HJSW:singulant_eom}: 
\begin{equation}\label{HJSW:singulant_trajectory}
\chi(t,0) = \chi(t=0,0) + (\Omega_I \pm i \Omega_R)\int_0^t dt' T(t',0).
\end{equation}
Since we cannot determine independently the initial values of the singulants, we have chosen to match the prediction \eqref{HJSW:singulant_trajectory} to the numerical results at $t = 0.5$. The final outcome of our analysis is that, for every singulant under consideration, the prediction of \rf{HJSW:singulant_trajectory}, which is represented as a solid line, does an excellent job in describing the time evolution of the singulants we have computed numerically.

\subsection{New model closer to holography}
\label{subsec:new_model}

\subsubsection{Physical motivation}

The analysis performed so far in the BRSSS and HJSW models revealed that the large-order behaviour of the gradient expansion is related to the sound channel nonhydrodynamic mode frequencies evaluated at $k=0$ through the singulant equation of motion. Schematically, \begin{equation}\label{eq:singulant_eom}
D\chi(t,x) = i \omega_\textrm{NH}(k=0)\Bigl\lvert_{T=T(t,x)}. 
\end{equation}
This is not a coincidence. As we mentioned in Sec.\,\ref{sec:singulants}, both the singulant equation of motion \eqref{eq:singulant_eom} as well as the equation that determines the mode frequencies are obtained through a linearisation procedure and, in the BRSSS and the HJSW models, the former equation can always be mapped onto the latter one evaluated at zero momentum. This rests crucially on the fact that the equation of motion for $\Pi_{\mu\nu}$ only involves comoving derivatives along a flow line. In light of this, one should not expect the relation between the singulant equation of motion and the sound channel nonhydrodynamic modes evaluated at zero momentum to hold when the equation of motion for $\Pi_{\mu\nu}$ features derivatives along directions orthogonal to $U$. 

As we will see later in Sec.\,\ref{sec:holography}, there is strong evidence that holography does not display the structure indicated in \rf{eq:singulant_eom}: derivatives in directions orthogonal to $U$ appear in the singulant equation of motion. Motivated by this, we introduce a new MIS-like model that also displays this feature. We achieve this by extending  \rf{HJSW:dynamical_constitutive_relation} to include derivatives along spacelike directions orthogonal to $U$. 

\subsubsection{Description of the model}

We work in $d=4$. Our new model is defined by the following the equation dictating the spacetime evolution of $\Pi_{\mu\nu}$,
\begin{align}\label{new_model_def}
&\left(\left(\frac{\mathcal{D}}{T}\right)^2 - \frac{c_{\cal L}}{T^2} \mathcal{L} + 2 \Omega_I\left(\frac{\mathcal{D}}{T}\right) + |\Omega|^2 \right)\Pi^{\mu\nu} = \nonumber \\ 
&- \eta |\Omega|^2 \sigma^{\mu\nu} - \frac{c_\sigma}{T} \mathcal{D}(\eta \sigma^{\mu\nu}),     
\end{align}
where the new term $\mathcal{L}\Pi_{\mu\nu}$ corresponds to the symmetric, transverse and traceless part of $(\Delta^{\rho\sigma}\mathcal{D}_\rho\mathcal{D}_\sigma) \Pi_{\mu\nu}$,  
\begin{equation}
\frac{1}{2}\left[\Delta_\mu^\alpha \Delta_\nu^\beta + \Delta_\nu^\alpha \Delta_\mu^\beta - \frac{2}{3}\Delta_{\mu\nu} \Delta^{\alpha\beta}\right] (\Delta^{\rho\sigma}\mathcal{D}_\rho\mathcal{D}_\sigma) \Pi_{\alpha\beta}.  \label{gHJSW_symmetric_transverse_traceless_laplacian}
\end{equation}
Note that, in principle, adding this term is only allowed by hyperbolicity once we go beyond the first-order equation of motion obeyed by $\Pi^{\mu\nu}$ in BRSSS theory. In this work, we will not establish that the initial value problem associated to the conservation equation and \rf{new_model_def} is well-posed at the fully nonlinear level. Despite this, we emphasise that (i) as we discuss in detail in Appendix~\ref{app:E} in the Supplemental Material there exists a parametric regime for $\eta/s$, $c_{\cal L}$, $c_\sigma$, and $\Omega$ such that linearised perturbations around thermal equilibrium are causal and stable in a general reference frame\footnote{In particular, it is possible to have causal and stable sound modes for $c_\sigma = 0$, unlike in the HJSW model.} and (ii) when performing numerical simulations of the nonlinear problem within such parametric regime, we have seen no issues arise. 

We conclude our presentation by clarifying that \rf{new_model_def} is not the most general equation of motion for $\Pi^{\mu\nu}$ featuring second-order derivatives that one can write down. For instance, the term $\mathcal{D}^{\langle\mu}\mathcal{D}_\alpha \Pi^{\alpha\nu\rangle}$ can certainly be incorporated into the left-hand side. Since \rf{new_model_def} already serves the purpose we outlined, the question of finding out what the most general version of this model is will not concern us here, although it might be relevant for the phenomenological modelling of scenarios like Ref.~\cite{Attems:2018gou}, where MIS or BRSSS approaches are known to fail.

\subsubsection{Gradient expansion}

In model \eqref{new_model_def}, the recursion relations obeyed by $\Pi_\star^{(n)}$ are related to the HJSW ones in a straightforward manner. To wit, $\Pi_\star^{(1)}$ and $\Pi_\star^{(2)}$ obey the HJSW recursion relations while, for $n > 2$, one has that 
\begin{align}\label{hybrid:recursion_relation}
&|\Omega|^2 T^2 \Pi_\star^{(n)} = [\textrm{HJSW}] - 7 c_{\cal L} (U^\alpha U^\beta \nabla_\alpha Z_\beta) (Z^\mu \nabla_\mu)\Pi_\star^{(n-2)} \nonumber \\
&+c_{\cal L} (Z^\mu \nabla_\mu)^2 \Pi_\star^{(n-2)} + 2 c_{\cal L} \left(\frac{1}{3}\theta^2 + 3 (U^\alpha U^\beta \nabla_\alpha Z_\beta)^2\right.\nonumber\\
&\left.-2 Z^\mu \nabla_\mu (U^\alpha U^\beta \nabla_\alpha Z_\beta) \right)\Pi_\star^{(n-2)},  
\end{align}
where $[\textrm{HJSW}]$ represents the right-hand side of \rf{HJSW:recursion_relations_n=3} and $Z = Z^\mu\partial_\mu$ is a longitudinal unit-normalised vector field orthogonal to $U = U^\mu \partial_\mu$. 

As in the BRSSS and HJSW cases, in this model we also find that the gradient expansion is factorially divergent for all our choices of parameters, initial data and spacetime point. To characterise the large-order behaviour of the gradient expansion, we turn now to the question of the singulant trajectory. 

\subsubsection{Singulant trajectory}

According to the factorial-over-power ansatz and \rf{hybrid:recursion_relation}, the singulant equation of motion in the new model is 
\begin{equation}\label{hybrid:singulant_eom}
U(\chi)^2 {-} c_{\cal L} Z(\chi)^2 {-} 2\Omega_I T U(\chi) {+} |\Omega|^2 T^2 {=} 0 .     
\end{equation}
where $U(\chi) = U^\mu\partial_\mu \chi$ and $Z(\chi) = Z^\mu \partial_\mu \chi$. This equation features derivatives along $Z$ and, as a consequence, $U(\chi)$ stops being related to the nonhydrodynamic sound modes at zero spatial momentum.

\begin{figure}[h]
\begin{center}
\includegraphics[width=\columnwidth]{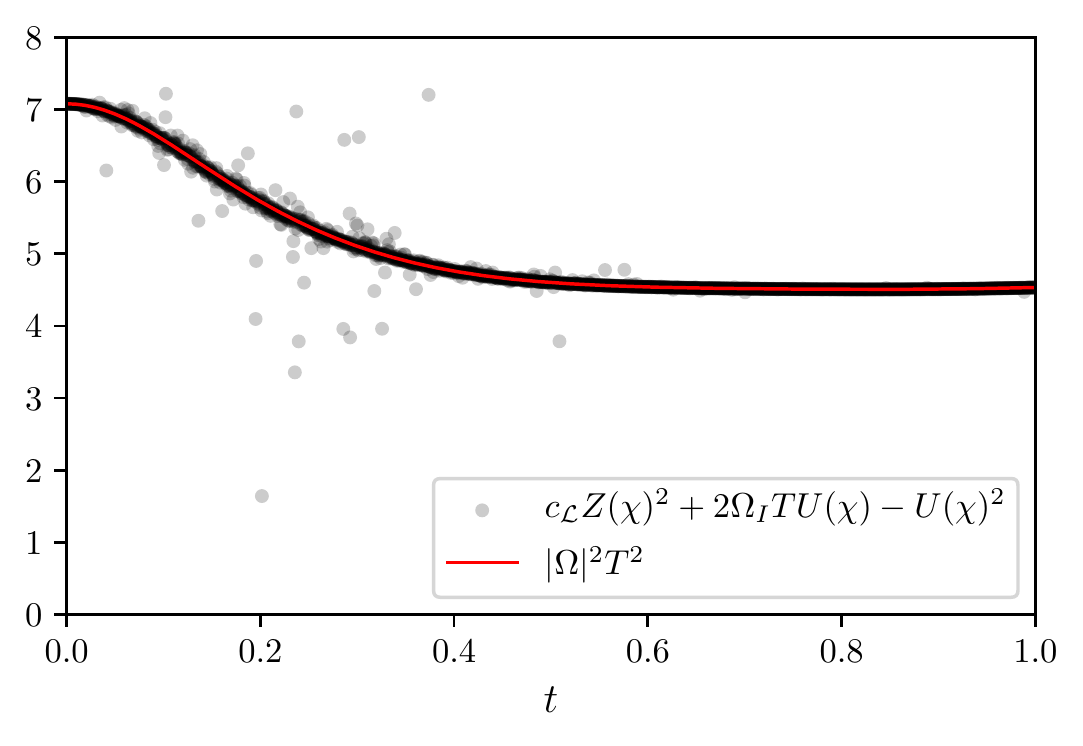}
\caption{Numerical confirmation of the singulant equation of motion Eq.~\eqref{hybrid:singulant_eom} along a flow line in the model \rf{new_model_def}. This model is distinguished from BRSSS and HJSW by the appearance of transverse derivatives in the singulant equation (controlled by $c_\mathcal{L}$), a feature it shares with holography and our main motivation for studying it. The values of $Z(\chi)^2, U(\chi), U(\chi)^2$ are extracted from the large-order behaviour of $\Pi_\star^{(n)}$ and its derivatives, as described in the text. Here the black disks are rendered partially transparent to convey their density.}
\label{fig:hybrid_singulants}
\end{center}
\end{figure}

We now turn to verifying that \rf{hybrid:singulant_eom} governs singulant motion in our numerical examples. In order to solve \rf{hybrid:singulant_eom} it is no longer enough to specify integration constants per flow line, due to the appearance of derivatives transverse to the flow line $Z(\chi)$. This term did not appear in the BRSSS and HJSW models. Now singulant initial data must be specified for some portion of a Cauchy surface, preventing us from testing \rf{hybrid:singulant_eom} by fitting one or two complex numbers. To sidestep this issue, we attempt to extract $Z(\chi)^2, U(\chi), U(\chi)^2$ for the dominant singulant and test that \rf{hybrid:singulant_eom} holds as an algebraic relation at each spacetime point. To extract $Z(\chi)^2, U(\chi), U(\chi)^2$ we utilise the following relations,
\be
U(\chi)\sim \frac{U\left(\Pi_\star^{(n-1)}\right)}{\Pi_\star^{(n)}},\quad 
U(\chi)^2\sim \frac{U^2\left(\Pi_\star^{(n-2)}\right)}{\Pi_\star^{(n)}},
\ee
and similarly for $Z(\chi)^2$ as $n\to \infty$, and confirm \rf{hybrid:singulant_eom} in Fig.\,\ref{fig:hybrid_singulants}, up to some scatter associated with this numerical procedure. Note that in this figure the new term $Z(\chi)^2$ makes a contribution of similar magnitude as those of $U(\chi), U(\chi)^2$.

\subsubsection{Singulant trajectory and linear response theory}

Despite the fact that the singulant trajectory is not controlled by the nonhydrodynamic sound modes at zero momentum, the lesson that the singulant equation of motion can be understood in terms of a linear response theory problem stands. Let us set the hydrodynamic fields to their values at thermal equilibrium in the rest frame of the fluid, 
\begin{equation}
T \to T_0, \quad U^\mu \partial_\mu  \to \partial_t, 
\end{equation}
and consider linearised perturbations of $\Pi_\star$, $\delta\Pi_\star$, around this static state. At leading order, one finds that $\delta\Pi_\star$ satisfies 
\begin{equation}\label{hybrid:delta_Pi_position_space}
\left(\partial_t^2 - c_{\cal L}\partial_x^2 + 2 \Omega_I T_0 \partial_t + |\Omega|^2 T_0^2 \right)\delta\Pi_\star = 0.  
\end{equation}
For a plane-wave perturbation of the form \begin{equation}
\delta\Pi_\star(t,x) = \hat{\delta\Pi_\star}e^{-i(\omega t-k x)}, 
\end{equation} 
\rf{hybrid:delta_Pi_position_space} reduces to a second-order polynomial equation for $\omega$, 
\begin{equation}\label{hybrid:dispersion_relation_singulant_problem}
-\omega^2 + c_{\cal L} k^2 - 2 i \Omega_I T_0 \omega + |\Omega|^2 T_0^2 = 0,
\end{equation}
and the map 
\begin{equation}\label{hybrid:singulant_map}
\omega \to -i U(\chi), \quad k \to \pm i Z(\chi), \quad T_0 \to T
\end{equation}
transforms \rf{hybrid:dispersion_relation_singulant_problem} into the singulant equation of motion~\eqref{hybrid:singulant_eom}. 

The above analysis is nothing but a particular illustration of the general results from Sec.\,\ref{sec:singulants}.  A crucial observation that has to be emphasised is that, just as we foresaw at the end of that section, the roots of \rf{hybrid:dispersion_relation_singulant_problem}, which determine the singulant trajectory, \emph{do not} correspond to the sound channel nonhydrodynamic modes when the momentum is finite. The latter (including the hydrodynamic modes) are given by the roots of the following fourth-order polynomial in $\omega$, 
\begin{equation}\label{hybrid:dispersion_relation}
\begin{split}
&\left(-\omega^2 - 2 i T_0 \Omega_I \omega + c_{\cal L} k^2 + T_0^2 |\Omega|^2\right)\left(\omega^2 - \frac{1}{3}k^2\right) \\
&+ \frac{4}{3} \frac{\eta}{s} \left(T_0 |\Omega|^2 - i c_\sigma \omega \right) i \omega k^2 = 0, \end{split}
\end{equation}
and only include the roots of \rf{hybrid:dispersion_relation_singulant_problem} when $k=0$. 

Despite the fact that they do not correspond to nonhydrodynamic modes, the roots of \rf{hybrid:dispersion_relation_singulant_problem} still allow for a physical interpretation. This physical interpretation follows from the observation that, due to rotational invariance, the sound channel dispersion relation is constrained to take the form \cite{Grozdanov:2019uhi}
\begin{equation}\label{hybrid:master_sound-channel-dispersion-relation}
\omega^2 + i \omega k^2 \gamma_s(\omega,k^2) + k^2 H(\omega, k^2) = 0,      
\end{equation}
where $\gamma_s$ is a momentum-dependent sound attenuation length. Indeed, the sound channel dispersion relation \eqref{hybrid:dispersion_relation} can be put in the form \eqref{hybrid:master_sound-channel-dispersion-relation} with the identifications $H(\omega,k^2) = -\frac{1}{3}$ and \begin{equation}
\gamma_s(\omega,k^2) = \frac{\frac{4}{3} \frac{\eta}{s} \left(T_0 |\Omega|^2 - i c_\sigma \omega \right)}{-\omega^2 - 2 i T_0 \Omega_I \omega + c_{\cal L} k^2 + T_0^2 |\Omega|^2}.  
\end{equation}
It is immediate to see that the roots of \rf{hybrid:dispersion_relation_singulant_problem} correspond precisely to the poles of $\gamma_s$. It is important to keep in mind that, for a general CFT in the linear response regime, $\gamma_s$ can be defined without resorting to  \rf{hybrid:master_sound-channel-dispersion-relation} by the relation 
\begin{equation}\label{gamma_s_def}
\delta\hat{\Pi}_\star(\omega,k) = \frac{2}{3}\mathcal{E}_0 \gamma_s(\omega,k) i k \delta\hat{u}(\omega,k),      
\end{equation}
where $\mathcal{E}_0$ is the equilibrium energy density and $\delta\hat{u}$ represents an infinitesimal perturbation of the fluid velocity in the sound channel: 
\begin{equation}
U^\mu \partial_\mu = \partial_t + \delta\hat{u} e^{-i\omega t+ i k x} \partial_x, \quad |\delta\hat{u}|\ll 1.  
\end{equation}
With the definition \eqref{gamma_s_def}, the equation that the sound channel modes have to obey always takes the form \eqref{hybrid:master_sound-channel-dispersion-relation} with $H = -1/3$. 

We conclude our analysis with two comments. The first and most important one is that the connection between the singulant equation of motion and the poles of $\gamma_s$ implied by the map \eqref{hybrid:singulant_map} is not restricted to the case at hand: it applies to the BRSSS and HJSW models as well. However, in the BRSSS and HJSW cases, the poles of $\gamma_s$ are independent of the spatial momentum $k$. In this situation, these poles have to coincide with the sound channel nonhydrodynamic modes evaluated at zero momentum. This observation is in agreement with our analysis of the singulant equation of motion in the BRSSS and HJSW theories. 

The second comment is that, when one restricts to Bjorken flow \cite{Bjorken:1982qr}, the singulant equation of motion is always insensitive to the difference between the nonhydrodynamic sound modes and the poles of $\gamma_s$. The reason is that for Bjorken flow $\chi$ is a function of the proper time $\tau = \sqrt{t^2-x^2}$ alone. This implies that $Z(\chi)  = 0$ and, according to the map \eqref{hybrid:singulant_map}, that the relevant poles of $\gamma_s$ are the ones with $k=0$, which agree with the zero-momentum sound channel nonhydrodynamic modes. 

\section{Holography}\label{sec:holography}

\subsection{Introduction}

In this section, we examine the large-order behaviour of the gradient expansion for a longitudinal flow in the context of holography.\footnote{See Refs.\,\cite{Withers:2018srf,Grozdanov:2019kge,Grozdanov:2019uhi,Amoretti:2019kuf,Abbasi:2020ykq,Jansen:2020hfd,Arean:2020eus,Ammon:2020rvg,Asadi:2021hds,Baggioli:2021ujk,Grozdanov:2021gzh,Jeong:2021zsv,Donos:2021pkk,Liu:2021qmt,Huh:2021ppg,Ammon:2021pyz,Cartwright:2021qpp} for related work in the context of the gradient expansion in linearised holography.} We do not attempt to perform the numerical computation of the gradient expansion evaluated on a particular state. Rather, we take advantage of the asymptotic ansatz \eqref{singulants:factorial-over-power_ansatz} to prove that a large-order factorial growth is allowed. We remind the reader that finding out that the asymptotic ansatz \eqref{singulants:factorial-over-power_ansatz} is consistent in holography does not imply that the gradient expansion is necessarily factorially divergent; rather, this observation just demonstrates that such large-order behaviour is, in principle, possible. 

Our study builds on previous developments in the AdS/CFT context. The first one is the original construction of fluid/gravity duality \cite{Bhattacharyya:2008jc}, which we employ to build the gradient-expanded constitutive relations at the fully nonlinear level. The second one is the analysis of the exact constitutive relations of the microscopic CFT in the linear response regime~\cite{Bu:2014ena,Bu:2014sia}. The bridge between both approaches is the linearisation enacted by the asymptotic ansatz \eqref{singulants:factorial-over-power_ansatz}.

The key result that we show in this section is as follows. If a large-order factorial growth is present, then the singulant equation of motion in holography is given by the poles of the momentum-dependent sound attenuation length, $\gamma_s$, under a map analogous to that which we saw in the MIS-like models \eqref{hybrid:singulant_map}.

\subsection{Longitudinal flows in holography}

To construct the geometry dual to a longitudinal flow, we follow Ref.~\cite{Bhattacharyya:2008mz} and put forward the following metric ansatz, 
\begin{equation}\label{hol:metric_ansatz}
\begin{split}
ds^2{=}&{-}2 U_\mu(x) dx^\mu (dr {+} \mathcal{V}_\nu(r,x) dx^\nu) \\
&{+} \mathcal{G}_{\mu\nu}(r,x)dx^\mu dx^\nu,     
\end{split}
\end{equation}
where $x^\mu$ represents the boundary coordinates, $U^\mu$ as elsewhere in the text is the unit-normalised fluid velocity and $\mathcal{G}_{\mu\nu}$ is transverse, $\mathcal{G}_{\mu\nu} U^\nu = 0$. 

It is convenient to employ a curvilinear coordinate system to describe the longitudinal flow. We focus on longitudinal flows in four-dimensional Minkowski spacetime. We take our boundary coordinates to be  
\begin{equation}
x^\mu = (\tau, \sigma, x_\perp^{(1)}, x_\perp^{(2)}),   \end{equation}
where $\tau$ and $\sigma$ are, respectively, a timelike and a spacelike coordinate that parameterise the longitudinal plane. We choose this curvilinear coordinate system in such a way that the boundary metric is diagonal, \begin{equation}\label{hol:boundary_metric}
dh^2 = - e^{2 a(\tau,\sigma)}d\tau^2 + e^{2 b(\tau,\sigma)}d\sigma^2 + d\vec{x}_\perp^2,  
\end{equation}
and the fluid velocity reads 
\begin{equation}
U^\mu \partial_\mu = e^{-a(\tau,\sigma)}\partial_\tau. 
\end{equation}
The orthonormal longitudinal vector field $Z$ is thus given by
\begin{equation}
Z^\mu \partial_\mu = e^{-b(\tau,\sigma)}\partial_\sigma. 
\end{equation}
Imposing flatness of the 
boundary metric~\eqref{hol:boundary_metric} leads to an equation linking $a$ and $b$. Finally, to comply with the symmetry restrictions of the flow, we take 
\begin{subequations}
\begin{equation}
\mathcal{V}_\mu dx^\mu = V_\tau d\tau + V_\sigma d\sigma, 
\end{equation}
\begin{equation}
\begin{split}
\mathcal{G}_{\mu\nu}dx^\mu dx^\nu = \Sigma^2 e^{-2 B}d\sigma^2 + \Sigma^2 e^{B}d\vec{x}_\perp^2,     
\end{split}
\end{equation}
\end{subequations}
where $V_\tau$, $V_\sigma$, $\Sigma$, and $B$ are functions of $\tau$, $\sigma$, and $r$. The transversality condition $\mathcal{G}_{\mu\nu} U^\nu = 0$ is automatically satisfied since $\mathcal{G}_{\tau\tau} = 0$. 

The energy-momentum tensor of the dual CFT is dictated by the holographic dictionary \cite{deHaro:2000vlm} in terms of the coefficients of the series expansion of the metric around the asymptotic boundary located at $r\to\infty$. In our coordinate system for a given quantity $f$, we denote the coefficient of $r^{-n}$ in the near-boundary (large-$r$) expansion as $f_n$. A straightforward computation shows that\footnote{We work with the normalisation $L=4 \pi G =1$.}
\begin{equation}\label{holo:Tdd}
\langle T_{\mu\nu}\rangle = t_{\mu\nu}, 
\end{equation}
where the only nonzero components of $t_{\mu\nu}$ are given by \begin{align}
&t_{\tau\tau} = \frac{3}{2}e^a V_{\tau,2}, \quad t_{\tau\sigma} = e^a V_{\sigma,2}, \nonumber \\ &t_{\sigma\sigma} = e^{2 b}\left(- 2 B_4 + \frac{1}{2}e^{-a}V_{\tau,2} + \Phi \right), \\
&t_{x_\perp^{(i)}x_\perp^{(i)}} = B_4 + \frac{1}{2}e^{-a}V_{\tau,2} - \frac{1}{2}\Phi \nonumber, \end{align}
and we have defined 
\begin{equation}
\begin{split}
\Phi{=}&\frac{4}{3}\Sigma_0^3 e^ {-b}U(b){+} \frac{2}{3}\Sigma_0^2 e^{{-}\frac{2b}{3}}U(b)^2\\
&{+}\frac{4}{9}\Sigma_0 e^{{-}\frac{b}{3}}U(b)^3{+}\frac{5}{81}U(b)^4.     
\end{split}
\end{equation}
To conclude our analysis, we impose the Landau frame condition by demanding that the fluid velocity $U = e^{-a} \partial_\tau$ is the only timelike eigenvector of $T^\mu_\nu$. This can only be achieved provided that 
\begin{equation}\label{hol:Landau_frame_condition}
V_{\sigma,2} = 0. 
\end{equation}
Once the Landau frame condition \eqref{hol:Landau_frame_condition} is imposed, the eigenvalues of $T^\mu_\nu$ are in one-to-one correspondence with its diagonal components. One has that the energy density $\mathcal{E}$, the longitudinal pressure $P_\parallel$ and the transverse pressure $P_\perp$ are given by   
\begin{equation}
\begin{split}
&\mathcal{E} = \frac{3}{2} e^{-a} V_{\tau,2}, \\ &P_\parallel = - 2 B_4 + \frac{1}{3}\mathcal{E} + \Phi, \\
&P_\perp = B_4 + \frac{1}{3}\mathcal{E} - \frac{1}{2}\Phi.
\end{split}
\end{equation}
Hence, $\Pi_\star = B_4 - \frac{1}{2}\Phi$. This relation can be simplified further by recalling that the metric ansatz \eqref{hol:metric_ansatz} is invariant under
\begin{align}
r \to r + f(x),& \quad \mathcal{V}_\mu(r,x) \to \mathcal{V}_\mu(r+f,x) - \partial_\mu f, \nonumber \\ &\mathcal{G}_{\mu\nu}(r,x) \to \mathcal{G}_{\mu\nu}(r+f,x), 
\end{align}
and that, in the $r \to \infty$ limit, 
\begin{equation}
\Sigma(r,x) = r + \Sigma_0(x) + \ldots,     
\end{equation}
implying that $\Sigma_0(x)$ can be tuned to any desired value by a suitable choice of $f(x)$. In our case, the gauge choice
\begin{equation}\label{hol:zero_Phi}
\Sigma_0 = -\frac{1}{6}e^{\frac{b}{3}}U(b), 
\end{equation}
sets $\Phi = 0$ and implies that $\Pi_\star$ can be directly identified with $B_4$: 
\begin{equation}\label{hol:B_Pi_relation}
\Pi_\star = B_4.     
\end{equation}
The existence of this relation is one of the main advantages of our metric ansatz and coordinate choice, as it implies that the gradient expansion of $\Pi_\star$ can be obtained from the near-boundary behaviour of the gradient expansion of $B$ in a straightforward manner. 

\subsection{Gradient expansion}

In order to construct the gradient expansion of $\Pi_\star$ we follow the strategy we described in Sec.\,\ref{sec:singulants}. As it is standard in fluid/gravity duality, the first step is decomposing the Einstein equations into dynamical equations and constraint equations. The dynamical equations can be employed to express $\Pi_{\mu\nu}$ as a functional of $\mathcal{E}$ and $U$; they can be interpreted as the AdS/CFT counterpart of the equation of motion for $\Pi_{\mu\nu}$ in MIS-like theories. In our case, the dynamical equations are 
\begin{equation}
\label{eq.dyneinst}
E_{rr},\,\,E_{r\sigma},\,\,E_{\sigma\sigma},\,\,E_{x_\perp^{(1)}x_\perp^{(1)}}.
\end{equation}
On the other hand, the constraint equations enforce the conservation of the energy-momentum tensor of the dual CFT. 

Once the dynamical equations have been identified, we take the metric \eqref{hol:metric_ansatz} and introduce the bookkeeping parameter $\epsilon$ by means of a homogeneous rescaling of the boundary coordinates: 
\begin{equation}\label{hol:rescaling}
\tau \to \frac{\tau}{\epsilon}, \quad \sigma \to \frac{\sigma}{\epsilon}, \quad \vec{x}_\perp \to \frac{\vec{x}_\perp}{\epsilon} \quad r \to r.  
\end{equation}
Then, we express $V_\tau$, $V_\sigma$, $\Sigma$, and $B$ as asymptotic series in $\epsilon$, 
\begin{equation}\label{hol:metric_gradient_expansion}
\begin{split}
&V_\tau = \sum_{n=0}^\infty V_\tau^{(n)} \epsilon^n, \quad V_\sigma = \sum_{n=0}^\infty V_\sigma^{(n)} \epsilon^n, \\
&\Sigma = \sum_{n=0}^\infty \Sigma^{(n)} \epsilon^n, \quad  B = \sum_{n=0}^\infty B^{(n)}\epsilon^n, 
\end{split}
\end{equation}
insert them into the dynamical Einstein equations, and demand that they are a solution order by order in an expansion around $\epsilon=0$. This procedure transforms the dynamical equations into recursion relations for $V_\tau^{(n)}$, $V_\sigma^{(n)}$, $\Sigma^{(n)}$, and $B^{(n)}$. The zeroth-order solution reads 
\begin{equation}
\begin{split}\label{hol:zeroth-order_solution}
&V_\tau^{(0)} = - \frac{1}{2} e^{a} \left(r^2 - \frac{r_h^4}{r^2}\right), \quad V_\sigma^{(0)} = 0, \\
&\Sigma^{(0)} = e^{\frac{1}{3}b}r, \quad B^{(0)}= -\frac{2}{3}b,  
\end{split}
\end{equation}
where we taken into account that the bulk spacetime is asymptotically AdS with boundary metric \eqref{hol:boundary_metric}. Note that $a$, $b$ and $r_h$ are functions of $\tau$ and $\sigma$ only.

The gradient expansion of $\Pi_\star$ can be obtained from Eq.~\eqref{hol:metric_gradient_expansion} through the relation \eqref{hol:B_Pi_relation}, provided that appropriate boundary conditions are satisfied by the $n$th order solution. These boundary conditions are divided into two sets, infrared and ultraviolet, depending on whether they are imposed at the event horizon of the zeroth-order solution
\eqref{hol:zeroth-order_solution}, located at $r=r_h$, or at the asymptotic boundary, located at $r=\infty$. Our infrared boundary conditions are regularity of $V_\tau^{(n)}$, $V_\sigma^{(n)}$, $\Sigma^{(n)}$, and $B^{(n)}$ at $r=r_h$. Since we are working in Eddington-Finkelstein coordinates, demanding regularity of our metric functions at $r= r_h$ corresponds to imposing infalling boundary conditions. As for the ultraviolet boundary conditions: 
\begin{enumerate}
\item We demand that the metric is asymptotically AdS$_5$. Since this boundary condition has already been fully taken into account by the zeroth-order solution \eqref{hol:zeroth-order_solution}, it follows that  \begin{equation}\label{hol:bc_1}
\lim_{r\to\infty} \left\{\frac{V_\tau^{(n)}}{r^2}, \frac{V_\sigma^{(n)}}{r^2},\frac{\Sigma^{(n)}}{r},B^{(n)}\right\} = 0,   
\end{equation}
for $n \geq 1$. We note that, when imposing Eq.~\eqref{hol:bc_1}, the dynamical Einstein equations imply that, in the $r\to\infty$ limit, $V_{\sigma}^{(n)} = O(r^{-2})$ for $n > 2$ and $B^{(n)} = O(r^{-4})$ for $n > 0$. This observation will be important later. 
\item We require that the Landau frame condition \eqref{hol:Landau_frame_condition} is obeyed at every order in the gradient expansion. This requires that 
\begin{equation}\label{hol:bc_2}
V_{\sigma,2}^{(n)} = 0,      
\end{equation}
for $n \geq 0$. When combined with the last observation performed in point (1) above, Eq.\,\eqref{hol:bc_2} implies that $V_{\sigma}^{(n)} = O(r^{-3})$ as $r\to\infty$ for $n > 2$. 
\item We demand that the energy density as computed from the zeroth-order solution, $\frac{3}{4}r_h^4$, agrees with the actual energy density of the system $\mathcal{E}$. This requires that 
\begin{equation}
V_{\tau,2}^{(n)} = 0,      
\end{equation}
for $n\geq1$. $r_h$ is thus related to the effective temperature $T$ as $\pi T = r_h$. 
\item We enforce the condition \eqref{hol:zero_Phi}. Since this relation is first order in gradients, we must impose it at the level of $\Sigma^{(1)}$. This implies that 
\begin{equation}\label{hol:bc_Sigma}
\Sigma_0^{(n)} = 0, 
\end{equation}
for $n \neq 1$. 
\end{enumerate}

\subsection{Large-order behaviour of the gradient expansion}

In order to assess the large-order behaviour of the gradient expansion \eqref{hol:metric_gradient_expansion}, we assume that the singulant fields corresponding to $V_\tau$, $V_\sigma$, $\Sigma$, and $B$ are equal and, subsequently, find that this assumption is self-consistent. Hence, we take 
\begin{equation}
V_\tau^{(n)}(r,\tau,\sigma) \sim \sum_q \bar{V}_{\tau,q}(r,\tau,\sigma) \frac{\Gamma(n + \alpha_q(r,\tau,\sigma))}{\chi_q(\tau,\sigma)^{n + \alpha_q(r,\tau,\sigma)}},     
\end{equation}
with analogous expressions holding for $V_\sigma$, $\Sigma$, and $B$. In the end, this restriction is equivalent to the assumption that the gradient expansion of the full spacetime metric \eqref{singulants:metric} takes the asymptotic form \eqref{factorial_ansatz_metric}.

To simplify the subsequent argument we introduce $\Psi$, which stands for any of the functions $V_\tau$, $V_\sigma$, $\Sigma$~and~$B$. According to the general analysis presented in Appendix~\ref{app:A} in the Supplemental Material---and whose most important take-home points were mentioned in Sec.~\ref{sec:singulants}---the terms in the dynamical equations that set the singulant equation of motion are of the form 
\begin{equation}
\partial_r^{p} \partial_{\mu_1} \ldots \partial_{\mu_m} \Psi,     
\end{equation}
where $p$, $m$ are non-negative integers. Since the dynamical equations are second order in spacetime derivatives, one has that $p+ m \leq 2$. Furthermore, it is very important to stress that, due to the functional form of the dynamical equations, the large-order ansatz \eqref{factorial_ansatz_metric} implies that $\chi$ is independent of the radial coordinate $r$. 

Let us discuss $E_{rr}$ first. One finds that, at leading order in $n$ as $n\to\infty$, 
\begin{equation}
\partial_r^2 \bar{\Sigma} = 0.     
\end{equation}
This relation, when combined with the boundary conditions spelled out before, entails that 
\begin{equation}
\bar{\Sigma} = 0.     
\end{equation}
Taking into account this fact, the remaining dynamical equations imply that, at leading order in $n$ as $n \to \infty$, 
\begin{subequations}
\begin{equation}\label{hol:D2_large-order}
\partial_r^2 \bar{V}_\sigma + \frac{\partial_r \bar{V}_\sigma}{r}-\frac{4 \bar{V}_\sigma}{r^2} + 2 e^b Z(\chi) \partial_r \bar{B} = 0,     
\end{equation}
\begin{equation}\label{hol:D3_large-order}
\begin{split}
&\partial_r^2 \bar{B}{+} \left(\frac{1}{r}{+}\frac{4r}{f^{(0)}}{-} \frac{2U(\chi)}{f^{(0)}}\right) \partial_r \bar{B} \\
&{-}\left(\frac{3 U(\chi)}{r f^{(0)}}{+}\frac{Z(\chi)^2}{3 r^2 f^{(0)}}\right) \bar{B}{-}\frac{2 e^{{-}b}Z(\chi)}{3r^2f^{(0)}}\partial_r \bar{V}_{\sigma} \\
&{-}\frac{2 e^{{-}b} Z(\chi)}{3 r^3 f^{(0)}}\bar{V}_\sigma = 0, 
\end{split}
\end{equation}
\begin{equation}\label{hol:D4_large-order}
\begin{split}
&\partial_r^2 \bar{V}_\tau{+}\frac{4\partial_r \bar{V}_\tau}{r}{+}\frac{2 \bar{V}_\tau}{r^2}{-}\frac{2 e^{a{-}b} Z(\chi)}{3 r^2}\partial_r \bar{V}_\sigma \\
&{-}\frac{2 e^{a{-}b} Z(\chi)}{3 r^3}\bar{V}_\sigma{-}\frac{e^{a} Z(\chi)^2}{3 r^2}\bar{B} = 0, 
\end{split}
\end{equation}
\end{subequations}
where $f^{(0)} = r^2 - r_h^4 r^{-2}$. Note that $\bar{V}_\tau$ decouples from $\bar{V}_\sigma$ and $\bar{B}$. 

Let us assume that the singulant field $\chi$ is known at the $\tau = 0$ slice. In order to integrate the singulant motion, we need to find $\partial_\tau \chi$. This can be achieved by solving the eigenvalue problem for $\partial_\tau \chi$ posed by Eqs.~\eqref{hol:D2_large-order} and \eqref{hol:D3_large-order} with boundary conditions given by Eqs.~\eqref{hol:bc_1} and \eqref{hol:bc_2}. We note that the singulant dynamics is ultralocal in the boundary coordinates: to compute $\partial_\tau \chi$ at a given spacetime point, the inputs that we need are the values of $\partial_\sigma \chi$ and $a$, $b$, and $r_h$---or, equivalently, the energy density and velocity of the fluid ---at the point. This ultralocality is in line with the general results discussed in Sec.\,\ref{sec:singulants}. Furthermore, in the next section, we also confirm that the eigenvalue problem spelled out above has a natural counterpart in linear response. 

\subsection{Singulant equation of motion and linear response theory}

In Refs.~\cite{Bu:2014ena,Bu:2014sia} it was shown that, in the linear response regime, it is possible to find the exact constitutive relations that express the dissipative tensor as a functional of the hydrodynamic fields in the Landau frame. Consider infinitesimal perturbations of the hydrodynamic fields and $\Pi_{\mu\nu}$ around a reference thermal state of energy density $\mathcal{E}_0$ and fluid velocity $U_0 = \partial_t$. These infinitesimal perturbations, which we denote as $\delta\mathcal{E}$, $\delta\vec{u}$, and $\delta\Pi_{\mu\nu}$, are defined as  
\begin{equation}
\begin{split}
\mathcal{E} = \mathcal{E}_0 + \delta\mathcal{E}, \quad U = \partial_t + \delta\vec{u}\cdot\vec{\partial}, \\
\Pi_{\mu\nu}dx^\mu dx^\nu = \delta \Pi_{ij} dx^i dx^j, 
\end{split}
\end{equation}
where 
\begin{equation}
\frac{\delta\mathcal{E}}{\mathcal{E}_0},\,|\delta\textbf{u}|,\,\frac{\delta \Pi_{ij}}{\mathcal{E}_0} \ll1,  
\end{equation}
and we have enforced the Landau frame condition. Latin indices range from one to three and refer to spatial directions of the boundary.  

The main result of Refs.~\cite{Bu:2014ena,Bu:2014sia} is obtaining the constitutive relations that express $\delta\Pi_{ij}$ as a functional of $\delta\mathcal{E}$ and $\delta u_i$ in closed form. In momentum space, 
\begin{equation}\label{hol:linear_constitutive-relation-BL}
\delta\hat{\Pi}_{ij} = -\eta(\omega,k^2)\hat{\sigma}_{ij} - \xi(\omega,k^2) \hat{\pi}_{ij},
\end{equation}
where $\delta\hat{\Pi}_{ij}$ is the Fourier transform of $\delta\Pi_{ij}$,
\begin{subequations}
\begin{equation}
\hat{\sigma}_{ij} = \frac{i}{2}\left(k_i \delta\hat{u}_j +k_j \delta\hat{u}_i - \frac{2}{3}\delta_{ij} k_l \delta\hat{u}^l \right), 
\end{equation}
\begin{equation}
\hat{\pi}_{ij} = -i\left(k_i k_j - \frac{1}{3}\delta_{ij}k_l k^l\right)k_m \delta\hat{u}^m, 
\end{equation}
\end{subequations}
and $\eta(\omega, k^2)$, $\xi(\omega, k^2)$ are momentum-dependent transport coefficients. As explained in Refs.~\cite{Bu:2014ena,Bu:2014sia}, these momentum-dependent transport coefficients are computed by solving a system of four coupled radial ODEs (ordinary differential equations) in a black brane background. 

For a sound wave $\delta\vec{u}$ is parallel to $\vec{k}$ and one can employ rotational invariance to set $\delta\hat{u}_i = \delta\hat{u}\,\delta_{i,1}$, $k_i = k\,\delta_{i,1}$ with no loss of generality. This implies that $\hat{\pi}_{ij} = -\frac{1}{2} k^2 \hat{\sigma}_{ij}$ and, hence, 
\begin{equation}
\delta\hat{\Pi}_\star = \delta\hat{\Pi}_{33} = \frac{1}{3}\left(\eta - k^2\xi\right) i k \delta\hat{u}. \end{equation}
Recalling the general definition of $\gamma_s$ given in Eq.~\eqref{gamma_s_def}, we have that 
\begin{equation}\label{hol:gamma_s-BL_relation}
2\mathcal{E}_0 \gamma_s = \eta - k^2\xi. 
\end{equation}
Equation~\eqref{hol:gamma_s-BL_relation} shows that $\gamma_s$ is a linear combination of the dynamical transport coefficients $\eta$ and $\xi$ originally defined in Refs.~\cite{Bu:2014ena,Bu:2014sia}, and indicates that the method put forward there can be straightforwardly modified to compute $\gamma_s$ directly. Both this computation and the results it leads to are described in detail in Appendix~\ref{app:C} in the Supplemental Material. For our purposes here, it suffices to mention the following. 
\begin{enumerate}
\item For fixed $k \in \mathbb R$, $\gamma_s$ is a meromorphic function of $\omega\in\mathbb C$, with an infinite number of simple poles, $\Omega_q^{(\pm)}(k)$, symmetric around the imaginary $\omega$ axis, $\Omega_q^{(+)}(k) = - \Omega_q^{(-)}(k)^*$. The imaginary part of $\Omega_q^{(\pm)}(k)$ is always negative. 
\item These simple poles can be computed by solving the following eigenvalue problem:  
\begin{subequations}
\begin{equation}\label{hol:residue_eq_1}
P''+\frac{P'}{r}-\frac{4P}{r^2} - 2 i k Q' = 0,     
\end{equation}
\begin{equation}
\begin{split}\label{hol:residue_eq_2}
&Q'' + \frac{f+4r^2-2ir\Omega_p}{r f}Q' + \frac{k^2-9 i \Omega_p r}{3 r^2 f}Q \\
&+ \frac{2 i k}{3 r^2 f}P' +\frac{2 i k}{3 r^3 f}P = 0,
\end{split}
\end{equation}
\end{subequations}
where $f= r^2 - \mu^4 r^{-2}$. The boundary conditions to be imposed on $P$ and $Q$ are regularity at the black brane horizon located at $r = \mu$ and the near-boundary behaviour $P = O(r^{-3})$, $Q=O(r^{-4})$ as $r \to \infty$. 
\end{enumerate}
The two observations above are sufficient to state the main result of this section: if the gradient expansion grows factorially and the singulant field is independent of the metric component under consideration, then the singulant equation of motion is determined by the poles of~$\gamma_s$, just as it was the case in the model discussed in Sec.~\ref{subsec:new_model}. This fact follows immediately from the realisation that the map, 
\begin{equation}
\begin{split}
&P \to \pm e^{-b}\bar{V}_\sigma,\quad Q \to \bar{B},\quad \mu \to r_h, \\
&\Omega_p \to - i U(\chi), \quad
k \to \pm i Z(\chi), 
\end{split}
\end{equation}
transforms Eqs.~\eqref{hol:residue_eq_1} and \eqref{hol:residue_eq_2} into Eqs.~\eqref{hol:D2_large-order} and \eqref{hol:D3_large-order} and that the infrared and ultraviolet boundary conditions obeyed by $P, Q$ and $V_\sigma^{(n)}, B^{(n)}$ are the same for $n > 2$. The existence of this map conforms to the general expectations put forward in Appendix~\ref{app:A} in the Supplemental Material and summarised in Sec.\,\ref{sec:singulants}.

\section{Kinetic theory}
\label{sec:kt}

In kinetic theory, we take a factorial ansatz for the distribution function itself, and derive constraints on the singulant from the Boltzmann equation. We cross-check our results in RTA kinetic theory for Bjorken flow, where we can compute precise numerical solutions. We also reanalyse the $1/w$ expansion in this new perspective, extending results from previous work on the large-order behaviour of moments \cite{Denicol:2016bjh, Heller:2016rtz, Heller:2018qvh}.

\subsection{Singulants of the distribution function}

The Boltzmann equation is
\begin{equation}\label{kt:boltzmann}
    \epsilon p^\mu \partial_\mu f(x,p) = C[f],
\end{equation}
which together with the ansatz \eqref{singulants:f} implies that $f^{(n)}$ is of order $n$ in gradients of $f^{(0)}$. With the factorial-over-power ansatz \eqref{factorial_ansatz_f} the resulting linearisation implies that up to subleading corrections in $1/n$,
\begin{equation}
(- p^\mu\partial_\mu \chi(x,p)) \frac{A(x,p)}{\chi^{n+\alpha}(x,p)} = \bold{C}\left[\frac{A}{\chi^{n+\alpha}}\right],
\end{equation}
where $\bold{C}$ is the linearised collision operator: 
\begin{equation}
\bold{C} = \left.\frac{\delta C[f]}{\delta f} \right|_{f=f^{(0)}}.
\end{equation}
Differently from the other theories, the factor $A/\chi^n$ does not cancel out, so the large-order equation seems to depend explicitly on $n$. 

Let us sidestep the issue of defining what $A/\chi^n$ converges to. In the limit, we see that the action of $\bold{C}$ must be identical to the action of $-p^\mu\partial_\mu \chi(x,p)$. The latter does not couple different values of $p$. Thus, the equation of motion for the singulant can be determined by finding those distribution functions where $\bold{C}$ effectively acts \emph{diagonally} in $p$. 

This is automatically fulfilled in the simplest RTA model,\footnote{However, additional terms must be included in order for the momentum dependent relaxation time to be consistent with conservation laws \cite{Rocha:2021zcw}.} where
\begin{equation}
    \bold{C} = \frac{p^\mu U_\mu}{\tau_\text{rel}(x,p)},
\end{equation}
leading to 
\begin{equation}
    p^\mu\left(\partial_\mu \chi(x,p) + U_\mu/\tau_\text{rel} \right) = 0.
\end{equation}
We can write the solution as
\begin{equation}\label{eq:rta_singulant_eom}
    \chi = -\int \frac{U^\mu}{\tau_\text{rel}} dx_\mu + \chi_\text{FS},
\end{equation}
where $\chi_\text{FS}$ is any solution to the free-streaming Boltzmann equation. 

\subsection{Moments}

Consider some moment, whose $n$th term arises from the $n$th term of the distribution function as
\begin{equation}
I^{\mu_1 \ldots \mu_j}_{n} = \int_p f^{(n)} p^{\mu_1}\dots p^{\mu_j}.
\end{equation}
With the singulant ansatz for $f^{(n)}$, this is 
\begin{equation}
I^{\mu_1 \ldots \mu_j}_{n} = \int_p  \frac{\Gamma(n+\alpha)}{\chi(x,p)^{n+\alpha}} A(x,p) p^{\mu_1}\dots p^{\mu_j}.
\end{equation}
To evaluate this in the limit when $n\rightarrow \infty$, we can try a saddle point integration. Then the leading contributions come from those points $p_s$ where $\nabla_p\ln\chi(x,p)|_{p=p_s}=0$. This heuristic argument suggests that the large-order behaviour of the moment is governed by the saddle point singulant $\chi_s(x) \equiv \chi(x,p_s)$. This implies that
\begin{equation}
I^{\mu_1 \ldots \mu_j}_{n} \sim \sum_{\text{saddles}} A_s(x) \frac{\Gamma(n+\alpha_s(x))}{\chi_s(x)^{n+\alpha_s(x)}}  p_s^{\mu_1}\dots p_s^{\mu_j},
\end{equation}
where $\alpha_s= \alpha(x,p_s) - j/2$. $A_s$ gets contributions from $A$, the Hessian of $\chi$, and the measure \cite{bleistein2012saddle}.

We also note with this argument, the tensor structure is not sensitive to $n$, which is consistent with the general ansatz in \rf{singulant_ansatz_general}.

\subsection{Gradient expansion in Bjorken flow}

We now consider Bjorken flow in RTA kinetic theory, where the singulant equation of motion can be verified using the results of Ref.~\cite{Heller:2018qvh}. With a relaxation time $\tau_\text{rel} = \gamma/T$ the Boltzmann equation is 
\begin{equation}\label{eq:boltzmann_rta_bjorken}
    \epsilon \frac{\gamma}{T} \partial_\tau f(\tau, v, p_T) = f_\text{eq}-f,
\end{equation}
where $p_T$ is the magnitude of the transverse momentum, $v = t P_L - z E$, and $f_\text{eq} = \exp{\left( -\sqrt{ \frac{v^2}{(\tau T)^2} + \frac{p_T^2}{T^2} } \right)}$. There is a simple recursion relation for $f^{(n)}$: 
\begin{subequations}
\begin{align}
    f^{(0)} &= f_\text{eq}, \\
    f^{(n+1)} &= -\frac{\gamma}{T} \partial_\tau f^{(n)},
\end{align}
\end{subequations}
and the singulant satisfies
\begin{equation}\label{eq:rta_bjorken_epsilon_singulant_eom}
\partial_\tau \chi(\tau,v,p_T) = \frac{1}{\tau_\text{rel}},
\end{equation}
which is just \rf{eq:rta_singulant_eom} specialised to this model.

Equation \eqref{eq:boltzmann_rta_bjorken} can be integrated to give an equation for the temperature $T(\tau)$ \cite{Baym:1984np, Florkowski:2013lya}, which can be used to get precise numerical solutions. Together with the recursion relation this allows access to large orders of $f^{(n)}$.
\begin{figure}[th]
    \centering
    \includegraphics[width=.95\linewidth]{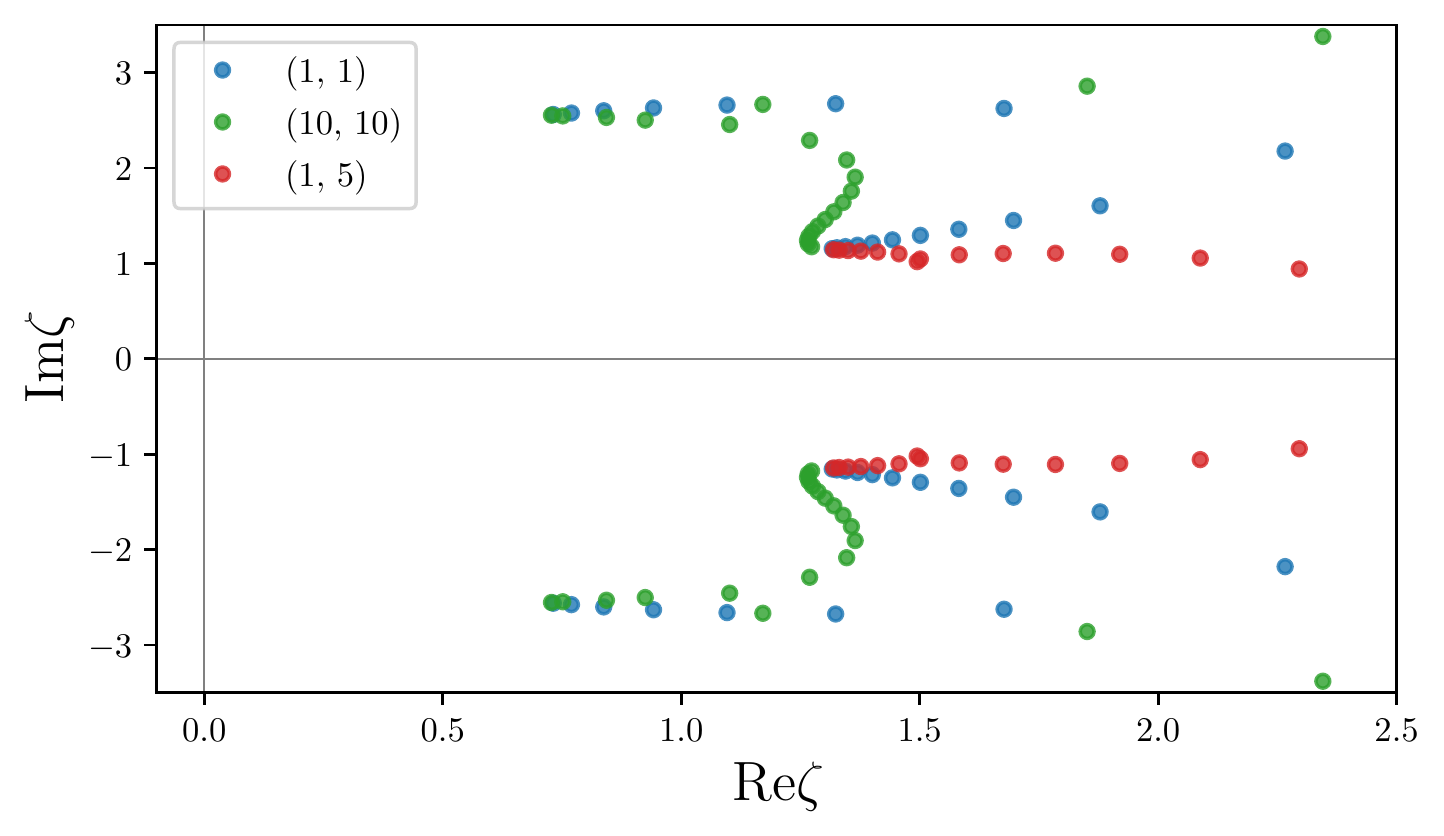}
    \includegraphics[width=.95\linewidth]{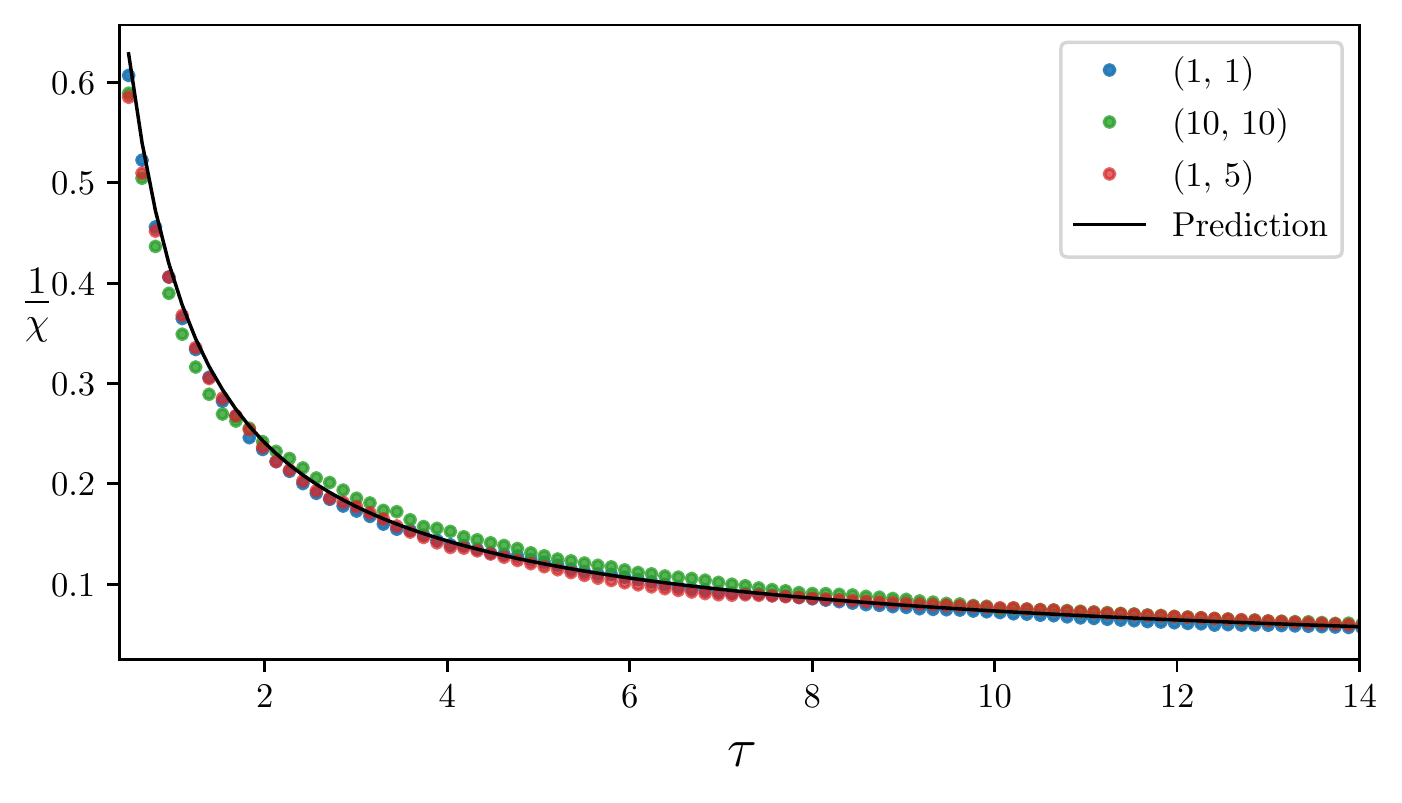}
    \caption{Singulants in the gradient expansion in RTA kinetic theory in Bjorken flow. \textbf{Top panel:} Borel plane at $\tau \approx 0.6$ for three values of the momentum $(v,p_T)$. One singulant seems to be approximately momentum independent, and is the dominant one. Another depends on their ratio. 
    \textbf{Bottom panel:} test of the singulant equation of motion. $\chi$ can be measured numerically at each point in time by calculating the gradient expansion. Alternatively, we can take the dominant singulant from the top panel and evolve it by \rf{eq:rta_bjorken_epsilon_singulant_eom}. The figure shows that these procedures agree.}
    \label{fig:rta_epsilon}
\end{figure}

In Fig.\,\ref{fig:rta_epsilon}, we verify that the singulant equation of motion is satisfied, and that there are momentum-dependent singulants.

\subsection{$1/w$ expansion revisited \label{eq.1overw}}

Another expansion, used in previous works \cite{Heller:2016rtz, Heller:2018qvh}, is to expand the pressure anisotropy 
\begin{equation}
    \mathcal{A} \equiv \frac{P_T - P_L}{P}
\end{equation}
in inverse powers of the variable $w \equiv \tau T$. The essential difference between these expansions is that the latter one uses the conservation equations to remove all appearances of $\partial_\tau T$; see Ref.\,\cite{Heller:2021oxl} for details. This expansion leads to different singulants from those in previous section.

\begin{figure}[h!]
    \centering
    \includegraphics[width=.95\linewidth]{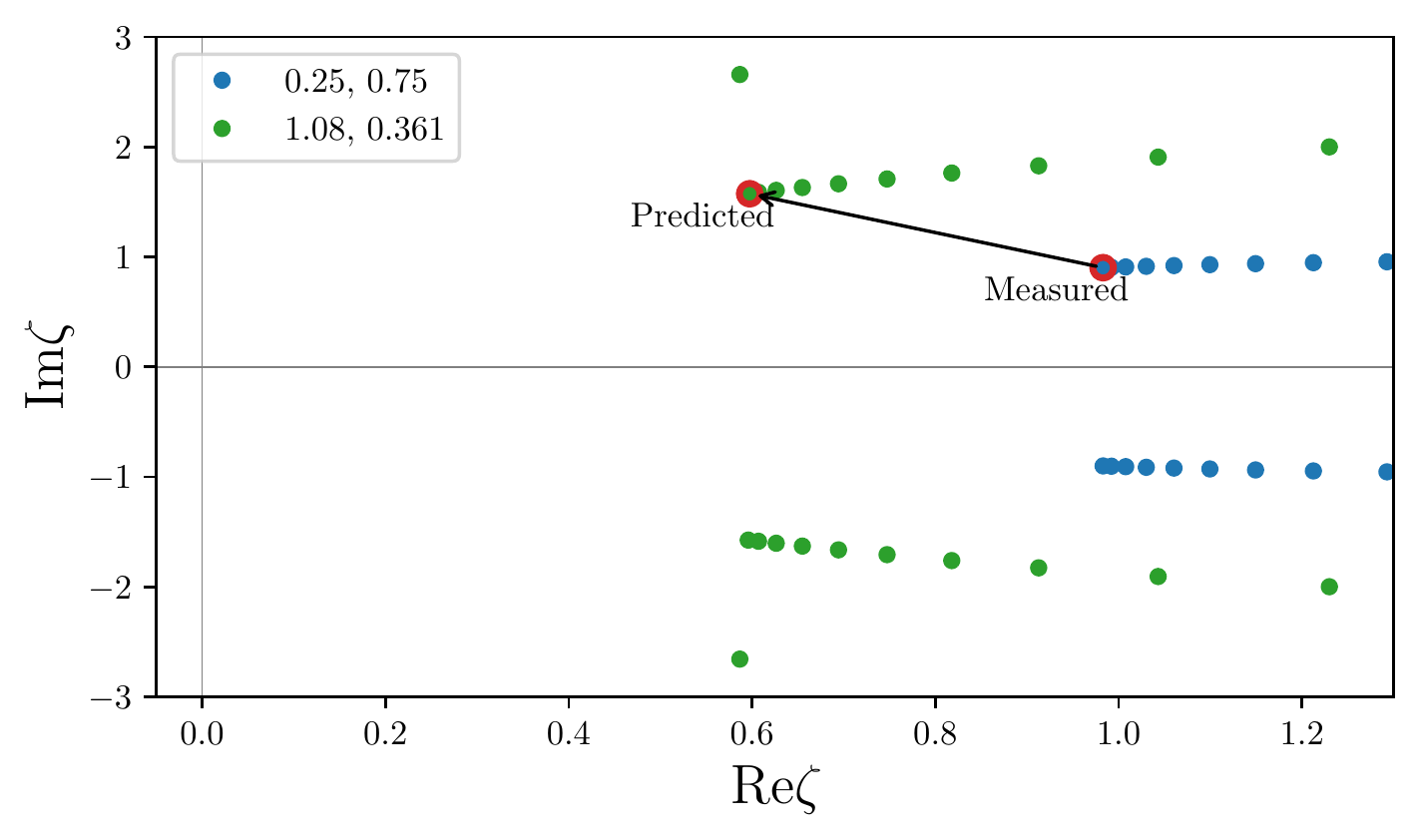}
    \caption{Borel plane for the distribution function in RTA kinetic theory in Bjorken flow, for the $1/w$ expansion. According to \rf{eq:kinetic_chi_solution}, the value of the singulant $\chi(\mathfrak{p}_1,\mathfrak{p}_2)$ can be predicted along a certain trajectory in the space of $(\mathfrak{p}_1,\mathfrak{p}_2)$. We verify this by calculating the singulants for two points on such a curve. These singulants are shown in the Borel plane and the two sets of parameters correspond to the blue and green points in the figure. From the singulant revealed by one set, we predict the location for the other parameters and confirm this prediction using the explicit large-order computation.}
    \label{fig:kinetic_borel}
\end{figure}

It is convenient to change variables to
\begin{subequations}\label{eq:rta_momentum_variables}
\begin{align}
    \mathfrak{p}_1 &\equiv \frac{v^2}{w^2},\\
    \mathfrak{p}_2 &\equiv \frac{p_T^2}{T^2},
\end{align}
\end{subequations}
and we now expand as
\begin{equation}
    f = \sum_{n=0}^\infty \frac{\tilde{f}_n(\mathfrak{p}_1,\mathfrak{p}_2)}{w^n},
\end{equation}
with $\tilde{f}_0 = e^{-\sqrt{\mathfrak{p}_1 + \mathfrak{p}_2}}$. Here tildes denote series coefficients in the $1/w$ expansion. The pressure anisotropy is also expanded as
\begin{equation}
    \mathcal{A}(w) = \sum_{n=1}^\infty \frac{\tilde{a}_n}{w^n}.
\end{equation}
The Boltzmann equation leads to the following  recursion relation for the expansion coefficients: 
\begin{multline}
    \frac{\tilde{f}_{n+1}}{\gamma} = \frac{2}{3} n \tilde{f}_n + \frac{4}{3} \mathfrak{p}_1\partial_{\mathfrak{p}_2} \tilde{f}_n - \frac{2}{3} \mathfrak{p}_2 \partial_{\mathfrak{p}_2} \tilde{f}_n \\+ \frac{1}{18}\sum_{k=0}^n \tilde{a}_{n-k}(k \tilde{f}_k + 2 \mathfrak{p}_1 \partial_{\mathfrak{p}_1} \tilde{f}_k +2 \mathfrak{p}_2 \partial_{\mathfrak{p}_2} \tilde{f}_k).
\end{multline}
This recursion can be used to calculate several hundred orders of $\tilde{f}_n$ and $\tilde{a}_n$ analytically. The $\tilde{a}_n$ coefficients were observed to diverge factorially in Ref.\,\cite{Heller:2016rtz}. Numerical studies of these series show that the coefficients $\tilde{f}_n(\mathfrak{p}_1,\mathfrak{p}_2)$ diverge factorially, with a slope that depends on the values of $\mathfrak{p}_1$ and $\mathfrak{p}_2$. We are therefore led to the ansatz
\begin{equation}
    \tilde{f}_n = \frac{\Gamma(n+\alpha)}{\chi(\mathfrak{p}_1,\mathfrak{p}_2)^{n+\alpha}} A(\mathfrak{p}_1,\mathfrak{p}_2),
\end{equation}
with a momentum-dependent singulant. The recursion relation implies that $\chi$ satisfies
\begin{equation}\label{eq:kinetic_chi_equation}
    \frac{3}{2\gamma} = \chi -2 \mathfrak{p}_1\partial_{\mathfrak{p}_1}\chi + \mathfrak{p}_2\partial_{\mathfrak{p}_2} \chi,
\end{equation}
where the terms involving the coefficients of the pressure anisotropy have dropped out. This equation has the solutions 
\begin{equation}\label{eq:kinetic_chi_solution}
    \chi(\mathfrak{p}_1,\mathfrak{p}_2) = \frac{3}{2\gamma} + \sqrt{\mathfrak{p}_1} G\left( \mathfrak{p}_2\sqrt{\mathfrak{p}_1}\right),
\end{equation}
where $G$ is an arbitrary function. $G$ is in principle determined by matching a sum of singulants to the zeroth order of the perturbative series, but we do not know how to do this in practice. However, even without knowing the function $G$, the values of $\chi(\mathfrak{p}_1,\mathfrak{p}_2)$ on a curve where $\mathfrak{p}_2\sqrt{\mathfrak{p}_1}$ is constant can be calculated from any other. We verified this using the explicit large-order computation; see Fig.\,\ref{fig:kinetic_borel}. 

For the pressure anisotropy, let us apply the saddle point argument from above. Equation \eqref{eq:kinetic_chi_equation} immediately implies that $\chi_* = \frac{3}{2\gamma}$ at a saddle point. Thus, even though $G$ is unknown, and therefore the saddle points $p_*$ cannot be determined, the value of the singulant at the saddle point is uniquely fixed. However, it is also known that there are contributions with a nonzero imaginary part \cite{Heller:2016rtz}, which do not correspond to physical excitations. We do not know why they do not show up in this analysis, but a possibility is that they come from end point contributions of the saddle point integral. 

Finally, note that introducing ${\cal A}(w)$ simplifies a lot of analysis in the Bjorken flow and makes some features manifest. For example, it led to the notion of hydrodynamic attractors. Therefore, it has been an interesting and potentially important question if ${\cal A}(w)$ acquires a natural generalisation when symmetries of the flow are relaxed. While generalisation of ${\cal A}$ can be trivially obtained as long as the local rest frame exists, the $w$ clock variable is less obvious. Earlier attempts in this direction include~\cite{Romatschke:2017acs,Denicol:2020eij} and our work adds singulants as a particularly natural possibility. The reason for it is twofold: the leading singulant governs the asymptotics of the gradient expansion and, through resurgence, also the most significant exponentially decaying term supplementing optimally truncated (see the next section) hydrodynamic constitutive relations, as $w$ also does.

\section{Optimal truncation of the gradient expansion}
\label{sec:optimal}

Having established that the gradient expansion \eqref{singulants:Pi_series} is factorially divergent in MIS-like theories, the natural question that arises is what its practical usefulness is. There are two different ways in which one can employ a factorially divergent gradient expansion: a fixed-order truncation or an optimal truncation. In this section, we explore the second option. We remove the bookkeeping parameter $\epsilon$ by setting it to one. 

The optimal truncation of the gradient expansion at a given spacetime point is the partial sum 
\begin{equation}
S^{(n_\textrm{opt})}(t,x) =\sum_{n=1}^{n_\textrm{opt}} \Pi_\star^{(n)}(t,x) \end{equation}
closest to the actual value of $\Pi_\star(t,x)$. Note that, due to this definition, the order of optimal truncation $n_\textrm{opt}$ is expected to be spacetime dependent. 

Our main objective in this section is to put forward a criterion for estimating $n_\textrm{opt}$ that relies exclusively on the gradient expansion itself. Our choice is the following. Let $|\chi_d(t,x)|$ correspond to the absolute value of the dominant singulant at a given spacetime point. We propose to estimate the order of optimal truncation at that spacetime point by the relation 
\begin{equation}\label{n_o_def}
n_\textrm{opt,est}(t,x) = [|\chi_d(t,x)|], 
\end{equation}
where the brackets instruct us to take the integer part of the quantity they enclose. 
\begin{figure}[h!]
\centering
\includegraphics[width=0.97\linewidth]{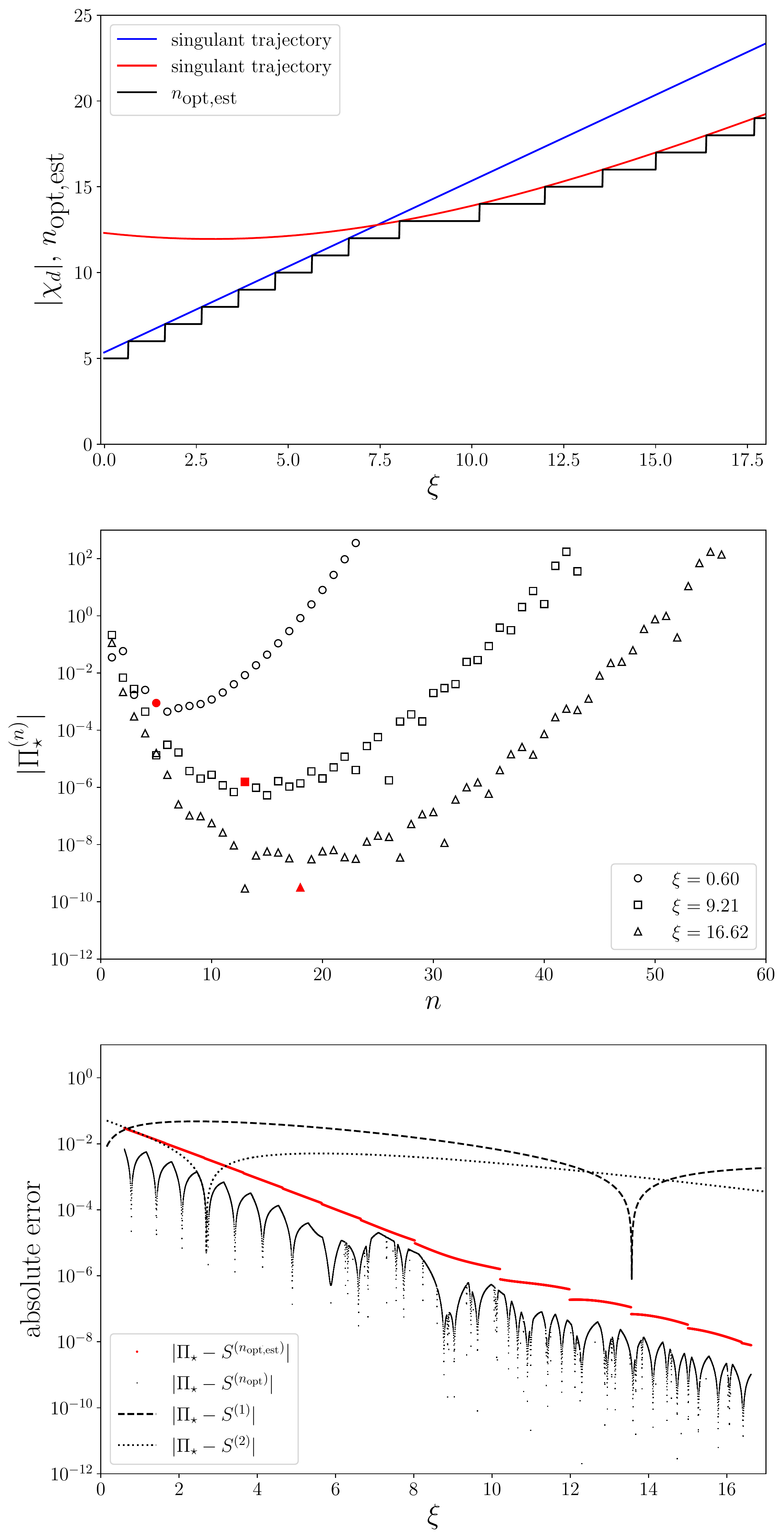}
\caption{\textbf{Upper panel}: evolution of the norm of the dominant singulant $|\chi_d|$ as a function of $\xi$ along the flow line located at $x=0$ for initial data \eqref{opt:id}. A real singulant, whose trajectory is denoted by the solid blue line, dominates at early times. At late times, a complex-conjugated singulant pair, whose trajectory corresponds to the solid red line, takes over. The discontinuous solid black line corresponds to $n_\textrm{opt,est}$, as given by Eq.\,\eqref{n_o_def}. \textbf{Middle panel}: coefficients of the gradient expansion at three selected times. The ones associated to the corresponding $n_\textrm{opt,est}$ have been highlighted in red. \textbf{Lower panel}: absolute error of the optimal truncation (black crosses) and the truncation at order $n_\textrm{opt,est}$ (red dots). Crucially, the latter error provides an upper bound for the former that displays the same time dependence. The dashed (dotted) black line represents the absolute error of the first-order (second-order) order truncation. \label{fig:optimal_truncation}}
\end{figure}

To explore the consequences of \rf{n_o_def}, let us consider the case in which the large-order behaviour of the gradient expansion is of the form \eqref{factorial_ansatz} and that $|\chi_d| \gg 1$. In this situation, it is immediate to demonstrate that $\Pi_\star^{(n_\textrm{opt,est})}$ is the smallest coefficient of the gradient expansion, 
\begin{equation}
|\Pi_\star^{(n_\textrm{opt,est})}(t,x)| < |\Pi_\star^{(n)}(t,x)| \quad \forall n \neq n_\textrm{opt,est}, 
\end{equation}
and, furthermore, that this smallest coefficient is exponentially suppressed in $|\chi_d|$: 
\begin{equation}
|\Pi_\star^{(n_\textrm{opt,est})}| \sim \sqrt{2\pi} P |\chi_d|^{-\frac{1}{2}}e^{-|\chi_d|}.  
\end{equation}
The standard expectation is that $S^{(n_\textrm{opt,est})}$ provides a representation of the actual $\Pi_{\star}$ which is accurate up to such exponentially small term: 
\begin{equation}\label{error_expectation}
|\Pi_\star - S^{(n_\textrm{opt,est})}| = O\left( e^{-|\chi_d|}\right), \quad |\chi_d| \gg 1.     
\end{equation}
In the case that the dominant singulant is not real but rather corresponds to a complex-conjugated pair, the norm of $\Pi_{\star}^{(n)}$ displays additional oscillations at large $n$ with frequency $\arg \chi$,
\begin{equation}\label{error_expectation_complex}
|\Pi_\star^{(n)}| \sim 2|P|\cos((n+\alpha)\arg \chi - \arg P) \frac{\Gamma(n + \alpha)}{|\chi|^{n+\alpha}},   
\end{equation}
where we assumed that $\alpha$ is real. In this situation, the definition \eqref{n_o_def} singles out the gradient expansion coefficient for which the envelope of these oscillations is the smallest. 

To test our estimate, we consider the following initial data in BRSSS theory:  
\begin{equation}\label{opt:id}
T(0,x) = 2 - \tanh(x)^2,\,\,\,u(0,x) = \Pi_\star(0,x) = 0, 
\end{equation}
which correspond to an initial overdensity in the effective temperature such that $\frac{|T(0,0)-T(0,\infty)|}{T(0,\infty)} = 1$. We have set $\mathcal{E}/T^4 =1$, $\eta/s = 1/(4\pi)$ and $\tau_{\Pi}T = (2-\log 2)/(2\pi)$. We evaluate the gradient expansion along the flow line passing through $x = 0$ at $t = 0$. Our results will be parameterised in terms of the dimensionless variable
\begin{equation}
\xi(t) = \int_0^t \frac{dt'}{\tau_\Pi(T(t',0))}.     
\end{equation}
In the upper panel of Fig.\,\ref{fig:optimal_truncation}, we represent the time evolution of the dominant singulant and the corresponding $n_{\textrm{opt},\textrm{est}}$ as we move along this flow line. This dominant singulant has been extracted with the help of the method put forward in our analysis of the BRSSS model in  Sec.\,\ref{sec:phenomenological_models}. The reader can find where the gradient expansion coefficient $\Pi_\star^{(n_{\textrm{opt},\textrm{est}})}$ picked by our procedure sits at three different times in the middle panel of Fig.\,\ref{fig:optimal_truncation}. 

Finally, in the bottom panel of Fig.\,\ref{fig:optimal_truncation}, we plot the time evolution of the error of our estimate for optimal truncation 
\begin{equation}
|\Pi_\star - S^{(n_{\textrm{opt},\textrm{est}})}|,     
\end{equation}
and compare it with the error of the actual optimal truncation as originally defined. The latter is always upper bounded by the former and, furthermore, both errors decrease exponentially as $\xi$ grows. For reference, we also include the absolute errors incurred by the first and second-order truncations of the gradient expansion.

\section{Outlook}
\label{sec:outlook}

We have established that there exists a deep connection between far-from-equilibrium nonlinear relativistic hydrodynamics and linear response around global thermal equilibrium. This connection is manifest in the large-order behaviour of the hydrodynamic gradient expansion and is naturally encoded in singulants. When viewing this duality from the gradient expansion side, one can regard singulants as emergent excitations coming from a dramatic reorganisation of the series at large orders. We have outlined the principles governing singulant dynamics in longitudinal flows for MIS-like models, holography and kinetic theory, and checked their validity by explicit numerical computations in a number of cases. 

In the course of this work we have also investigated the large-order behaviour of the gradient expansion and singulants in non-relativistic theories along the lines of \cite{Chao:2011cy, Brewer:2015ipa}, finding results analogous to the ones presented here. This is beyond the scope of the current paper and we hope to report on it elsewhere \cite{ourupcomingstuff}. This suggests that singulants are a useful language for hydrodynamics in general.

Our work opens new research directions which come with new technical and conceptual challenges to be addressed. On the technical front, it would be interesting to solve for nonlinear longitudinal flows as an initial value problem in holography and RTA kinetic theory and confirm the predicted behaviour of singulants by analysing the gradient expansion at large order (just as we have done here for MIS-like models). On the conceptual front, we singled out four questions that we believe are of particular importance and discuss them in the sections below.

\subsection{Singulants and initial conditions}

While we have succeeded in determining the time evolution of individual singulants, we have not discussed how the number of singulants and their initial values are related to the initial out-of-equilibrium state under consideration. The complete answer to this problem remains unknown to us; we have, however, made partial progress on this front by working in the linear response regime. In this case, the problem of finding initial conditions for the singulant fields can be solved, and we refer the reader to Appendix~\ref{app:D} in the Supplemental Material for a worked-out example in BRSSS theory.

Revisiting this question for nonlinear flows is of fundamental importance in order to understand how to promote the gradient expansion \eqref{intro:master} to a full-fledged transseries representation of the energy-momentum tensor. If there were a method to extract the values of the singulant fields at $t = 0$ from the initial data, then the singulant equation of motion would allow us to determine the dominant singulant at any given point \emph{without} having to compute the gradient expansion itself. In light of the results presented in Sec.\,\ref{sec:optimal}, the natural expectation that arises is that full knowledge of the spacetime profile of the dominant singulant would allow us to single out the spacetime region where relativistic hydrodynamics, when truncated at low order, will not be applicable---for instance, because $|\chi_d| \sim 1$ and no optimal truncation even exists.

Finally, it would be interesting to find out whether making progress on this front could shed light on the necessary conditions that the initial data have to obey for a factorial growth to be present, and the singulant ansatz \eqref{singulant_ansatz_general} to be applicable. Further discussion on necessary conditions for factorial growth can be found in \cite{Heller:2021oxl}.

\subsection{Singulants beyond longitudinal flows}

Another important question we have not discussed up to this point is whether the results presented in this work generalise beyond longitudinal flows. The main simplification brought by restricting ourselves to this class of fluid flows was that $\Pi_{\mu\nu}$ could be described in terms of a single scalar field, $\Pi_\star$. We have seen that a longitudinal flow is nothing but a nonlinear sound wave. Because of this, and the linearisation entailed by the asymptotic ansatz \eqref{singulants:factorial-over-power_ansatz}, the singulant equation of motion was determined by a linear response theory problem formulated in the sound channel. 
In the absence of symmetry constraints, a $d$-dimensional conformal energy-momentum tensor is associated with a  $\Pi_{\mu\nu}$ that encompasses $d(d{-}1)/2{-}1$ degrees of freedom. It is natural to wonder whether, in this completely general situation, the leading large-order behaviour of $\Pi_{\mu\nu}^{(n)}$ can be expressed as a linear superposition of contributions defined in independent channels\footnote{We omit the sum over independent singulant contributions within a given channel for presentational simplicity.} 
\begin{equation}\label{singulants:channel_decomposition}
\Pi_{\mu\nu}^{(n)} \sim \sum_c \frac{\Gamma(n+\alpha)}{\chi_{c}^{n+\alpha}} A_{c,\mu\nu}, 
\end{equation}
and, furthermore, if these conjectured channels can be put into one-to-one correspondence with the tensor, vector, and scalar channels that arise when decomposing an infinitesimal perturbation of the energy-momentum tensor around thermal equilibrium. 

In Appendix~\ref{app:B} in the Supplemental Material we provide the first steps to address this question in the context of MIS-like theories. The analysis presented there shows that, for a decomposition along the lines of Eq.\,\eqref{singulants:channel_decomposition} to be possible, one would need to work with a basis for the tangent space at a given point that depends explicitly on the singulant fields themselves. A fully fledged numerical analysis beyond the realm of longitudinal flows in these models along the lines of Ref.\,\cite{Heller:2021oxl} is required to find out whether this conclusion is correct. This computation will be numerically more costly than the longitudinal flow one, but we expect it to be feasible. 

We would like to point out that, when it comes to this issue, kinetic theory stands aside the other models we have considered. In the kinetic theory case, we know that the large-order behaviour of the gradient expansion of the distribution function does not decompose into channels. It would be interesting to understand whether, in spite of this, computing the second-order moments to get the energy-momentum tensor leads naturally to a channel decomposition. 

While this discussion might sound a bit technical, it does have interesting physical implications. The key reason for this is the idea of rheology, which aims to improve hydrodynamic predictions by making transport coefficients depend on gradients of fluid variables, see in particular~\cite{Lublinsky:2007mm} for a pioneering effort in this direction. Recently it was revisited in~\cite{Romatschke:2017vte,Romatschke:2017acs,Blaizot:2017ucy,Behtash:2018moe,Behtash:2019txb} by including contributions from transients in redefined transport coefficients. The true potential of the latter approach depends on correctly accounting for transient effects, which in our language are simply the singulant fields. In particular, the extent to which this is feasible in general depends crucially on understanding the detailed tensor structure of the nonhydrodynamic sectors of the transseries representing the energy-momentum tensor. Our work opens this possibility.

\subsection{Singulants in other gradient expansions \label{sec.othergradients}}

Finally, we would like to point out that  \rf{intro:PiSeries}, while defining classical relativistic hydrodynamics as an effective field theory, is not the only gradient expansion one can work with. Our analysis of the $1/w$ expansion in RTA kinetic theory already illustrates that singulants provide novel insights beyond the realm of the gradient expansion \eqref{intro:PiSeries}, and we expect this observation to be completely general. 

To further illustrate this point, let us stay within the realm of longitudinal flows and consider a scenario in which, besides $\Pi_\star$, both $\mathcal{E}$ and $U$ are gradient-expanded, 
\begin{equation}\label{grad_exp_alt}
\begin{split}
\mathcal{E} = \sum_{n=0}^\infty &\mathring{\mathcal{E}}^{(n)} \epsilon^n,\quad U = \sum_{n=0}^\infty \mathring{U}^{(n)} \epsilon^n, \\
&\Pi_\star = \sum_{n=1}^\infty \mathring{\Pi}_\star^{(n)} 
\epsilon^n,    
\end{split}
\end{equation}
with the conservation equation $\nabla_\mu T^{\mu\nu} = 0$ being solved in a small-$\epsilon$ expansion. In this alternative gradient expansion, $\mathring{\mathcal{E}}^{(0)}$ and $\mathring{U}^{(0)}$ represent the solution of ideal hydrodynamics associated with the initial spatial profiles of the energy density and the velocity field, while higher-order terms encode dissipative corrections to the ideal solution which vanish at $t=0$. We refer to the gradient expansion \eqref{grad_exp_alt} as the \emph{ideal} expansion. In the context of Bjorken flow, the ideal expansion is precisely the expansion in inverse powers of the proper time $\tau$, and the $1/w$ expansion arises as a (partial) resummation of it. 

Let us assume that the large-order behaviour of $\mathring{\mathcal{E}}^{(n)}$, $\mathring{U}^{(n)}$, and $\mathring{\Pi}_\star^{(n)}$ are governed by a common singulant field. The general rules we put forward in Appendix~\ref{app:A} in the Supplemental Material entail that the singulant equation of motion is connected to a linear response theory problem through the map \eqref{master_map}. In this regard, the fundamental differences between the gradient expansions \eqref{intro:master} and \eqref{grad_exp_alt} are twofold. First, the role that the microscopic $\mathcal{E}$ and $U$ had in the map associated to the gradient expansion \eqref{intro:master} is taken over by $\mathring{\mathcal{E}}^{(0)}$ and $\mathring{U}^{(0)}$ when the ideal expansion is considered. Second, the particular linear response theory problem associated to the ideal expansion is different from the one relevant for the gradient expansion \eqref{intro:master}: rather than being determined by the poles of $\gamma_s$, the singulant equation of motion in the ideal expansion is set by the transient sound channel modes. The reason is that, to compute the ideal expansion, the conservation equation is also gradient expanded. 

Another expansion to which one can apply some of the techniques developed in this work is the \emph{Cauchy data expansion}. In this case one employs the conservation equations to systematically replace  longitudinal derivatives in the constitutive relations by transverse ones. The Cauchy data expansion is known up to third order in gradients at the fully nonlinear level \cite{Grozdanov:2015kqa,Diles:2019uft}, and its large-order behaviour was analysed in Ref.\,\cite{Heller:2020uuy} for conformal fluids in the context of linearised hydrodynamics, where it only involves spatial derivatives---hence our nomenclature. The aforementioned replacement, while inconsequential from the effective field theory perspective, makes a difference when the constitutive relations are evaluated on a fluid flow. Given the results of Ref.\,\cite{Heller:2020uuy}, we expect the Cauchy data expansion to be factorially divergent at the nonlinear level, and it would be interesting to explore in detail the link between  singulant dynamics and linear response theory in this case. 

Finally, we want to mention that our general expectation is that different frame choices, i.e., different choices of collective fields, would result in singulants with different dynamics. This should not come as a surprise, given the previous discussion on the ideal expansion. In fact, as the causal completions of relativistic hydrodynamics recently put forward by Bemfica, Disconzi, Noronha \cite{Bemfica:2017wps} and Kovtun \cite{Kovtun:2019hdm} (BDNK) illustrate, there exist phenomenological models for which even the very existence of the gradient expansion \eqref{intro:master} is contingent on the choice of collective fields. To make meaningful comparisons between gradient expansions across different models, one has to keep the choice of collective fields invariant. In this regard, it might be worthwhile to perform a field redefinition in the BDNK theories (or the theory put forward in Ref.~\cite{Noronha:2021syv}) to put their gradient-expanded constitutive relations in the Landau frame and check whether the singulant dynamics in longitudinal flows is still controlled by the poles of $\gamma_s$. 

\subsection{Physical interpretation of singulants}

For systems near thermal equilibrium a particularly useful theoretical language is provided by linear response theory. To the leading order in the amplitude of distortions away from equilibrium, the dynamics of the system is determined by the properties of retarded two-point functions of local operators. Singularities of these correlators in Fourier space determine the dynamics of the system. In particular, black hole quasinormal modes in holography arise as single pole singularities of boundary correlators. MIS-like theories also have such single pole singularities, whereas in kinetic theory it is known that this singularity structure is richer and includes also branch points.

In holography (with a straightforward generalisation to MIS-like theories), short-lived quasinormal modes describe the decay of non-hydrodynamic degrees of freedom with the subsequent approach to equilibrium being governed by a few long-lived quasinormal modes realising linearised hydrodynamics. It is fair to say that quasinormal mode picture underlies much of progress achieved in studying strongly coupled systems in the course of the past two decades. However, outside the realm of linear response theory and in the absence of high degree of symmetries, it has never been clear what object, if any, arises as a natural generalisation of a quasinormal mode. We advocate that singulants can fill this role.

They are quasinormal-mode-like because they are excitations which describe non-hydrodynamic decay, reflected in the fact that in all examples considered in this paper, the value of $|\chi|$ grows after enough time has passed. We have not observed any examples where $\chi$ orbits the origin of the Borel plane, which is fundamentally tied to this interpretation. At the same time, they do not call for a proximity of global equilibrium. They differ from standard quasinormal modes in the sense that they apply arbitrarily far from equilibrium.

Finally, for singulants to fill the role of quasinormal modes for far-from-equilibrium systems, then what is their imprint on a given flow? We will now outline a class of measurements that can be performed, at least in principle. The data we assume is a measurement of the energy-momentum tensor $T_{\mu\nu}$ in a given spacetime region. Given this data one can solve the eigenvalue problem $T_{\mu\nu}U^\nu = -{\cal E} U_\mu$ at each spacetime point to obtain the fields ${\cal E}$ and $U$. Taking derivatives of ${\cal E}$ and $U$ one can assemble the structures we have given in detail and directly observe singulant dynamics, just as we have done through numerical simulation in the preceding sections.

In the special case of Bjorken flow, see Sec.~\ref{eq.1overw}, the hydrodynamic gradient expansion is the same for all nonequilibrium states. Therefore, another way of accessing the singulant dynamics in this case is to construct ${\cal A}(w)$ for two different flows and subtract them from one other. In this case the hydrodynamic sector of the transseries cancels completely, leaving singulant physics visible to the naked eye~\cite{Spalinski:2018mqg}.

As we discussed in section \ref{sec.othergradients} one has the freedom to vary the type of gradient expansion. This leads to different singulant dynamics which can be observed by constructing combinations of derivatives of ${\cal E}$ and $U$ appropriate for the gradient expansion in question. In this way, the freedom to choose the gradient expansion leads to an infinite family of related self-consistent predictions.
  
\begin{acknowledgments}
It is a pleasure to thank In\^es Aniceto for comments on the manuscript. We would also like to thank our anonymous referees for their insightful comments and questions. The Gravity, Quantum Fields and Information group at AEI was supported by the Alexander von Humboldt Foundation and the Federal Ministry for Education and Research through the Sofja Kovalevskaja Award. A. S. acknowledges financial support from Grant No. CEX2019-000918-M funded by Ministerio de Ciencia e Innovación (MCIN)/Agencia Estatal de Investigación (AEI)/10.13039/501100011033 and from the Polish National Science Centre Grant No. 2018/29/B/ST2/02457. M. S. is supported by the National Science Centre, Poland, under Grants No. 2018/29/B/ST2/02457 and No. 2021/41/B/ST2/02909. B. W. is supported by a Royal Society University Research Fellowship. Finally, we would like to thank the organisers and participants of the \emph{Applicable resurgent asymptotics: towards a universal theory} program at the Isaac Newton Institute for Mathematical Sciences for stimulating atmosphere and for an opportunity to present for the first time some of the early results behind this work. 
\end{acknowledgments}

\bibliographystyle{bibstyl}
\bibliography{literature}

\onecolumngrid
\clearpage
\begin{center}
\textbf{\large Supplemental Material}
\vspace{1em}
\end{center}
\twocolumngrid
\appendix
\setcounter{page}{1}

\section{General results on the singulant equation of motion}\label{app:A}

\subsection{Linearisation of the constitutive relations: a gradient expansion approach}

In this section we show how the constitutive relations become effectively linear when using the factorial-over power-ansatz in the large-order limit. For definiteness, we focus on the case of MIS-like theories.

Consider a factorially growing series 
\begin{subequations}
\begin{align}\label{eq:large_order_relations}
    f = \sum_{k=0}^\infty f_{(n)} \epsilon^n = \sum_{k=0}^\infty \frac{\Gamma(n+\alpha)}{\chi^{n+\alpha}(x)} \epsilon^n,
\end{align}
\end{subequations}
where for simplicity we suppress the $x$-dependence of both $A$ and $\alpha$. To leading order in large $n$, it is straightforward to show that the following relations hold
\begin{align}
(f^k)_{(n)} &\sim k f_{(0)}^{k-1} f_{(n)}, \\
(\epsilon^{k} \partial^{\alpha_1} \cdots \partial^{\alpha_k} f)_{(n)} &\sim f_{(n)} (-1)^{k}\partial^{\alpha_1}\chi\cdots \partial^{\alpha_k}\chi, 
\end{align}
where the $(n)$ instructs us to take the $n$-th order coefficient of the expansion in $\epsilon$ of the function. 

These relations have consequences for the gradient expanded constitutive relations. We take the ansatz for $\Pi^{\mu\nu}$ as 
\begin{equation}
\Pi^{\mu\nu} = \sum_{n=1}^\infty \epsilon^n \Pi^{\mu\nu}_{(n)}, \quad \Pi^{\mu\nu}_{(n)} = \frac{\Gamma(n+\alpha(x))}{\chi(x)^{n+\alpha(x)}} A^{\mu\nu}(x),
\end{equation}
and assume that the equation of motion for $\Pi^{\mu\nu}$ takes the form
\begin{equation}
    0 = F^{\mu\nu}[\Pi^{\alpha\beta}, T, U^\alpha,\partial^\gamma \Pi^{\alpha\beta},\partial^\alpha T,\dots],
\end{equation}
where $F^{\mu\nu}$ can be expanded in gradients as
\begin{equation}
F^{\mu\nu}=\sum_n \epsilon^n F_{(n)}^{\mu\nu}. 
\end{equation}
In the large-order limit, due to the relations above, only certain terms will contribute to $F_{(n)}^{\mu\nu}$.

Since the gradient expansion for $\Pi^{\mu\nu}$ starts at order $\epsilon$, \rf{eq:large_order_relations} implies that terms which are nonlinear in $\Pi^{\mu\nu}$ are suppressed at large $n$. The same holds for terms where $\Pi^{\mu\nu}$ is multiplied with gradients of the hydrodynamic fields $T$ and $U^\mu$. 

However, terms which consist solely of gradients of the latter fields, such as $\partial^n T$, present a problem. If they scale as $\Gamma(n)$, they should be included, otherwise not. We assume here that these terms do not contribute, which is the case in e.g. MIS-like models, where the gradient expansion is truncated at some order in those gradients.

With this assumption, only terms of the form $f_{\alpha\beta,\alpha_1\dots\alpha_k}^{\mu\nu}(T,U^\gamma) \epsilon^k \partial^{\alpha_1}\dots\partial^{\alpha_k} \Pi^{\alpha\beta}$ contribute. We then have
\begin{multline}
0 = F_{(n)}^{\mu\nu} = \left( \sum_{k=0}^\infty f_{\alpha\beta,\alpha_1\dots\alpha_k}^{\mu\nu}(T,U^\gamma) \epsilon^k \partial^{\alpha_1}\dots\partial^{\alpha_k} \Pi^{\alpha\beta} \right)_{(n)} \\ \sim \Pi^{\alpha\beta}_{(n)}\sum_{k=0}^\infty (-1)^{|k|} f_{\alpha\beta,\alpha_1\dots\alpha_k}^{\mu\nu}(T,U^\gamma) \partial^{\alpha_1}\chi\cdots \partial^{\alpha_k}\chi \\
= \Pi^{\alpha\beta}_{(n)} f_{\alpha\beta}^{\mu\nu}(T,U^\gamma,\partial^\gamma\chi).
\end{multline}

\subsection{Linearisation of the constitutive relations: a transseries approach}

Let us assume that the equations of motion that determine $\Pi_\star$ in the model under consideration take the form of nonlinear PDE system for $N$ fields
\begin{equation}
\boldsymbol{\varphi} = \{\varphi_1, \ldots \varphi_N\},
\end{equation}
that depend on variables $X^A$. Let us divide the variables $X^A$ into Minkowski space coordinates, $x^\mu$, and additional coordinates, $\xi^a$. In the MIS-like models discussed in Sec.~\ref{sec:phenomenological_models}, the latter variables are absent while, in holography, there is only one additional coordinate that parameterises the radial direction. We define 
\begin{equation}
\boldsymbol{\Phi} = \{\boldsymbol{\varphi},\partial_a \boldsymbol{\varphi}, \partial_a \partial_b \boldsymbol{\varphi}, \ldots\}, \end{equation}
and 
\begin{equation}
\boldsymbol{\Psi} = \{\boldsymbol{\Phi},\partial_\mu \boldsymbol{\Phi}, \partial_\mu \partial_\nu \boldsymbol{\Phi}, \ldots\}.   
\end{equation}
Finally, we assume that the nonlinear PDE system takes the form 
\begin{equation}\label{master_constitutive_relation}
\mathcal{L}(\boldsymbol{\Psi}, \boldsymbol{\alpha}, \partial_\mu \boldsymbol{\alpha}, \ldots) = 0,     
\end{equation}
where we have allowed for the existence of $M$ background fields $\boldsymbol{\alpha} = \{\alpha_1,\ldots,\alpha_M\}$ that depend on the Minkowski coordinates alone. 

Let us introduce the formal parameter $\epsilon$ by performing the rescaling 
\begin{equation}\label{general_rescaling}
x^\mu \to \frac{x^\mu}{\epsilon}, \quad \xi^a \to \xi^a, 
\end{equation}
and assume that, in terms of $\epsilon$, the fields $\boldsymbol{\varphi}$ can be written as a transseries in $\epsilon$, 
\begin{equation}\label{transseries_ansatz}
\varphi_i = \varphi_i^{\{0\}} + \sum_{q} e^{-\frac{\chi_q}{\epsilon}}\epsilon^{\gamma_q}\varphi_i^{\{1,q\}} + \ldots 
\end{equation}
where $\ldots$ denotes higher-order transseries sectors. $\varphi_i^{\{0\}}$, $\varphi_i^{\{1,q\}}$ are formal power series in $\epsilon$, 
\begin{equation}
\varphi_i^{\{0\}} = \sum_{n=0}^\infty \varphi^{\{0\}}_{i\,(n)} \epsilon^n, \quad \varphi_i^{\{1,q\}}=\sum_{n=0}^\infty \varphi^{\{1,q\}}_{i\,(n)} \epsilon^n,       
\end{equation}
and we assume that $\chi_q$ only depend on the Minkowski coordinates. Under these circumstances, the transseries ansatz \eqref{transseries_ansatz} translates directly into an analogous transseries ansatz for $\Phi_i \in \boldsymbol{\Phi}$. Note that $\boldsymbol{\alpha}$ are background fields for which no transseries representation is provided. 

Let us perform the rescaling \eqref{general_rescaling} at the level of the equation of motion \eqref{master_constitutive_relation}. We get that 
\begin{equation}\label{master_constitutive_relation_rescaled}
\mathcal{L}(\boldsymbol{\Phi}, \epsilon \partial_\mu \boldsymbol{\Phi}, \ldots, \boldsymbol{\alpha}, \epsilon\partial_\mu\boldsymbol{\alpha}, \ldots) = 0.     
\end{equation}
The coefficients $\varphi^{\{0\}}_{i\,(n)}$,$\varphi^{\{1,q\}}_{i\,(n)},\ldots$ are fixed by the requirement that the transseries ansatze \eqref{transseries_ansatz} solve  \eqref{master_constitutive_relation_rescaled} order by order in $\epsilon$ in a $\epsilon \to 0$ expansion. One finds the formal expansion 
\begin{align}\label{master_series_expansion}
\mathcal{L}(\boldsymbol{\Phi}, \epsilon \partial_\mu \boldsymbol{\Phi}, \ldots, &\boldsymbol{\alpha}, \epsilon\partial_\mu\boldsymbol{\alpha}, \ldots) = \mathcal{L}|_0 + \frac{\delta \mathcal L}{\delta \Phi_i}\Bigr|_0(\Phi_i-\Phi_{i\,(0)}^{\{0\}}) + \nonumber \\
&\frac{\delta \mathcal L}{\delta( \partial_{\mu_1}\ldots\partial_{\mu_n})\Phi_i}\Bigr|_0\epsilon^n\partial_{\mu_1}\ldots\partial_{\mu_n}\Phi_i +\nonumber\\
&\frac{\delta \mathcal L}{\delta( \partial_{\mu_1}\ldots\partial_{\mu_n})\alpha_i}\Bigr|_0\epsilon^n\partial_{\mu_1}\ldots\partial_{\mu_n}\alpha_i + \ldots
\end{align}
where the Einstein summation convention has been invoked. The vertical bar with the `0' subscript means that the expression to its left is to be evaluated at $(\boldsymbol{\Phi}^{\{0\}}_0, 0, \ldots, 0, \boldsymbol{\alpha}, 0, \ldots, 0)$, where $\boldsymbol{\Phi}^{\{0\}}_0 = \{\Phi^{\{0\}}_{i\,(0)}\}$. Finally, the term `$\ldots$' denotes higher-order contributions that play no role in the subsequent analysis.

Our focus is on coefficient of $e^{-\frac{\chi_q}{\epsilon}}$ in expression \eqref{master_series_expansion}. There most crucial property that we need to take into account is that 
\begin{equation}\label{crucial_property}
\begin{split}
&\epsilon^n\partial_{\mu_1}\ldots\partial_{\mu_n} \left[e^{-\frac{\chi_q}{\epsilon}}\epsilon^{\gamma_q}\left(\Phi^{\{1,q\}}_{i\,(0)} + O(\epsilon)\right)\right] = \\ &(-1)^n (\partial_{\mu_1}\chi_q)\ldots(\partial_{\mu_n}\chi_q)\left[e^{-\frac{\chi_q}{\epsilon}}\epsilon^{\gamma_q}\left(\Phi^{\{1,q\}}_{i\,(0)} + O(\epsilon)\right)\right]. 
\end{split}
\end{equation}
Hence, according to \eqref{crucial_property}, the first and last terms explicitly displayed in equation \eqref{master_series_expansion} make no contribution at leading order in $\epsilon$, while from the second and third ones we get our final result 
\begin{equation}\label{master_singulant_eom}
\frac{\delta \mathcal L}{\delta \Phi_i}\Bigr|_0 +  \frac{\delta \mathcal L}{\delta( \partial_{\mu_1}\ldots\partial_{\mu_n})\Phi_i}\Bigr|_0 (-1)^n (\partial_{\mu_1}\chi_q)\ldots(\partial_{\mu_n}\chi_q) = 0, 
\end{equation}
which is nothing but the singulant equation of motion.

With the hindsight provided by equation \eqref{master_singulant_eom}, let us elaborate further on the meaning of expression \eqref{master_series_expansion}. In order to obtain \eqref{master_series_expansion}, we treated every term in $\boldsymbol{\Phi}, \epsilon \partial_\mu \boldsymbol{\Phi}, \ldots$, $\epsilon \partial_\mu \boldsymbol{\alpha}, \ldots$ vanishing as $\epsilon \to 0$ as infinitesimal, without taking into account how fast this vanishing happens. Hence, there are terms explicitly kept in expression \eqref{master_series_expansion} which are subleading with respect to omitted terms represented by `$\ldots$'. Despite this, it is immediate to see that, since we focus solely on the coefficient multiplying $\epsilon^{\gamma_q}e^{-\frac{\chi_q}{\epsilon}}$, accounting for the correct hierarchy of the terms appearing in \eqref{master_series_expansion} is inconsequential for us. For instance, the possible second-order terms contribute a coefficient multiplying $\epsilon^{\gamma_q}e^{-\frac{\chi_q}{\epsilon}}$ that:  
\begin{itemize}
\item for $(\Phi_i - \Phi_{i\,(0)}^{\{0\}})(\Phi_j - \Phi_{j\,(0)}^{\{0\}})$ and $(\Phi_i - \Phi_{i\,(0)}^{\{0\}})\epsilon^n \partial_{\mu_1}\ldots\partial_{\mu_n}\Phi_j$ is at least $O(\epsilon)$, 
\item for $\epsilon^{n+m}(\partial_{\mu_1}\ldots\partial_{\mu_n}\Phi_i)(\partial_{\nu_1}\ldots\partial_{\nu_m}\Phi_j)$ is at least $O(\epsilon^{\textrm{min}(n,m)})$, 
\item for $(\Phi_i - \Phi_{i\,(0)}^{\{0\}})\epsilon^n \partial_{\mu_1}\ldots\partial_{\mu_n}\alpha_j$ and $\epsilon^{n+m}(\partial_{\mu_1}\ldots\partial_{\mu_m}\Phi_i)(\partial_{\nu_1}\ldots\partial_{\nu_n}\alpha_j)$ is at least $O(\epsilon^n)$. 
\end{itemize}
Hence, the second-order terms contribute a coefficient that it at least $O(\epsilon)$ and, as a consequence, do not alter equation \eqref{master_singulant_eom}. 

\subsection{Singulant equation of motion and linear response theory}

The singulant equation of motion \eqref{master_singulant_eom} is isomorphic to the result of the following procedure. First, we consider linearised perturbations of $\Phi_i$ around a $x^\mu$-independent reference state $\Phi_{i,0}$, 
\begin{equation}
\Phi_i = \Phi_{i,0} + \lambda\,\delta\Phi_i, \quad \delta\Phi_i=\delta\hat{\Phi}_i e^{i k_\mu x^\mu}, \quad \lambda\to0,
\end{equation}
with $k_\mu = (-\omega, \vec{k})$. Second, we take the background fields to be $x^\mu$-independent, $\alpha_{i,0} \in \mathbb R$. Finally, we assume that $\boldsymbol{\Phi}_0$ and $\boldsymbol{\alpha}_0$ are such that $\mathcal{L}(\boldsymbol{\Phi}_0,\ldots,  \boldsymbol{\alpha}_0,\ldots)=0$. 

To find the equation of motion for the perturbations, we expand in $\lambda$, 
\begin{equation}
\begin{split}
\mathcal{L} = &\lambda \left[\frac{\delta \mathcal{L}}{\delta\Phi_i}\Bigr|_0 \delta\Phi_i{+} \frac{\delta \mathcal{L}}{\delta(\partial_{\mu_1}\ldots\partial_{\mu_n}\Phi_i)}\Bigr|_0 \partial_{\mu_1}\ldots\partial_{\mu_n}\delta\Phi_i \right] \nonumber \\
&{+} O(\lambda^2). 
\end{split}
\end{equation}
Taking into account that, for $\delta\Phi_i=\delta\hat{\Phi}_i e^{i k_\mu x^\mu}$,  
\begin{equation}
\partial_{\mu_1}\ldots\partial_{\mu_n}\delta\Phi_i = (i k_{\mu_1})\ldots(i k_{\mu_n})\delta\Phi_i,     
\end{equation}
we get 
\begin{equation}\label{master_fluctuation_eom}
\frac{\delta\mathcal{L}}{\delta\Phi_i}\Bigr|_0 + \frac{\delta \mathcal{L}}{\delta(\partial_{\mu_1}\ldots\partial_{\mu_n}\Phi_i)}\Bigr|_0 (i k_{\mu_1})\ldots(i k_{\mu_n})=0.     
\end{equation}
Equations \eqref{master_singulant_eom} and \eqref{master_fluctuation_eom} are mapped into one another by 
\begin{equation}\label{master_map}
- \partial_\mu \chi_q \leftrightarrow i k_\mu, \quad \Phi_{i\,(0)}^{\{0\}} \leftrightarrow \Phi_{i,0}, \quad \alpha_i \leftrightarrow \alpha_{i,0}.     
\end{equation}
This is the map we referred to for the first time in Sec.~\ref{sec:singulants}. 

\section{Decomposing the large-order behaviour of the gradient expansion into channels}
\label{app:B}

In this Appendix, we discuss whether the large-order behaviour of $\Pi_{\mu\nu}$ allows for a channel decomposition in a general conformal fluid. For simplicity, we restrict our discussion to the case of MIS-like models where the equation of motion for $\Pi_{\mu\nu}$ is at most second order in spacetime derivatives. 

According to the analysis presented in Appendix \ref{app:A}, the building blocks we have at our disposal to construct the singulant equation of motion in this case are: 
\begin{itemize}
\item Tensors: $A^{\mu\nu}$. 
\item Vectors: $k_{\perp}^\mu \equiv  \Delta^{\mu\nu}\partial_\nu\chi$. 
\item Scalars: $T$, $U(\chi)$, $k_\perp \cdot k_\perp$. \end{itemize}
Note that each building block is a field in spacetime. The most general singulant equation of motion one can write with these ingredients is 
\begin{align}\label{chi_eom_ansatz}
&c(T, U(\chi), k_\perp \cdot k_\perp) A^{\mu\nu} {+}f\left[\frac{1}{2}\left(A^{\mu\alpha}k_{\perp\alpha} k_\perp^{\nu}{+} \mu \leftrightarrow \nu\right)\right. \nonumber \\
&\left.{-}\frac{1}{d{-}1} \Delta^{\mu\nu} (A^{\alpha\beta} k_{\perp,\alpha} k_{\perp,\beta}) \right]{=} 0,   
\end{align}
where, with no loss of generality, we have taken the coefficient $f$ to be a positive real number, and  
\begin{align}
&c(T, U(\chi), k_\perp \cdot k_\perp) = c_{0,0}T^2 + c_{1,0}T U(\chi) \nonumber \\ 
&+ c_{2,0} U(\chi)^2 + c_{0,2}(k_\perp \cdot k_\perp).     
\end{align}
The coefficients $c_{i,j}$ and $f$ define the MIS-like theory being considered; for instance, in the models explored in this work, $f=0$. On the other hand, the $T$-dependent functions multiplying $c_{i,j}$ are fixed by dimensional analysis due to conformal invariance. 

Let us define $d{-}2$ vector fields $Z_i$ such that 
\begin{equation}
Z_i \cdot U = Z_i \cdot k_\perp = 0, \quad  Z_i \cdot Z_j = \delta_{ij}.    
\end{equation}
At a given spacetime point, we decompose $A^{\mu\nu}$ as 
\begin{equation}\label{A_decomposition}
A^{\mu\nu} = A_{TT}^{\mu\nu} + A_{TL}^{\mu\nu} + A_{LL}^{\mu\nu}, \end{equation}
with
\begin{subequations}
\begin{equation}
A_{LL}^{\mu\nu} = a_{LL}\left(k_\perp^\mu k_\perp^\nu -\frac{1}{d-1}\Delta^{\mu\nu} k_\perp^2 \right),     
\end{equation}
\begin{equation}
A_{TL}^{\mu\nu} = \sum_{i=1}^{d{-}2}a^{(i)}_{TL} \frac{k_\perp^{\mu}Z_i^{\nu} + k_\perp^{\nu}Z_i^{\mu}}{2}, 
\end{equation}
\begin{equation}
A_{TT}^{\mu\nu} = \sum_{j \leq i}a_{TT}^{(i,j)} Z_i^\mu Z_j^\nu,   
\end{equation}
\end{subequations}
where the matrix $a_{TT}^{(i,j)}$ is symmetric in $i,j$ and has vanishing trace. Note that $A_{LL\mu}^{\mu} = A_{TL\mu}^{\mu} = A_{TT\mu}^{\mu}=0$. Our decomposition parameterises the $d(d-1)/2{-}1$ independent degrees of freedom that in principle comprise $A^{\mu\nu}$ as $\alpha_{LL}$, $\alpha_{TL}^{(i)}$, and $\alpha_{TT}^{(i,j)}$. 

If we introduce the decomposition 
\eqref{A_decomposition} into the ansatz for the singulant equation of motion \eqref{chi_eom_ansatz}, we get that 
\begin{equation} 
c A_{TT}^{\mu\nu} {+} \left(c {+} \frac{1}{2}f k_\perp^2\right)A_{TL}^{\mu\nu} {+} \left(c {+}\frac{d{-}2}{d{-}1}f k_\perp^2\right)A_{LL}^{\mu\nu} {=} 0.   \end{equation}
At the spacetime point under consideration, it seems consistent to have a large-order behaviour of the form\footnote{We omit the sum over independent singulant contributions within each channel.}
\begin{equation}\label{Pi_decomposition}
\Pi^{\mu\nu} \sim \sum_{c=LL,LT,TT} A_c^{\mu\nu}\frac{\Gamma(n + \alpha_c)}{\chi_c^{n+\alpha_c}}, 
\end{equation}
provided that 
\begin{subequations}
\begin{equation}
c(T, U^\mu \partial_\mu \chi_{TT}, \Delta^{\mu\nu} \partial_\mu \chi_{TT} \partial_\nu \chi_{TT}) = 0,     
\end{equation}
\begin{align}
&c(T, U^\mu \partial_\mu \chi_{TL}, \Delta^{\mu\nu} \partial_\mu \chi_{TL} \partial_\nu \chi_{TL}) \nonumber \\
&+ \frac{f}{2} \Delta^{\mu\nu} \partial_\mu \chi_{TL} \partial_\nu \chi_{TL}  = 0,     
\end{align}
\begin{align}
&c(T, U^\mu \partial_\mu \chi_{LL}, \Delta^{\mu\nu} \partial_\mu \chi_{LL} \partial_\nu \chi_{LL}) \nonumber \\
&+ \frac{(d{-}2)f}{d{-}1} \Delta^{\mu\nu} \partial_\mu \chi_{LL} \partial_\nu \chi_{LL}  = 0.     
\end{align}
\end{subequations}
The nontrivial question is whether the decomposition \eqref{Pi_decomposition}, which we argued is valid at a given spacetime point, is self-consistent across the whole spacetime. This issue has to be decided by the constraints placed upon $A^{\mu\nu}$ by the subleading terms in the large $n$ expansion of the recursion relations. Hopefully, our $A^{\mu\nu}$ decomposition will be compatible with these constraints, provided that $\alpha_{LL}$, $\alpha_{TL}^{(i)}$, and $\alpha_{TT}^{(i,j)}$ obey a particular equation of motion that allows us to fix them. 

There seems to be a natural map to linear response theory, where the LL channel corresponds to sound and the TL channel to shear; in this case, it should be the case that $A_{TT} = 0$, since hydrodynamics does not capture the tensor channel. 

\section{Computation of $\gamma_s$ in holography}\label{app:C}

\subsection{Gravity dual of a sound wave}

In this Appendix, we explain how to compute $\gamma_s$ in holography by adapting the approach of Refs.~\cite{Bu:2014ena,Bu:2014sia}. Our goal is constructing the five-dimensional geometry dual to a sound wave of infinitesimal amplitude propagating on a background thermal state. Since this sound wave is a longitudinal flow, we take the metric ansatz \eqref{hol:metric_ansatz} and write 
\begin{subequations}
\begin{equation}
U_\mu dx^\mu = -dt + \lambda\,\delta u\,dx,     
\end{equation}
\begin{equation}
\mathcal{V}_\mu dx^\mu = \left(-\frac{1}{2}f(r) + \lambda\,\delta V_t\right) dt + \lambda\,\delta V_x dx,   
\end{equation}
\begin{equation}
\begin{split}
&\mathcal{G}_{\mu\nu}dx^\mu dx^\nu = r^2 (dx^2 {+} d\vec{x}_\perp^2)\\
&+\lambda r^2 \left[2\left({-}\delta B{+}\frac{\delta \Sigma}{r} \right)dx^2{+}\left(\delta B{+}2\frac{\delta \Sigma}{r} \right) d\vec{x}_\perp^2\right],    
\end{split}
\end{equation}
\end{subequations}
where $f = r^2 - \mu^4 r^{-2}$ is the blackening factor of the background black brane. We parameterise the longitudinal plane with Minkowski coordinates $t$ and $x$. The perturbations $\delta V_t$, $\delta V_x$, $\delta \Sigma$, and $\delta B$ are functions of $r$, $t$, and $x$, while $\lambda$ is a bookkeeping parameter that reminds us that these perturbations are infinitesimally small. 

Due to the translational invariance of the background black brane, it is convenient to work in momentum space. Hence, we take 
\begin{equation}
\delta V_t(r,t,x) = \delta\hat{V}_t(r,\omega, k) e^{- i \omega t + i k x},    
\end{equation}
and similarly for $\delta V_x$, $\delta \Sigma$, and $\delta B$. 

With our choices, the Einstein equations read 
\begin{equation}
E_{\mu\nu} = E_{\mu\nu}^{(0)} + \lambda e^{-i\omega t + i k x} \delta\hat{E}_{\mu\nu}, \end{equation}
at leading order in $\lambda$. As dynamical equations, we take the following linear combinations of $\delta\hat{E}_{\mu\nu}$, 
\begin{subequations}
\begin{equation}
D_1 = \delta\hat{E}_{rr},     
\end{equation}
\begin{equation}
D_2 = 2 \delta\hat{E}_{rx},     
\end{equation}
\begin{equation}
D_3 = \frac{2}{3r^2f}(\delta\hat{E}_{xx}-\delta\hat{E}_{yy}), 
\end{equation}
\begin{equation}
D_4 = - \frac{1}{3r^2}(\delta\hat{E}_{xx}+2\delta\hat{E}_{yy}+\delta\hat{E}_{rt}).  
\end{equation}
\end{subequations}
$D_1$ is simply 
\begin{equation}
\partial_r^2\delta\hat{\Sigma} = 0,     
\end{equation}
implying that 
\begin{equation}
\delta\hat\Sigma = A_1 r + A_2. 
\end{equation}
The requirement that the boundary metric is the Minkowski metric entails that $A_1=0$. Furthermore, it is always possible to shift $A_2$ by the coordinate change $r \to r + \lambda \Psi e^{-i\omega t+ i k x}$. Therefore, $A_2$ can always be set to zero and, as a consequence, we take $\delta\hat{\Sigma} = 0$ with no loss of generality. Once the latter condition is imposed, the remaining dynamical equations read 
\begin{subequations}
\begin{equation}\label{hol:linear_D2}  
\delta\hat{V}''_x+\frac{\delta\hat{V}'_x}{r}-\frac{4 \delta\hat{V}_x}{r^2} - 2 i k \delta\hat{B}' - \frac{3 i \omega \delta\hat{u}}{r} = 0,   
\end{equation}
\begin{equation}\label{hol:linear_D3}
\begin{split}
&\delta\hat{B}''{+}\frac{f{+}4r^2-2i\omega r}{r f}\delta\hat{B}'{+} \frac{k^2{-}9 i \omega r}{3 r^2 f}\delta\hat{B} \\
&{+}\frac{2 i k}{3 r^2 f}\delta\hat{V}_x'{+}\frac{2 i k}{3 r^3 f}\delta\hat{V}_x'{+}\frac{f{-}4r^2+2 i \omega r}{3r^3f} i k\delta\hat{u}=0,    
\end{split}
\end{equation}
\begin{equation}
\begin{split}
&\delta\hat{V}_t''{+}\frac{1{+}4r^2}{r^3}\delta\hat{V}_t'{+} \frac{2(1{+}r^2)}{r^4}\delta\hat{V}_t{+}\frac{ik(1{+}4r^2)}{6 r^4}\delta\hat{V}_x'\\
&{+}\frac{2 i k (1{+}r^2)}{3 r^5}\delta\hat{V}_x{+}\frac{k^2(1{+}r^2)}{3 r^4}\delta\hat{B}\\
&{-} \frac{r(1{+}4r^2)(2r{-}i\omega){+}(1{-}2r^2)f}{6 r^5} i k \delta\hat{u} = 0. 
\end{split}    
\end{equation}
\end{subequations}
Note that $\delta\hat{V}_t$ decouples from $\delta\hat{V}_x$ and $\delta\hat{B}$. According to the dynamical equations, the asymptotic series expansion of the metric is 
\begin{subequations}
\begin{equation}
\delta\hat{V}_t = \frac{1}{3} i k \delta\hat{u}r - \frac{\omega k \delta\hat{u}}{3} + \delta\hat{V}_{t,2}r^{-2}+\ldots,    
\end{equation}
\begin{equation}\label{hol:linear_ase-Vx}
\delta\hat{V}_x = -\frac{1}{2}\delta\hat{u}r^2 - i \omega \delta\hat{u} r + \frac{k^2\delta\hat{u}}{3} + \delta\tilde{V}_{x,2}r^{-2} +\ldots,    
\end{equation}
\begin{equation}\label{hol:linear_ase-B} 
\delta\hat{B} = - \frac{2}{3} i k \delta\hat{u}r^{-1} + \delta\hat{B}_4 r^{-4}+\ldots    
\end{equation}
\end{subequations}
where we have chosen the boundary conditions at $r =\infty$ in such a way that the bulk spacetime is asymptotically AdS. From these asymptotic series expansions, holographic renormalisation gives a dual energy-momentum tensor of the form  
$T_{\mu\nu} = T_{\mu\nu}^{(0)} + \lambda \delta \hat{T}_{\mu\nu} e^{-i\omega t+i k x}$, with 
\begin{equation}
T_{\mu\nu}^{(0)} = \textrm{diag}\left(\frac{3}{4}, \frac{1}{4},\frac{1}{4},\frac{1}{4}\right)\mu^4,     
\end{equation}
and 
\begin{subequations}
\begin{align}
&\delta\hat{T}_{tt} = \frac{3}{2} \delta\hat{V}_{t,2}, \\
&\delta\hat{T}_{tx} = \delta\hat{V}_{x,2}-\frac{1}{2}\mu^4 \delta\hat{u}, \\     
&\delta\hat{T}_{xx} = -2\delta\hat{B}_4 + \frac{1}{2}\delta\hat{V}_{t,2}, \\ 
&\delta\hat{T}_{yy}=\delta\hat{T}_{zz} = \delta\hat{B}_4 + \frac{1}{2}\delta\hat{V}_{t,2},     
\end{align}
\end{subequations}
with the remaining components vanishing. The Landau frame condition demands that, up to $O(\lambda)$, the fluid velocity $U^\mu = (1, \lambda \delta\hat{u}e^{-i\omega t + i k x}, 0, 0)$ is a timelike eigenvector of $T^{\mu}_\nu$. This condition can only be satisfied provided that 
\begin{equation}\label{hol:linear_Landau}
\delta\hat{V}_{x,2} = - \frac{1}{2}\mu^4 \delta\hat{u}.     
\end{equation}
Once that the relation above has been imposed, we can finally identify 
\begin{equation}\label{hol:Pi_*_B_relation}
\mathcal{E}_0 = \frac{3}{4}\mu^4, \quad \delta\hat{\mathcal{E}} = \frac{3}{2}\delta\hat{V}_{t,2}, \quad \delta\hat{\Pi}_\star = \delta\hat{B}_4. 
\end{equation}
Recalling Eq.~\eqref{gamma_s_def}, we observe that $\gamma_s$ can be obtained directly from the near-boundary series expansion of $\delta\hat{B}$. 

The poles of $\gamma_s$ play a central role in the analysis presented in Sec.~\ref{sec:holography}. To compute them, we plug the ansatz 
\begin{equation}\label{hol:linear_pole-ansatz}
\delta\hat{V}_x = \frac{P}{\omega - \Omega_p(k)}, \quad \delta\hat{B} = \frac{Q}{\omega - \Omega_p(k)}, 
\end{equation}
into Eqs.~\eqref{hol:linear_D2} and \eqref{hol:linear_D3} and expand around $\omega = \Omega_p(k)$. The leading order result corresponds precisely to Eqs.~\eqref{hol:residue_eq_1} and \eqref{hol:residue_eq_2} in the main text. These equations, on the other hand, are nothing but  Eqs.~\eqref{hol:linear_D2} and \eqref{hol:linear_D3} with $\delta\hat{u} = 0$ after the identification $\Omega_p \to \omega$, $Q \to \delta\hat{B}$ and $P\to\delta\hat{V}_x$. This is physically consistent: at a pole of $\gamma_s$,  Eq.~\eqref{gamma_s_def} predicts that $\hat{\Pi}_\star$ can be finite even if $\delta\hat{u} = 0$. Taking into account the asymptotic series expansions \eqref{hol:linear_ase-Vx} and \eqref{hol:linear_ase-B} and the Landau frame condition \eqref{hol:linear_Landau}, one recovers the near-boundary behaviour quoted in the main text: $P = O(r^{-3})$ and $Q = O(r^{-4})$ as $r\to\infty$. 

\subsection{Numerical results}

To compute $\gamma_s$, we need to solve the coupled ODEs \eqref{hol:linear_D2} and \eqref{hol:linear_D3}. We proceed as follows. First, we take into account the $r\to\infty$ series expansions \eqref{hol:linear_ase-Vx} and \eqref{hol:linear_ase-B} and perform the field redefinitions, 
\begin{subequations}\label{gamma_s:field_redefinition}
\begin{equation}
\delta\hat{V}_x(r){=}{-}\frac{r^2}{2}\delta\hat{u}{-}i \omega r \delta\hat{u}{+}\frac{k^2}{3}\delta\hat{u}{-}\frac{\mu^4}{2 r^2}\delta\hat{u}{-}\frac{a(r)}{r^3} \delta\hat{u},    
\end{equation}
\begin{equation}
\delta\hat{B}(r) = - \frac{2 i k}{3r}\delta\hat{u} + \frac{i k \mu^4 b(r)}{2 r^4} \delta\hat{u},     
\end{equation}
\end{subequations}
where the Landau frame condition \eqref{hol:linear_Landau} has been imposed. Note that, according to Eq.~\eqref{hol:Pi_*_B_relation} and the definition \eqref{gamma_s_def}, 
\begin{equation}\label{gamma_s:gamma_s_b_relation}
\gamma_s = b(r = \infty).     
\end{equation}
The field redefinition \eqref{gamma_s:field_redefinition} transforms equations \eqref{hol:linear_D2} and \eqref{hol:linear_D3} into 
\begin{subequations}\label{gamma_S:final_equations}
\begin{equation}
u^2 a'' + 7 u a' + 5 a + k^2 u b' + 4 k^2 b = 0,    
\end{equation}
\begin{equation}
\begin{split}
&3 u (1-u^4) b'' + (15-27 u^4 + 6 i \omega  u)b' \\
&+ (k^2 u - 48 u^3 + 15 i \omega) b + 4u^2 a' + 8 u a = -4, 
\end{split}
\end{equation}
\end{subequations}
where we have set $\mu = 1$ and traded the radial coordinate $r$ for  
\begin{equation}
u = \frac{1}{r}.     
\end{equation}
In Eqs.\,\eqref{gamma_S:final_equations}, the primes denote derivatives with respect to $u$. The reader should keep in mind that, due to the choice $\mu = 1$, $\omega$ and $k$ are measured in units of $\pi T$. 

To solve \eqref{gamma_S:final_equations}, we resort to pseudospectral methods. We introduce a Chebyshev–Gauss–Lobatto grid in the variable $u$, transforming the two coupled ODEs \eqref{gamma_S:final_equations} into a single algebraic equation. There are no boundary conditions to be imposed on $a(u)$ and $b(u)$ on the discretisation grid, neither at the asymptotic boundary, located at $u=0$, nor at the black hole horizon, located at $u=1$. The reason is that the correct $u \to 0$ behaviour has been already taken into account by the field redefinitions \eqref{gamma_s:field_redefinition}, while the requirement of $a(u)$ and $b(u)$ to be regular at $u = 1$ is automatically incorporated by our choice of spectral decomposition. 

\begin{figure}[h!]
\begin{center}
\includegraphics[width=0.75\linewidth]{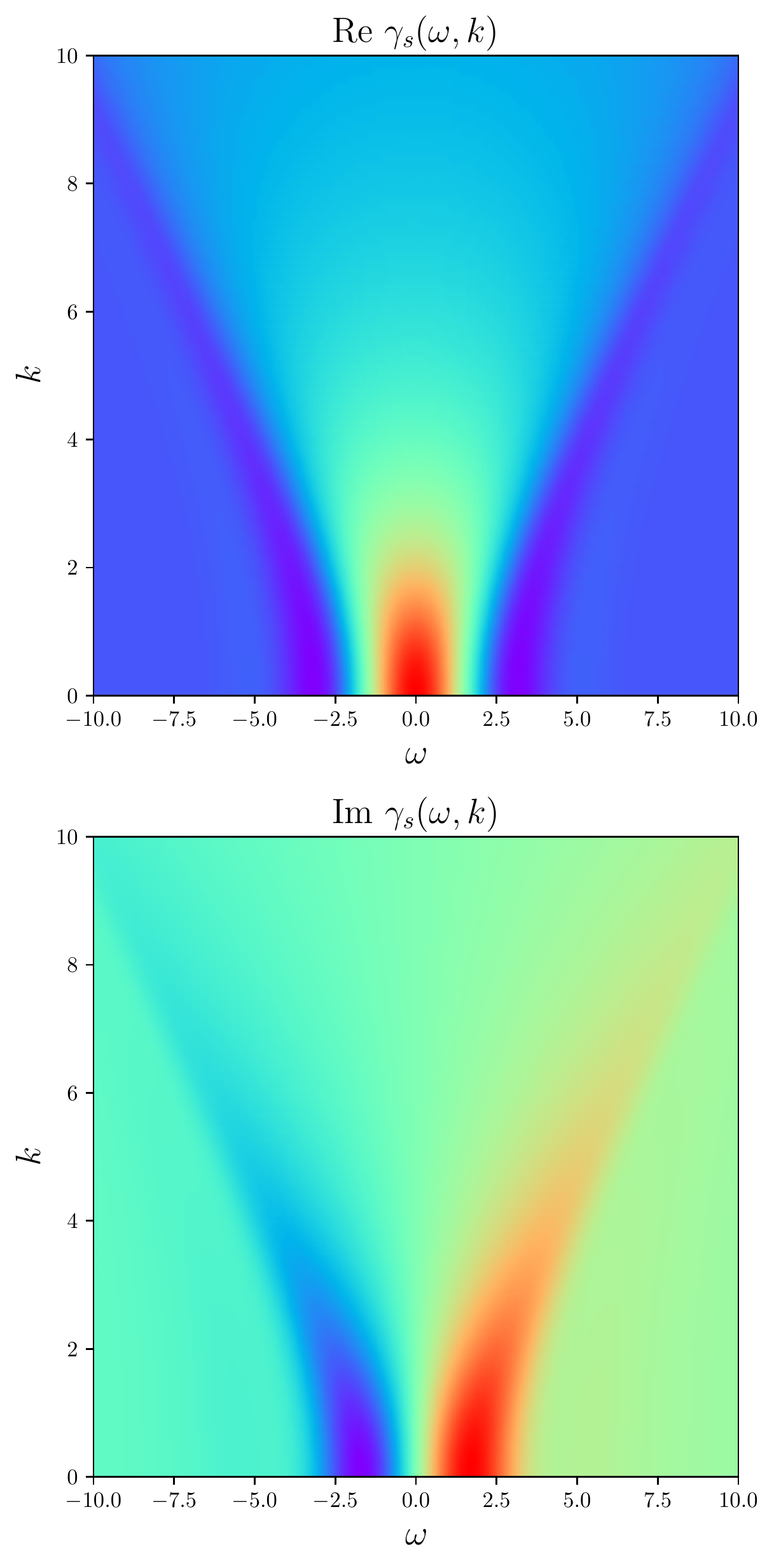}
\caption{Real (upper panel) and imaginary (bottom panel) parts of $\gamma_s$, for $\omega \in [10,10]$ and $k \in [0,10]$. $\omega$ and $k$ are measured in units of $\pi T = 1$. These numerical results have been obtained from a collocation grid of 60 points, with $\omega$ and $k$ forming a square lattice of spacing $0.125$.}
\label{fig:gamma_s_1}
\end{center}
\end{figure}

Once the numerical solution corresponding to a given pair $(\omega, k)$ is known, we can read the corresponding $\gamma_s$ by employing the relation \eqref{gamma_s:gamma_s_b_relation}. In Fig.~\ref{fig:gamma_s_1}, we plot $\gamma_s$ for $\omega, k \in \mathbb R$. Due to the symmetry of equations \eqref{gamma_S:final_equations} under $k \to - k$, we restrict ourselves to $k \geq 0$. We see that, for a given $k$, the real part of $\gamma_s$ is even in $\omega$, while the imaginary part is odd. We refer the reader to Refs.~\cite{Bu:2014ena,Bu:2014sia} for a discussion of earlier numerical results on $\eta$ and $\xi$---linked to $\gamma_s$ by Eq.~\eqref{hol:gamma_s-BL_relation}---for real $\omega$ and $k$. 

The relation between the singulant trajectory and the poles of $\gamma_s$ we uncovered in Sec.~\ref{sec:holography} entails that our main interest does not lie in the behaviour of $\gamma_s$ for real wave-vectors, but on its singularities in the complex $\omega$-plane. 

\begin{figure}[h]
\begin{center}
\includegraphics[width=0.75\linewidth]{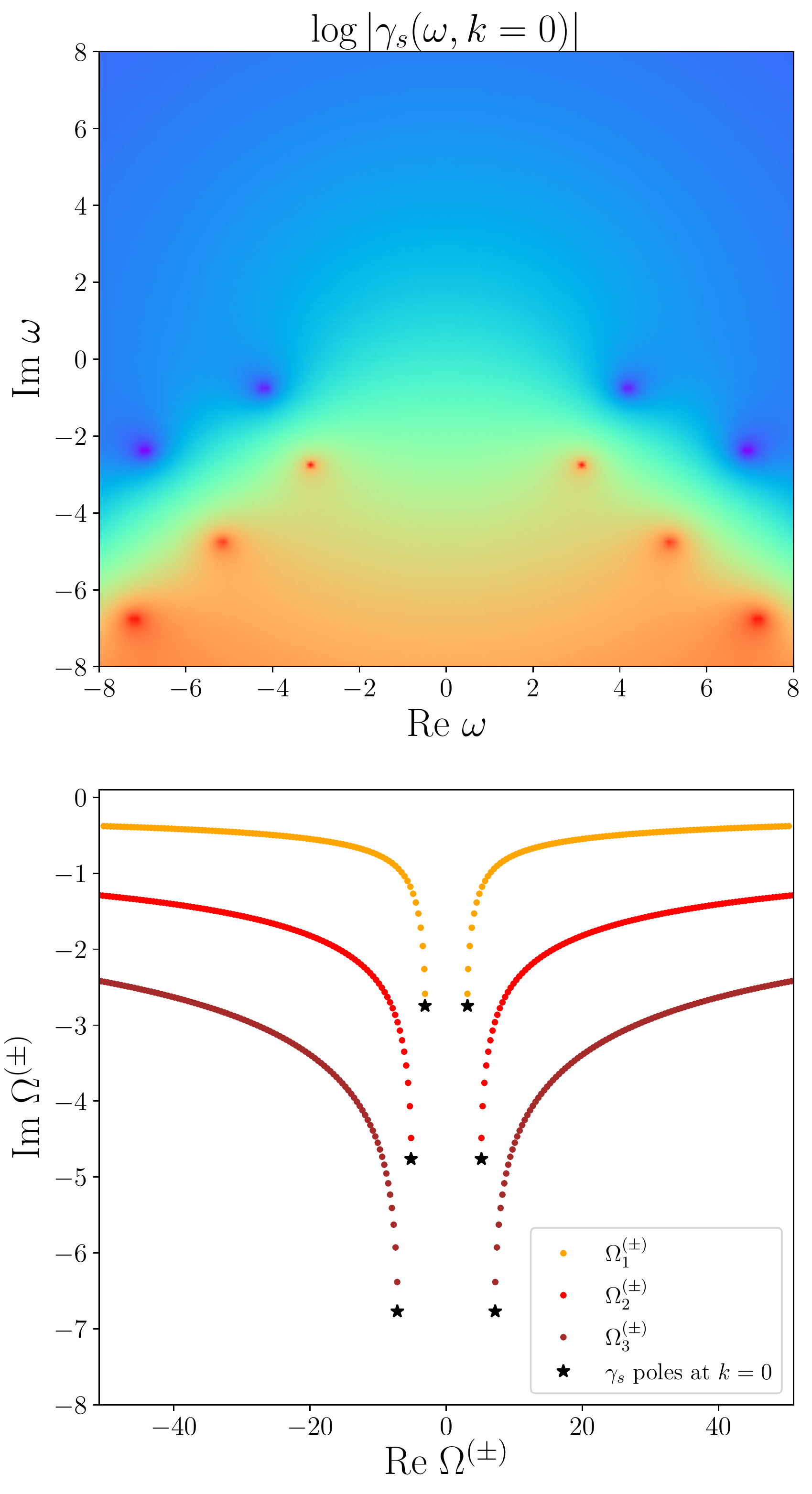}
\caption{\textbf{Upper panel}: $\log(|\gamma_s|)$ as a function of $\omega$ for $k=0$. The function $\gamma_s$ has been obtained using a collocation grid of $60$ points, with the real and imaginary parts of the frequency forming a square lattice of spacing $0.125$. \textbf{Bottom panel}: trajectories described by $\Omega_1^{(\pm)}$ (orange), $\Omega_2^{(\pm)}$ (red) and $\Omega_3^{(\pm)}$ (brown) in the complex $\omega$-plane as $k$ is increased from 0 to 50 in units of 0.5. The poles at $k=0$ are represented by the black stars. We have employed a collocation grid of 80 points in order to get these results.}
\label{fig:gamma_s_3}
\end{center}
\end{figure}

We start by demonstrating that $\gamma_s$ is not an entire function of $\omega$ at fixed $k$. To this aim, in the top panel of Fig.~\ref{fig:gamma_s_3} we represent the (logarithm of) the norm of $\gamma_s$ as a function of $\omega \in \mathbb C$ for $k=0$. Our results show that $\gamma_s$, while being analytic in the upper-half complex $\omega$-plane, has poles in the lower one. At zero spatial momentum, these poles coincide with the nonhydrodynamic sound channel quasinormal modes.

The poles $\Omega_p^{(\pm)}$ describe trajectories in the complex $\omega$-plane as $k$ varies. While, for a given $k$, $\Omega_p^{(\pm)}$ could be found from Eqs.~\eqref{gamma_S:final_equations} by means of a binary search algorithm that minimises $-\log(|\gamma_s|)$, it is more efficient to compute them through Eqs.~\eqref{hol:residue_eq_1} and \eqref{hol:residue_eq_2}. To solve these equations numerically, we perform the field redefinitions  
\begin{equation}
P(r) = \frac{a(r)}{r^3}, \quad Q(r) = \frac{\mu^4}{2 r^4}i k b(r),    
\end{equation}
which ultimately transform them into Eqs.~\eqref{gamma_S:final_equations} with vanishing source terms on the right-hand-side. The pseudospectral discretisation proceeds in the same way as described before, resulting in a generalised algebraic eigenvalue problem whose solution returns $\Omega_p^{(\pm)}$. In the bottom panel of Fig.~\ref{fig:gamma_s_3}, we plot the trajectories followed by the three lowest-lying poles in the complex $\omega$-plane. At large $k$, we always find that $\textrm{Re } \Omega_p^{(\pm)} \to \pm k$, while $\textrm{Im } \Omega_p^{(\pm)}$ approaches zero, in such a way that $\Omega_p^{(\pm)}$ approach the real $\omega$ axis from below.  

\section{Mapping initial data to singulants: an explicit example}\label{app:D}

In this Appendix, we illustrate how the singulant field can be completely determined in the linear response regime. For definiteness, we focus on conformal BRSSS theory, which has been discussed at the fully nonlinear level in Sec.\,\ref{sec:phenomenological_models}. 

To access the linear response regime, we place ourselves in the fluid rest frame, write 
\begin{subequations}
\begin{equation}
T(t,x) = T_0 + \lambda\,\delta T(t,x),    
\end{equation}
\begin{equation}
u(t,x) = \lambda\, \delta u(t,x),   
\end{equation}
\begin{equation}
\Pi_\star(t,x) = \lambda\,\delta \Pi_\star(t,x),      
\end{equation}
\end{subequations}
and take the $\lambda \to 0$ limit. At leading order in $\lambda$, the recursion relations simplify to 
\begin{equation}
\delta \Pi_\star^{(1)}{=} \frac{2}{3}\eta(T_0) \partial_x \delta u,\, \delta \Pi_\star^{(n+1)} {=}{-}\tau_\Pi(T_0) \partial_t \delta \Pi_\star^{(n)}.     
\end{equation}
The equations of motion for BRSSS theory in the linear response regime can be solved in closed-form by working in momentum space. One finds that 
\begin{equation}
\delta u(t,x) = \int_{\mathbb R} dk\, e^{ikx} \sum_{q=+,-,\textrm{NH}} \delta u_q(k) e^{- i \omega_q(k) t} \label{u_linear_response_sol}
\end{equation}
where $\pm$ and $\textrm{NH}$ refer respectively to the contributions of the two hydrodynamic modes and the nonhydrodynamic one. $\delta u_{+}(k)$, $\delta u_{-}(k)$ and $\delta u_\textrm{NH}(k)$ can be expressed in terms of the Fourier transforms of the initial data, $\delta T(0,x)$, $\delta u(0,x)$ and $\delta\Pi_\star(0,x)$. Expressions analogous to \eqref{u_linear_response_sol} hold for $\delta T$ and $\delta \Pi_\star$ as well. 

Given the form of the linearised recursion relations, it is immediate to see that contributions of individual modes decouple. Focusing on the $q$th contribution one finds that, in Fourier space,
\begin{subequations}
\begin{equation}
\delta\Pi_\star^{(1)}(k)=\frac{2}{3}\eta(T_0)ik\delta u_q(k),  
\end{equation}
\begin{equation}
\delta\Pi_\star^{(n+1)}(k)=i \tau_\Pi(T_0) \omega_q(k) \delta\Pi_\star^{(n)}(k).     
\end{equation}
\end{subequations}
The recursion relations above are immediately solved by 
\begin{equation}\label{recursion_relations_solution_Fourier}  
\delta\Pi_\star^{(n)}(k) = \frac{2}{3}\eta(T_0) i^{n} \tau_\Pi(T_0)^{n-1}\omega_q(k)^{n-1} k \delta u_q(k).  
\end{equation}
Let us assume that the analytical continuation of the Borel transform of the gradient expansion,
\begin{equation}
\delta\Pi_\star^{(B)}(t,x;\epsilon) = \sum_{n=1}^\infty \frac{\delta\Pi_\star^{(n)}(t,x)}{n!}\epsilon^n,     
\end{equation}
has a well-defined Fourier transform, $\delta\Pi_\star^{(B)}(t,k;\epsilon)$. Then, we can employ \eqref{recursion_relations_solution_Fourier} to compute the contribution of the $q$th mode to $\delta\Pi_\star^{(B)}(t,k;\epsilon)$, $\delta\Pi_{\star,q}^{(B)}(t,k;\epsilon)$, with the final result that
\begin{equation}\label{Borel_Fourier_component}
\delta\Pi_{\star,q}^{(B)}(t,k;\epsilon){}=\frac{2\eta(T_0) (e^{i \epsilon \tau_\Pi(T_0) \omega_q(k)}{-}1)}{3 \tau_\Pi(T_0) \omega_q(k)} k \delta u_q(k).
\end{equation}
Our main working hypothesis is that, when the Fourier integral  
\begin{equation}
\int_{\mathbb R} dk\, e^{ikx} \sum_{q=+,-,\textrm{NH}} \delta\Pi_{\star,q}^{(B)}(t,k;\epsilon) e^{-i\omega_q(k) t} \label{Borel_Fourier}
\end{equation}
ceases to exist for a particular $\epsilon$, the original analytically continued Borel transform $\delta\Pi_\star^{(B)}(t,x;\epsilon)$ becomes singular. Hence, it is natural to identify the $\epsilon$'s at which this phenomenon takes place with the singulants controlling the large-order behaviour of the gradient expansion.  
\begin{figure}[h]
\begin{center}
\includegraphics[width=\linewidth]{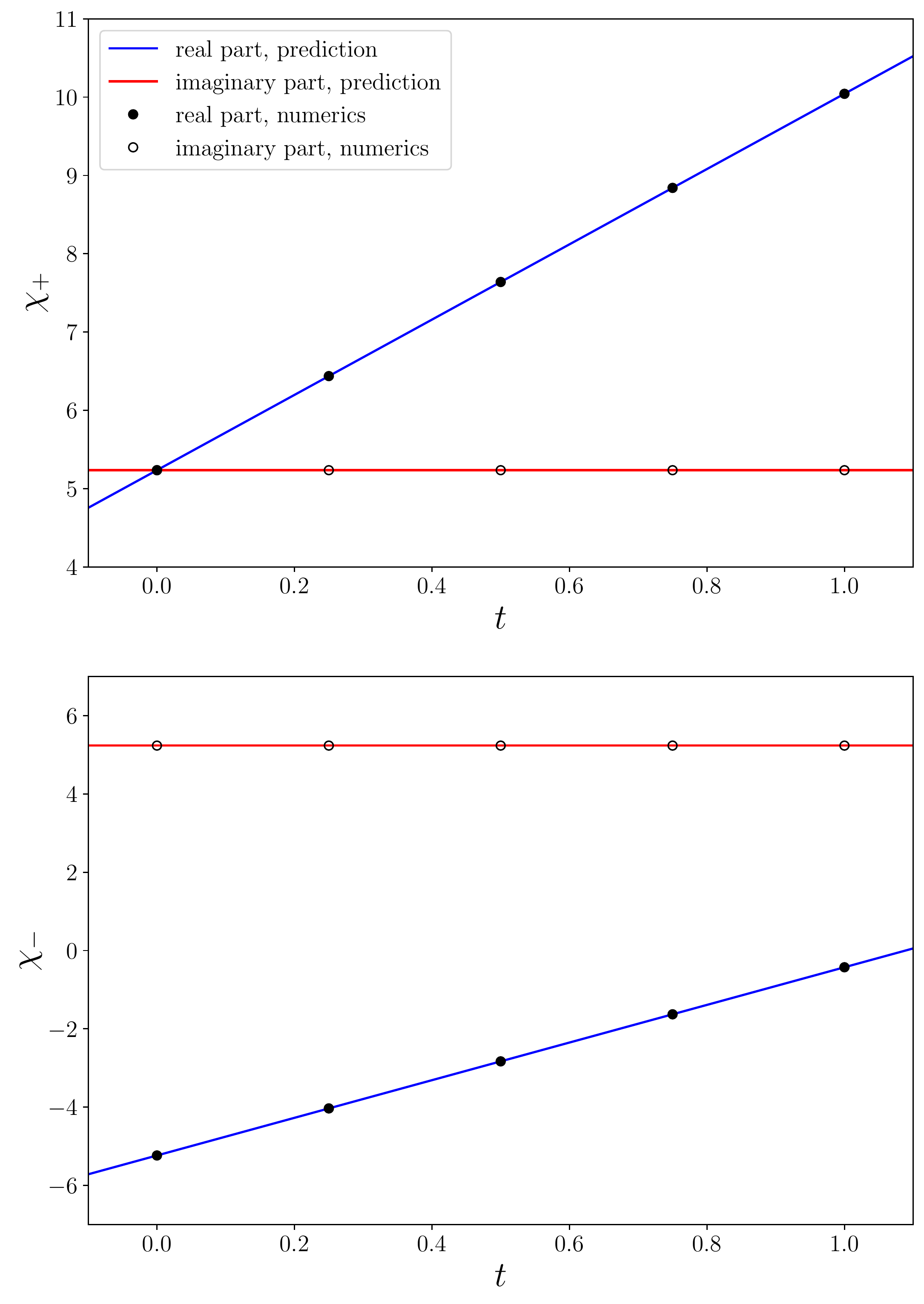}
\caption{For $\alpha = 1$, $\frac{\eta}{s} = \frac{1}{4\pi}$ and $\tau_{\Pi,0} = \frac{2-\log{2}}{2\pi}$, comparison between the analytic prediction \eqref{singulants_LID_linear_response} and a numerical computation of the singulants $\chi_+$ (upper panel) and $\chi_-$ (bottom panel). Filled (open) circles and blue (red) curves correspond, respectively, to the real (imaginary) part of the singulants computed numerically and the prediction \eqref{singulants_LID_linear_response}. }
\label{fig:singulant_comparison}
\end{center}
\end{figure}

In the light of expressions \eqref{Borel_Fourier_component} and \eqref{Borel_Fourier}, the convergence of the Fourier integral \eqref{Borel_Fourier} depends on two inputs: 
\begin{itemize}
\item The large-$k$ behaviour of the initial data, which determines the large-$k$ behaviour of $\delta u_q(k)$. 
\item The large-$k$ behaviour of the mode frequencies. 
\end{itemize}
Regarding the latter, it is straightforward to show that, as $k \to \infty$,
\begin{subequations}\label{BRSSS_w_large-k}
\begin{equation}\omega_+(k) = c_{UV} k + \ldots,\end{equation}
\begin{equation}\omega_-(k) = - c_{UV} k + \ldots,\end{equation} 
\begin{equation}\omega_\textrm{NH}(k) = - i \frac{T_0}{4 \frac{\eta}{s} + \tau_{\Pi,0}} + \ldots,\end{equation}
\end{subequations}
with asymptotic group velocity 
\begin{equation}
c_{UV} = \sqrt{\frac{4 \frac{\eta}{s} + \tau_{\Pi,0}}{3 \tau_{\Pi,0}}}.     
\end{equation}
Hence, for initial data which decay faster than any power-law as $k\to\infty$, any divergence of the Fourier integral \eqref{Borel_Fourier} has to be traced back to the contributions of the hydrodynamic modes, $\omega_\pm(k)$.  

For definiteness, let us consider initial data such that
\begin{equation}
\delta u(0,x) = \delta \Pi_\star(0,x) = 0    
\end{equation}
It follows that
\small
\begin{align}
&\delta u_+(k){=}\frac{(i T_0{+}\tau_{\Pi,0} \omega_+(k))k \delta T(0,k)}{T_0  \tau_{\Pi,0} (\omega_+(k){-} \omega_-(k))(\omega_+(k){-}\omega_\textrm{NH}(k))}, \\
&\delta u_-(k){=}\frac{(i T_0{+}\tau_{\Pi,0} \omega_-(k))k \delta T(0,k)}{T_0 \tau_{\Pi,0} (\omega_-(k){-} \omega_+(k))(\omega_-(k){-}\omega_\textrm{NH}(k))}, \\
&\delta u_\textrm{NH}(k){=}\frac{(i T_0{+}\tau_{\Pi,0} \omega_\textrm{NH}(k))k \delta T(0,k)}{T_0 \tau_{\Pi,0} (\omega_\textrm{NH}(k){-} \omega_+(k))(\omega_\textrm{NH}(k){-}\omega_{-}(k))},
\end{align}
\normalsize
where $\delta T(0,k)$ is the Fourier transform of the temperature perturbation at $t=0$. 

We will take a Lorentzian as initial data for $\delta T$, 
\begin{equation}\label{LID}
\delta T(0,x) = \frac{\alpha}{\pi (x^2 + \alpha^2)}, \quad \delta T(0,k) = \frac{1}{2\pi} e^{-\alpha |k|}.     
\end{equation}
The reason is that, for these initial data, a numerical computation shows that the gradient expansion in the linear response regime diverges as $n!$. In the light of \eqref{Borel_Fourier_component}, the Fourier integral \eqref{Borel_Fourier} ceases to be well-defined whenever the argument of 
\begin{equation}
\exp(i (\epsilon \tau_\Pi(T_0) - t) \omega_q(k) + i k x - \alpha |k|)    
\end{equation}
vanishes as $|k| \to \infty$ along the real axis. Recalling the large-$k$ behaviour of the hydrodynamic mode frequencies given in Eq.\,\eqref{BRSSS_w_large-k}, we find that this vanishing happens precisely when $\epsilon = \chi_+, \chi_-, \chi_+^\star, \chi_-^\star$, with 
\begin{equation}\label{singulants_LID_linear_response}
\chi_\pm = \frac{T_0 t}{\tau_{\Pi,0}} \pm \frac{T_0 x}{\tau_{\Pi,0} c_{UV}}+ i \frac{\alpha T_0}{\tau_{\Pi,0} c_{UV}}. 
\end{equation}
Expression \eqref{singulants_LID_linear_response} teaches us two important lessons: 
\begin{itemize}
\item The time-derivative of the singulants behaves as the factorial-over-power ansatz predicts, $\partial_t \chi_\pm = \frac{1}{\tau_\Pi(T_0)}$. 
\item The singulants at $t=0$ depend on $x$, the initial data (through $\alpha$) and the large-$k$ behaviour of the hydrodynamic mode frequencies (through $c_{UV}$). 
\end{itemize}
In Fig. \ref{fig:singulant_comparison}, we test the prediction \eqref{singulants_LID_linear_response} against the results of an explicit computation of the singulants, finding perfect agreement. We have computed the gradient expansion by evaluating numerically the inverse Fourier transform of $\delta\Pi_\star^{(n)}(k)$, and extracted the singulants from the poles of the Padé approximant of the Borel transform. 

We conclude by pointing out that the argument presented here can be straightforwardly generalised to find analytic expressions for the singulants associated with initial data of the form \eqref{LID} in the other MIS-like models we have considered in this work. We have checked that, for these models, these analytic expressions are also compatible with the results of the corresponding numerical computation. 

\section{Causality and stability of the new MIS-like model in the linear response regime}\label{app:E}

\noindent In this section, we discuss the causality and stability of the new MIS-like model put forward in Sec.~\ref{subsec:new_model}, at the level of linearised perturbations around global thermal equilibrium. We assume that our fluid is placed in four-dimensional Minkowski space, parameterised by coordinates $X^\mu = (t,x,y,z)$. 

We will work in a general reference frame, and respectively denote by $T_0$ and $U_0$ the temperature and unit-normalised fluid velocity of the background equilibrium state. Introducing a wave-vector 
\begin{equation}
k^\mu = (\omega, \textbf{k}) = (\omega, k_x, k_y, k_z),     
\end{equation}
the linearised perturbations we consider take the form 
\begin{subequations}\label{new_model:linearised_perturbations}
\begin{equation}
T = T_0 + \lambda~\delta T e^{i k \cdot X},     
\end{equation}
\begin{equation}
U = U_0 + \lambda~\delta U e^{i k \cdot X},     
\end{equation}
\begin{equation}
\Pi_{\mu\nu} = \lambda~\delta \Pi_{\mu\nu} e^{i k \cdot X},     
\end{equation}
\end{subequations}
with $\lambda \in \mathbb R^+$, $\lambda \ll 1$. At $O(\lambda)$, the conservation equation of the energy-momentum tensor and Eq.~\eqref{new_model_def}, when evaluated on infinitesimal plane-wave perturbations of the form \eqref{new_model:linearised_perturbations}, reduce to a linear system of algebraic equations. Rotational invariance demands that this system decomposes naturally into three different channels:
\begin{itemize}
\item[(i)] sound, $\delta T \neq 0$, $\delta U \parallel k_\perp$,   
\item[(ii)] shear, $\delta T = 0$, $\delta U \perp k_\perp$, 
\item[(iii)] tensor, $\delta T = \delta U = 0$, 
\end{itemize}
where $k_\perp$ is the part of the wave-vector $k$ orthogonal to the background fluid velocity $U_0$, 
\begin{equation}
k_\perp = k + (k \cdot U_0) U_0.      
\end{equation}
We discuss the dispersion relation associated to each channel below, and refer the reader to Ref.~\cite{Brito:2020nou} for a comprehensive discussion on how to perform the channel decomposition. For simplicity, we restrict ourselves to the case in which 
\begin{equation}
c_\mathcal{L} = 1, \quad c_\sigma = 0.  
\end{equation}
This agrees with the parameter choice performed in the main text. Furthermore, when performing explicit computations, we set 
\begin{equation}
U_0 = U_0^\mu \partial_\mu = \cosh(u_0) \partial_t + \sinh(u_0) \partial_x, \quad u_0 \in \mathbb{R}, 
\end{equation}
with no loss of generality. The local rest frame of the fluid corresponds to $u_0 = 0$. 

\subsection{Tensor channel}

\noindent In the tensor channel, the equations of motion allow for nontrivial solutions if and only if the wave-vector $k$ obeys the following dispersion relation, 
\begin{equation}\label{new_model:tensor_disp_rel_eq}
k^2 + 2 i \Omega_I T_0 (k \cdot U_0) + T_0^2 |\Omega|^2 = 0,    
\end{equation}
whose roots are 
\begin{equation}\label{new_model:tensor_disp_rel}
\begin{split}
&\omega_\pm{=}{\pm}\left(\textbf{k}^2{+}T_0^2\Omega_R^2{-}T_0^2 \Omega_I^2 \sinh(u_0)^2{+}2 i T_0 \Omega_I \sinh(u_0) k_x \right)^\frac{1}{2} \\
&{-}i T_0 \Omega_I \cosh(u_0).     
\end{split}
\end{equation}
These roots correspond to nonhydrodynamic modes. To establish whether these nonhydrodynamic modes behave in a way compatible with relativistic causality, we note that in the $|\textbf{k}|\to\infty$ limit with $k_x = \alpha |\textbf{k}|\to\infty$, $|\alpha| \leq 1$, one has that 
\begin{equation}\label{asymptotic_tensor}
\omega_\pm = \pm |\textbf{k}| - i T_0 \Omega_I (\cosh(u_0) \mp \alpha \sinh(u_0)) + \ldots,  
\end{equation}
in such a way that the group velocity is exactly one. The limit in which $|\textbf{k}| \to \infty$ with fixed $k_x$ is obtained from the expression above by setting $\alpha = 0$ and leads to identical conclusions. 

Since $\cosh(u_0) \pm \alpha \sinh(u_0) \geq 0$ for $|\alpha| \leq 1$, Eq.~\eqref{asymptotic_tensor} also shows that irrespectively of the background fluid velocity the tensor channel modes are asymptotically stable provided that $\Omega_I \geq 0$. 

We performed a numerical scan of $\textrm{Im}~\omega_\pm$ in the three-dimensional parameter space spanned by $|\textbf{k}|$, $\alpha$ and $u_0$---with the remaining parameters as in the numerical simulation discussed in the main text---and did not encounter any point where this quantity became positive. In particular, note that in the local rest frame one always has that $\textrm{Im}~\omega_\pm = - T_0 \Omega_I \leq 0$. Therefore, our stability results conform to the findings of Refs.~\cite{Bemfica:2020zjp,Brito:2020nou,Gavassino:2021owo} showing that stability is a Lorentz-invariant notion in a causal theory. 

\subsection{Shear channel}

\noindent Shear channel modes correspond to the roots of the following cubic polynomial in $k \cdot U_0$, 
\begin{equation}\label{new_model:shear_disp_rel_eq}
\begin{split}
&(k \cdot U_0)^3 - 2 i T_0 \Omega_I (k \cdot U_0)^2 - (k \cdot U_0) (k_\perp^2 + T_0^2 |\Omega|^2) \\
&+i \frac{\eta}{s}T_0 |\Omega|^2 k_\perp^2 = 0,      
\end{split}
\end{equation}
and are divided into one hydrodynamic mode and two nonhydrodynamic ones. 

In the $|\textbf{k}| \to \infty$ limit with $k_x = \alpha |\textbf{k}|$, $|\alpha| \leq 1$, these modes behave as 
\begin{subequations}\label{shear_modes_asymptotic}
\begin{equation}
\omega_1 = |\textbf{k}| - \frac{1}{2}i T_0 \left(2 \Omega_I - \frac{\eta}{s}|\Omega|^2\right) (\cosh(u_0) - \alpha \sinh(u_0))+ \ldots, 
\end{equation}
\begin{equation}
\omega_2 = {-}|\textbf{k}|{-}\frac{1}{2}i T_0 \left(2 \Omega_I{-} \frac{\eta}{s}|\Omega|^2\right) (\cosh(u_0){+}\alpha \sinh(u_0))+\ldots, 
\end{equation}
\begin{equation}
\omega_3 = \tanh(u_0) \alpha |\textbf{k}|{-}i T_0 \frac{\eta}{s} |\Omega|^2 \cosh(u_0)^{-1} + \ldots, 
\end{equation}
\end{subequations}
and, since $|\alpha|,~|\tanh(u_0)| \leq 1$, they are manifestly causal. Furthermore, provided that
\begin{equation}\label{shear_asymp_st}
\frac{\eta}{s} \geq 0, \quad 2 \Omega_I - \frac{\eta}{s}|\Omega|^2 \geq 0, 
\end{equation}
the three modes are asymptotically stable. These asymptotic stability conditions are the identical to the ones obtained in the local rest frame of the fluid, since $\cosh(u_0)\mp \alpha \sinh(u_0) \geq 0$ for $|\alpha| \leq 1$. Finally, we note that in the regime in which $|\textbf{k}| \to \infty$ with fixed $k_x$, the behavior of the shear channels modes is described by \eqref{shear_modes_asymptotic} with the replacements $\alpha \to 0,~ \alpha |\textbf{k}| \to k_x$. 

\begin{figure}[h]
\begin{center}
\includegraphics[width=\linewidth]{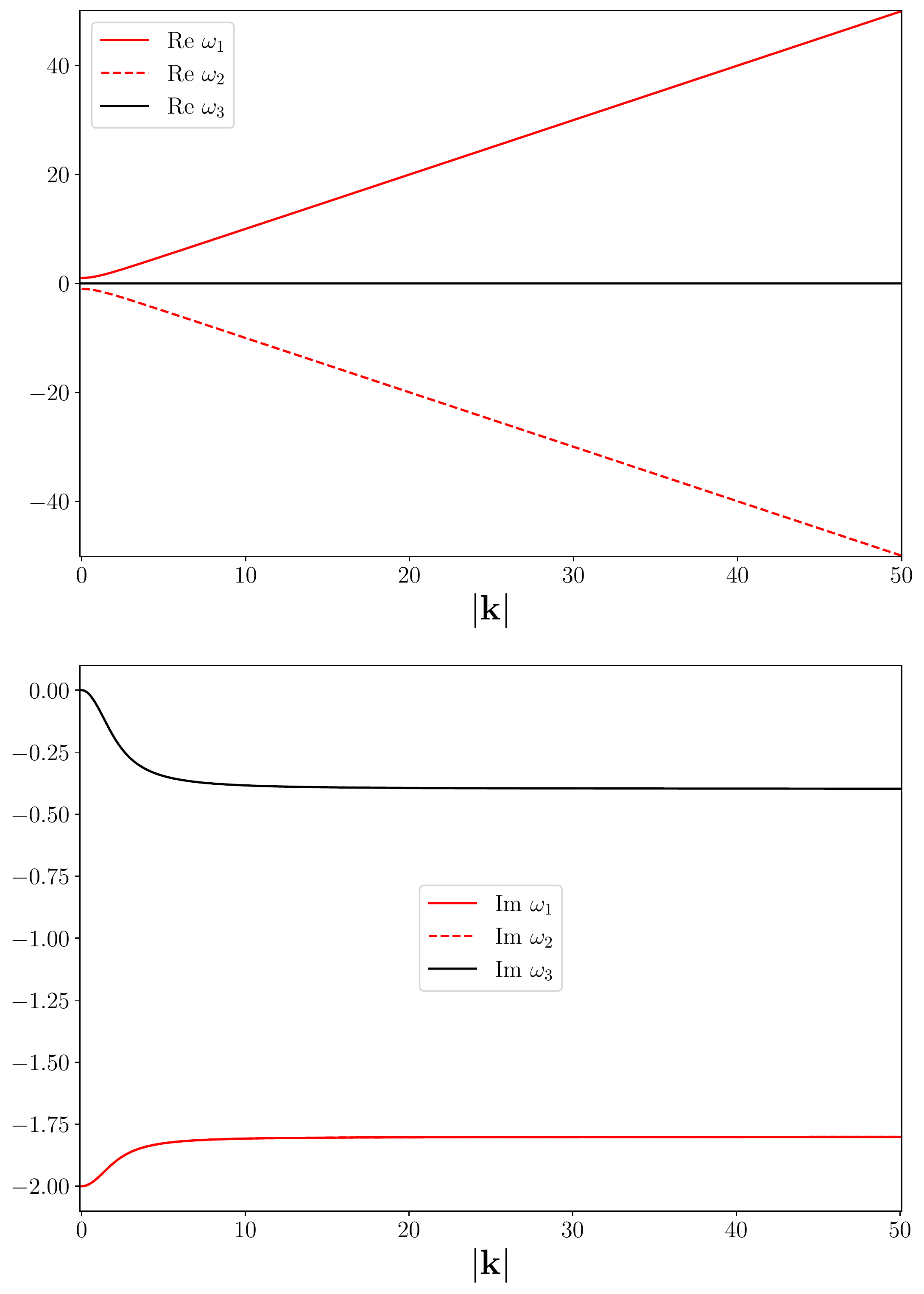}
\caption{Real (upper panel) and imaginary (bottom panel) parts of the shear channel modes in the new MIS-like model in the local rest frame of the fluid as a function of $|\textbf{k}|$. We have set $c_\mathcal{L}=1$, $c_\sigma=0$, $\frac{\eta}{s}= \frac{1}{4\pi}$, $\Omega_I = 2$ and $\Omega_R = 1$, and worked in $T_0=1$ units. The two nonhydrodynamic modes are associated to red curves, while the hydrodynamic mode is associated to black ones.}
\label{fig:new_model-shear}
\end{center}
\end{figure}

We performed a numerical scan of $\textrm{Im}~\omega_q$ in the three-dimensional parameter space spanned by $|\textbf{k}|$, $\alpha$ and $u_0$---with the remaining parameters as in the numerical simulation discussed in the main text---and did not encounter any point where this quantity became positive. The shear modes for our parameter choice, real $|\textbf{k}|$, and $u_0=0$ are plotted in Fig.~\ref{fig:new_model-shear}.

\subsection{Sound channel}

\noindent The sound channel modes are the roots of the following quartic polynomial in $k \cdot U_0$,
\begin{equation}
\begin{split}
&(- 3 (k\cdot U_0)^2 + k_\perp^2) (k^2  + 2 i \Omega_I T_0 (k \cdot U_0) + T_0^2 |\Omega|^2) \\
&+4 i \frac{\eta}{s} T_0 |\Omega|^2 (k\cdot U_0)k_\perp^2=0, 
\end{split}
\end{equation}
\begin{figure}[h]
\begin{center}
\includegraphics[width=\linewidth]{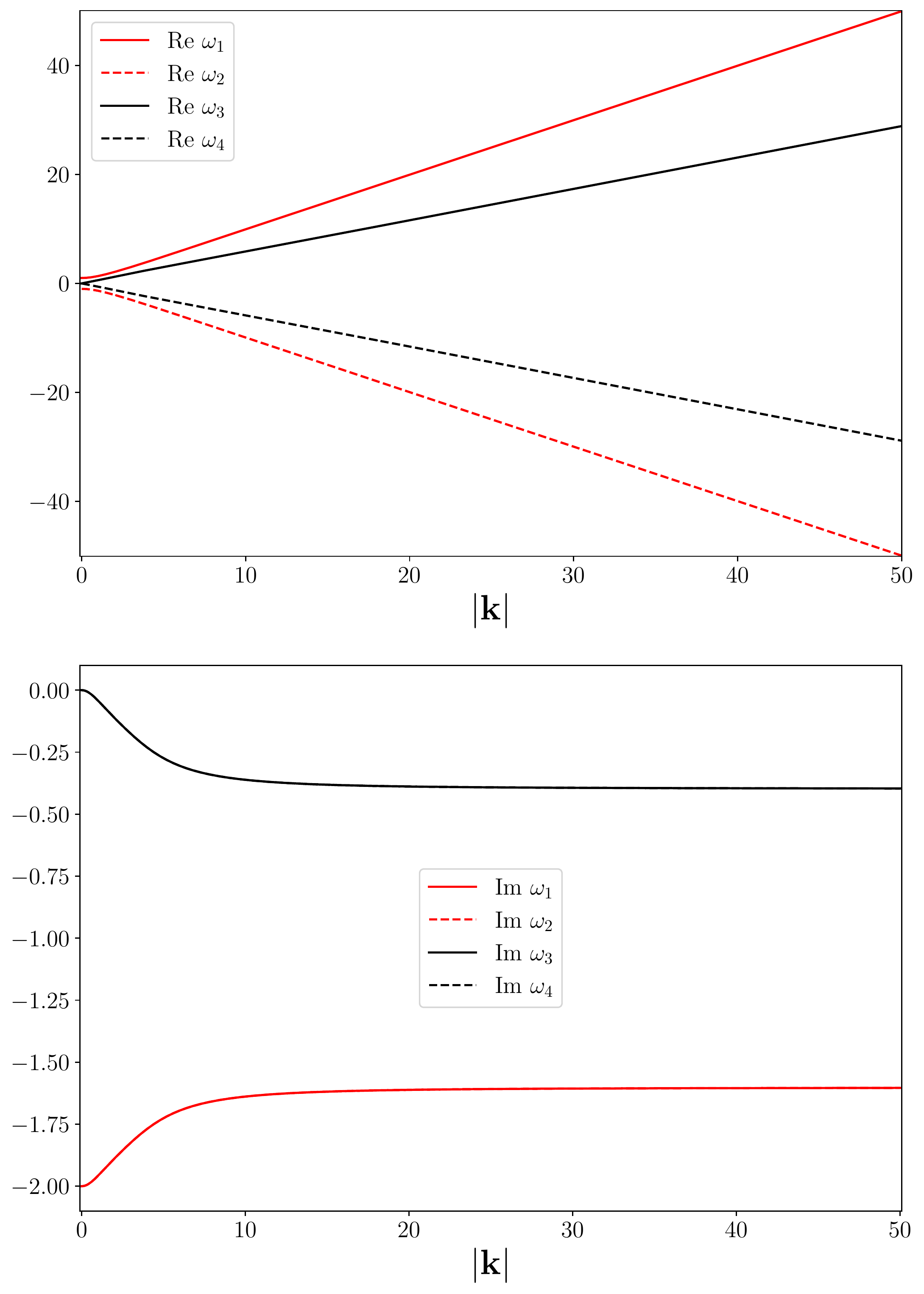}
\caption{Real (upper panel) and imaginary (bottom panel) parts of the sound channel modes in the new MIS-like model in the local rest frame of the fluid as a function of $|\textbf{k}|$. We have set $c_\mathcal{L}=1$, $c_\sigma=0$, $\frac{\eta}{s}= \frac{1}{4\pi}$, $\Omega_I = 2$ and $\Omega_R = 1$, and worked in $T_0=1$ units. The two nonhydrodynamic modes are associated to red curves, while the two hydrodynamic modes are associated to black ones.}
\label{fig:new_model-sound}
\end{center}
\end{figure}

\noindent and are divided into two hydrodynamic modes and two nonhydrodynamic ones. 
In the $|\textbf{k}|\to\infty$, $k_x = \alpha |\textbf{k}|$ limit with $|\alpha| \leq 1$, the sound channel modes behave as 
\begin{subequations}
\begin{equation}
\omega_1 = |\textbf{k}| - i T_0 \left(\Omega_I - \frac{\eta}{s}|\Omega|^2\right)(\cosh(u_0) - \alpha \sinh(u_0)) + \ldots,  
\end{equation}
\begin{equation}
\omega_2 = -|\textbf{k}| - i T_0 \left(\Omega_I - \frac{\eta}{s}|\Omega|^2\right)(\cosh(u_0) + \alpha \sinh(u_0)) + \ldots,
\end{equation}
\begin{equation}
\begin{split}
&\omega_3 = \frac{\alpha \sinh(2u_0){+}\beta}{2{+}\cosh(2u_0)} |\textbf{k}| \\
&{-}\frac{3 i T_0 \frac{\eta}{s}|\Omega|^2 \left(\cosh(u_0)\beta - \alpha\sinh(u_0)\right)}{(2{+}\cosh(2u_0))\beta} +\ldots,     
\end{split}
\end{equation}
\begin{equation}
\begin{split}
&\omega_4 = \frac{\alpha \sinh(2u_0){-}\beta}{2{+}\cosh(2u_0)} |\textbf{k}| \\
&- \frac{3 i T_0 \frac{\eta}{s}|\Omega|^2 \left(\cosh(u_0)\beta + \alpha\sinh(u_0)\right)}{(2{+}\cosh(2u_0))\beta}+\ldots, 
\end{split}
\end{equation}
\end{subequations}
where we have defined 
\begin{equation}
\beta = \sqrt{2{+}\alpha^2{+} (1{-}\alpha^2)\cosh(2u_0)}. 
\end{equation}
The modes $\omega_{1,2}$ are obviously causal. This is also the case for the modes $\omega_{3,4}$, since 
\begin{equation}
\Bigl|\frac{\alpha \sinh(2u_0){\pm}\beta}{2{+}\cosh(2u_0)}\Bigr| \leq 1
\end{equation}
for any $|\alpha|\leq 1$ and $u_0 \in \mathbb R$. 

Regarding asymptotic stability, the modes $\omega_{3,4}$ impose no new condition since 
\begin{equation}
\cosh(u_0)\beta\pm \alpha\sinh(u_0) \geq 0    
\end{equation}
for $|\alpha| \leq 1$ and $u_0 \in \mathbb R$ while the modes $\omega_{1,2}$ demand that 
\begin{equation}
\Omega_I - \frac{\eta}{s}|\Omega|^2 \geq 0.     
\end{equation}
Since $\Omega_I$ is a positive real number, the condition above is more stringent than the second entry in \eqref{shear_asymp_st} and overtakes it.

We performed a numerical scan analogous to the shear channel one and did not encounter any instances of $\textrm{Im}~\omega_q$ becoming positive. The sound modes for our parameter choice, real $|\textbf{k}|$, and $u_0=0$ are plotted in Fig.~\ref{fig:new_model-sound}.

\end{document}